\documentclass[aps,prb,onecolumn,showpacs,showkeys,footinbib,groupedaddress,eqsecnum]{revtex4-2}

\usepackage{bm}
\usepackage{amsmath}
\usepackage{amssymb}
\usepackage{graphicx}

\newcommand{\bvm}{\begin{verbatim}}
\newcommand{\evm}{\end{verbatim}}

\newcommand{\ocite}{\onlinecite}

\newcommand{\iy}{\infty}

\newcommand{\x}{\text}
\newcommand{\pd}{\partial}

\newcommand{\dg}{\dagger}

\newcommand{\ran}{\rangle}
\newcommand{\lt}{\left}
\newcommand{\rt}{\right}

\newcommand{\f}{\frac}
\newcommand{\tf}{\tfrac}

\newcommand{\sq}{\sqrt}
\newcommand{\lbl}{\label}

\newcommand{\cd}{\cdot}

\newcommand{\p}{\perp}
\newcommand{\n}{\nabla}

\newcommand{\nm}{\hat{0}}
\newcommand{\um}{\hat{1}}

\newcommand{\tm}{\times}

\newcommand{\eq}[1]{Eq.~(\ref{eq:#1})}
\newcommand{\eqs}[2]{Eqs.~(\ref{eq:#1}) and (\ref{eq:#2})}
\newcommand{\eqss}[3]{Eqs.~(\ref{eq:#1}), (\ref{eq:#2}), and (\ref{eq:#3})}
\newcommand{\eqsss}[4]{Eqs.~(\ref{eq:#1}), (\ref{eq:#2}), (\ref{eq:#3}), and (\ref{eq:#4})}
\newcommand{\eqsd}[2]{Eqs.~(\ref{eq:#1})-(\ref{eq:#2})}
\newcommand{\eqn}[1]{(\ref{eq:#1})}
\newcommand{\eqsn}[2]{(\ref{eq:#1}) and (\ref{eq:#2})}
\newcommand{\eqssn}[3]{(\ref{eq:#1}), (\ref{eq:#2}), and (\ref{eq:#3})}

\newcommand{\eqsdn}[2]{(\ref{eq:#1})-(\ref{eq:#2})}

\newcommand{\tabr}[1]{Tab.~\ref{tab:#1}}
\newcommand{\tabsr}[2]{Tabs.~\ref{tab:#1} and \ref{tab:#2}}

\newcommand{\secr}[1]{Sec.~\ref{sec:#1}}
\newcommand{\secsr}[2]{Secs.~\ref{sec:#1} and \ref{sec:#2}}
\newcommand{\secssr}[3]{Secs.~\ref{sec:#1}, \ref{sec:#2}, and \ref{sec:#3}}
\newcommand{\figr}[1]{Fig.~\ref{fig:#1}}
\newcommand{\figsr}[2]{Figs.~\ref{fig:#1} and \ref{fig:#2}}
\newcommand{\figsdr}[2]{Figs.~\ref{fig:#1}-\ref{fig:#2}}
\newcommand{\figssr}[3]{Figs.~\ref{fig:#1}, \ref{fig:#2}, and \ref{fig:#3}}

\newcommand{\appr}[1]{Appendix~\ref{app:#1}}

\newcommand{\spc}{\mbox{ }}

\newcommand{\beq}{\begin{equation}}
\newcommand{\eeq}{\end{equation}}
\newcommand{\beqar}{\begin{eqnarray}}
\newcommand{\eeqar}{\end{eqnarray}}
\newcommand{\beqarn}{\begin{eqnarray*}}
\newcommand{\eeqarn}{\end{eqnarray*}}
\newcommand{\ba}{\begin{array}}
\newcommand{\ea}{\end{array}}
\newcommand{\bwt}{\begin{widetext}}
\newcommand{\ewt}{\end{widetext}}

\newcommand{\rarr}{\rightarrow}
\newcommand{\larr}{\leftarrow}

\newcommand{\dx}{{\text d}}

\newcommand{\Ox}{{\text O}}

\newcommand{\ch}{\hat{c}}

\newcommand{\kh}{\hat{k}}

\newcommand{\ph}{\hat{p}}

\newcommand{\Ah}{\hat{A}}
\newcommand{\Bh}{\hat{B}}

\newcommand{\Hh}{\hat{H}}

\newcommand{\Jh}{\hat{J}}

\newcommand{\Mh}{\hat{M}}

\newcommand{\Sh}{\hat{S}}
\newcommand{\Th}{\hat{T}}
\newcommand{\Uh}{\hat{U}}

\newcommand{\Xh}{\hat{X}}

\newcommand{\psih}{\hat{\psi}}
\newcommand{\tauh}{\hat{\tau}}
\newcommand{\chih}{\hat{\chi}}

\newcommand{\Psih}{\hat{\Psi}}

\newcommand{\pr}{{\bar{p}}}

\newcommand{\Er}{{\bar{E}}}

\newcommand{\Ecr}{\bar{\mathcal{E}}}

\newcommand{\alr}{{\bar{\alpha}}}

\newcommand{\epsr}{{\bar{\varepsilon}}}

\newcommand{\Ec}{\mathcal{E}}

\newcommand{\Hc}{\mathcal{H}}

\newcommand{\Oc}{\mathcal{O}}

\newcommand{\Tc}{\mathcal{T}}

\newcommand{\Vc}{\mathcal{V}}

\newcommand{\Acr}{{\bar{\mathcal{A}}}}

\newcommand{\Vcr}{{\bar{\mathcal{V}}}}

\newcommand{\Hcth}{\hat{\tilde{\Hc}}}

\newcommand{\Hch}{\hat{\Hc}}

\newcommand{\sigb}{{\mbox{\boldmath{$\sigma$}}}}

\newcommand{\epst}{\tilde{\varepsilon}}

\newcommand{\nv}{{\bf 0}}

\newcommand{\kb}{{\bf k}}

\newcommand{\nb}{{\bf n}}
\newcommand{\pb}{{\bf p}}

\newcommand{\rb}{{\bf r}}

\newcommand{\vb}{{\bf v}}

\newcommand{\Bb}{{\bf B}}
\newcommand{\Cb}{{\bf C}}
\newcommand{\Db}{{\bf D}}

\newcommand{\Jb}{{\bf J}}

\newcommand{\Ob}{{\bf O}}

\newcommand{\Tb}{{\bf T}}

\newcommand{\pbr}{\bar{\bf p}}

\newcommand{\kbh}{\hat{\kb}}
\newcommand{\pbh}{\hat{\pb}}

\newcommand{\Jbh}{\hat{\Jb}}

\newcommand{\al}{\alpha}
\newcommand{\be}{\beta}
\newcommand{\ga}{\gamma}
\newcommand{\Ga}{\Gamma}
\newcommand{\de}{\delta}
\newcommand{\De}{\Delta}

\newcommand{\sig}{\sigma}
\newcommand{\Sig}{\Sigma}

\newcommand{\ka}{\varkappa}

\newcommand{\eps}{\varepsilon}
\newcommand{\e}{\epsilon}

\begin{document}
\title{Evolution of the surface states of the Luttinger semimetal under strain and inversion-symmetry breaking: Dirac, line-node, and Weyl semimetals}
\date{\today}

\author{Maxim Kharitonov$^{1,2}$}
\author{Julian-Benedikt Mayer$^1$}
\author{Ewelina M. Hankiewicz$^{1,3}$}
\affiliation{$^1$Institute for Theoretical Physics and Astrophysics,}
\affiliation{University of W\"urzburg, 97074 W\"urzburg, Germany}
\affiliation{$^2$Donostia International Physics Center (DIPC),
Manuel de Lardizabal 4, E-20018 San Sebastian, Spain}
\affiliation{$^3$W\"urzburg-Dresden Cluster of Excellence ct.qmat, Germany}

\begin{abstract}

The Luttinger model of a quadratic-node semimetal for electrons with the $j=\frac32$ angular momentum under cubic symmetry
is the parent, highest-symmetry low-energy model for a variety of topological and strongly correlated materials,
such as HgTe, $\alpha$-Sn, and iridate compounds.
Previously, we have theoretically demonstrated that the Luttinger semimetal exhibits surface states.
In the present work, we theoretically study the evolution of these surface states under symmetry-lowering perturbations:
compressive strain and bulk-inversion asymmetry (BIA).
This system is quite special in that each consecutive perturbation
creates a new type of a semimetal phase, resulting in a sequence of four semimetal phases,
where each successive phase arises by modification of the nodal structure of the previous phase:
under compressive strain, the Luttinger semimetal turns into a Dirac semimetal,
which under the linear-in-momentum BIA term turns into a line-node semimetal,
which under the cubic-in-momentum BIA terms turns into a Weyl semimetal.
We calculate the surface states within the generalized Luttinger model for these four semimetal phases
within a ``semi-analytical'' approach
and fully analyze the corresponding evolution of the surface states.
Importantly, for this sequence of four semimetal phases,
there is a corresponding hierarchy of the low-energy models describing the vicinities of the nodes.
We derive most of these models and demonstrate quantitative asymptotic agreement between the surface-state spectra
of some of them.
This proves that the mechanisms responsible for the surface states
are fully contained in the low-energy models within their validity ranges, once they are supplemented with proper boundary conditions,
and demonstrates that continuum models are perfectly applicable for studying surface states.

\end{abstract}


\maketitle

\begin{figure*}
\centering
\includegraphics[width=\linewidth]{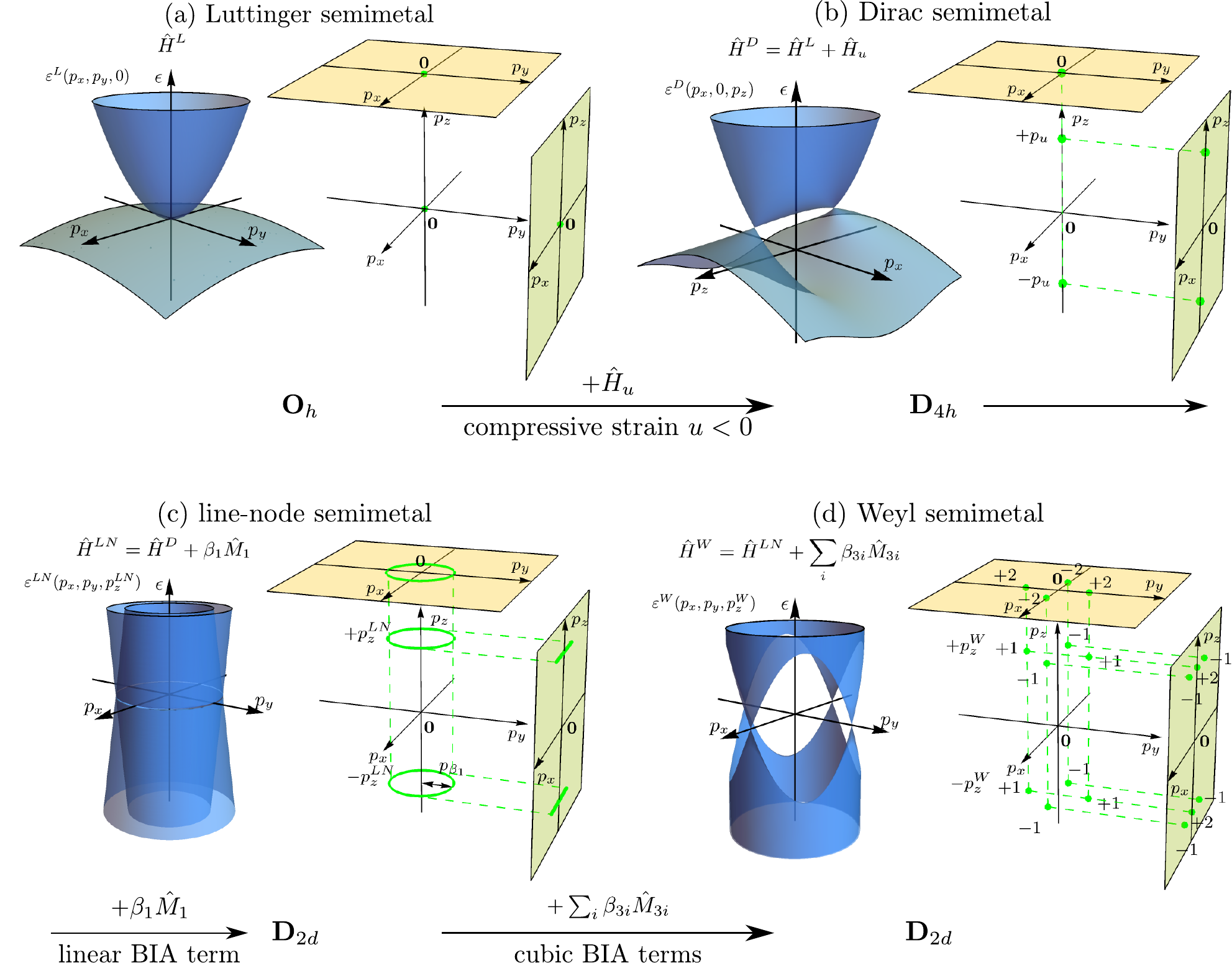}
\caption{
Bulk spectra of the four semimetal phases in the generalized Luttinger model
with compressive strain and bulk-inversion asymmetry (BIA) terms.
(a)~Unperturbed Luttinger semimetal [$\Hh^L(\pb)$, \eq{HL}] for cubic symmetry $\Ob_h$, in the absence of strain and BIA terms.
Double-degenerate bands $\eps_{\pm_b}^L(\pb)$ are given by \eq{eL}.
(b)~Dirac semimetal [$\Hh^D(\pb)$, \eq{HD}], obtained from the Luttinger semimetal by adding
compressive strain [$\Hh_u$, \eq{Hu}] along the $z$ direction. The symmetry is lowered to $\Db_{4h}$.
Double-degenerate bands $\eps_{\pm_b}^D(\pb)$ are given by \eq{eD}.
Compressive strain along the $z$ direction shifts the bands in energy for momentum $(0,0,p_z)$,
creating two Dirac points (green) at $p_z=\pm p_u$ [\eq{pu}].
(c)~Linear-node semimetal [$\Hh^{LN}(\pb)$, \eq{HLN}],
obtained by adding the linear BIA terms to the Dirac semimetal Hamiltonian.
Each Dirac point is transformed into a line node of an approximate ring shape.
The symmetry is lowered to $\Db_{2d}$.
(d)~Weyl semimetal, obtained by adding the cubic BIA terms to the line-node semimetal Hamiltonian [$\Hh^W(\pb)$, \eq{HW}].
The degeneracy of each line node is lifted everywhere expect for four Weyl points.
The zoom-in around $p_z=p_u$ is shown?
The Chern numbers $C^W=\pm 1$ of the Weyl points (green) are indicated.
The projections of the Dirac and Weyl points onto the surface-momentum planes are also shown.
The parameters of HgTe were used.
}
\label{fig:e}
\end{figure*}

\section{Introduction\lbl{sec:intro}}

A variety of materials termed Luttinger semimetals (LSMs) have been attracting a lot of attention
recently~\cite{Luttinger1956,Abrikosov1974,Balents,Herbut,Kondo2015,Ruan2016,Ruan2016-2,KharitonovLSM,Armitage2017,Mandal,
Goswami,Roy,Szabo,Cheng2017,Yao,Ghorashi2018,Ghorashi2019,Zhang2018,Mauri,Tchoumakov,Sim}.
The name is due to the model~\cite{Luttinger1956}, first derived by Luttinger, for the quartet of electron states with the angular momentum $j=\f32$
in the vicinity of the $\Gamma$ point $\pb=\nv$ in momentum space for a system with full cubic symmetry, with the point group $\Ob_h$.
This Luttinger model is the most general form of the single-particle Hamiltonian
to the lowest nonvanishing, quadratic order in momentum $\pb$, derived using the method of invariants.
The bulk spectrum of the Luttinger model consists of two double-degenerate quadratic bands;
at $\pb=\nv$, the quartet is unsplit (belongs to one spinful 4D irreducible representation of $\Ob_h$).
If the bands have curvatures of opposite signs, the model describes a quadratic-node semimetal [\figr{e}(a)]
and is called a Luttinger semimetal; and so is the material exhibiting such feature.
Since these requirements are quite general (only the presence of the $j=\f32$ states, $\Ob_h$ symmetry, and the range of the curvature parameters),
crystals with vastly different chemical composition and active electron states could be Luttinger semimetals.
Examples are $\al$-Sn with active $p$ orbitals and iridate compounds~\cite{Kondo2015}, like $\x{Pr}_2\x{Ir}_2\Ox_7$, with active $d$ orbitals.
Moreover, the symmetry $\Ob_h$ does not necessarily have to be exact, just dominant, for the system to manifest the properties of the Luttinger semimetal;
see below for the tetrahedral group $\Tb_d$, with HgTe being the prime example.

This special local feature of the band structure of the Luttinger semimetal leads to a variety
of physical effects~\cite{Luttinger1956,Abrikosov1974,Balents,Herbut,Kondo2015,Ruan2016,Ruan2016-2,KharitonovLSM,Armitage2017,Mandal,
Goswami,Roy,Szabo,Cheng2017,Yao,Ghorashi2018,Ghorashi2019,Zhang2018,Mauri,Tchoumakov,Sim,Scholz2018}.
Adding various single-particle symmetry-lowering perturbations modifies the quadratic node and introduces a wide range of nontrivial phases,
some of which are topological.
Also, as first shown by Abrikosov~\cite{Abrikosov1974},
electron interactions could lead to phases with spontaneously broken symmetries~\cite{Balents,Herbut,Goswami,Roy,Szabo,Tchoumakov,Sim}.
Some of the potential order parameters are such that the interaction-induced phase is also topological, at least at the mean-field level.
Interactions effects are expected to be strong in iridate compounds with active $d$ orbitals~\cite{Kondo2015}.

In particular, a number of other semimetal phases arise in the Luttinger semimetal
under compressive strain 
and the so-called bulk-inversion asymmetry (BIA) terms~\cite{Bir1974,Winkler2003,Ruan2016,Ruan2016-2}.
The latter are the terms in crystals with the tetrahedral point group $\Tb_d$
that describe the difference between the Hamiltonians with $\Tb_d$ and $\Ob_h$ point groups;
when small, it is instructive to consider them as perturbations. A prime example of such system is HgTe.
Under compressive strain, the Luttinger semimetal becomes a Dirac semimetal with a pair of linear nodes
in the still double-degenerate band structure
(for the opposite, tensile strain, the system is a 3D class-AII topological insulator~\cite{BrueneHgTe3DTI,BaumHgTe3DTI,BrueneHgTe3DTIb}).
Introducing only the lowest-order, linear-in-momentum BIA terms turns the Dirac semimetal into a line-node semimetal,
with each Dirac node transforming into an approximate ring line node.
This line-node semimetal is likely of accidental nature, as introducing the next-order, cubic-in-momentum BIA terms (which does not change the symmetry)
turns it into a Weyl semimetal, in which each line node is gapped out everywhere except at four Weyl points.
This evolution of the low-energy bulk spectrum is presented in \figr{e}.
The theoretical prediction of this Weyl-semimetal phase in the Luttinger model with such symmetry breaking terms has been made in Ref.~\ocite{Ruan2016}.

One of the primary interests in such topological semimetal phases are their surface states.
In Ref.~\ocite{Ruan2016}, the surface states were calculated numerically within the density-functional-theory (DFT) method.
On the other hand, the key transformations of the bulk spectrum due to perturbations arise
at low energies and are fully captured within the Luttinger model (\figr{e}).
Hence, a description of the surface states also within this minimal low-energy model would be desirable.
Indeed, such analysis is entirely possible, but the challenge is supplementing the bulk Hamiltonian
with proper boundary conditions (BCs) at the surface; determining possible BCs is a nontrivial and important problem.

Previously, we have derived~\cite{KharitonovLSM} the BCs for the Luttinger-semimetal model without perturbations
for the case when it arises as the low-energy limit
of the Kane model~\cite{Kane1957,Bir1974,Winkler2003} with hard-walls BCs, which have a clear physical interpretation.
The Kane model is a continuum model that, in addition to the $j=\f32$ quartet,
includes a doublet of $j=\f12$ states of opposite inversion parity;
it is commonly used for HgTe, $\al$-Sn, and similar materials.
And we have shown~\cite{KharitonovLSM} that for such BCs, the Luttinger semimetal model without any additional perturbations,
in fact, exhibits surface states.
In this work, we expand this study further, by theoretically exploring the evolution of these surface states
of the Luttinger semimetal under compressive strain and bulk-inversion asymmetry, which create the Dirac, line-node, and Weyl semimetals.

Besides the immediate practical goal of characterizing these surface-state structures,
this work also has a methodological goal of demonstrating the applicability
and advantages of employing minimal low-energy continuum models supplemented with proper BCs for that purpose.
Within such models, the surface states can be calculated
using the {\em ``semi-analytical''} approach, in contrast to the common numerical calculations for large finite-size systems,
for tight-binding lattice or ab-initio (DFT) models.
The surface states are calculated for a true half-infinite system and require only modest computational resources:
the computational complexity is determined by the number of degrees of freedom of the continuum model,
given by the product of the number of wave-function components and momentum order of the Hamiltonian.
For many simpler models, surface states can be found entirely analytically.
As a result, essentially arbitrary desired accuracy can be achieved,
which is important to resolve fine features of the surface states
(which will be washed out in finite-size calculations, when spatial quantization exceeds their scale),
such as the vicinity of the nodes of the bulk spectrum.
Perhaps most importantly, this analytical simplicity of the approach
allows one to clearly identify mechanisms and scales responsible for the surface states.
Such understanding can hardly be gained from more microscopic models,
like tight-binding or ab-initio, which can usually be analyzed only numerically.

For the system in question, the main general picture we demonstrate is that the hierarchy of the perturbation scales
(strain, linear BIA term, and cubic BIA terms) defines a hierarchy of successively smaller momentum and energy regions,
where each consecutive perturbation modifies both the bulk and surface-state spectrum of the previous semimetal phase in a qualitative way,
by creating a new nodal fine structure: from the Luttinger to the Dirac to the line-node to the Weyl semimetal.
Whereas outside of each such region, the spectrum retains the structure of the previous semimetal phase,
and there are crossovers between the behaviors in larger and smaller regions.
Since the Luttinger semimetal exhibits surface states already without any perturbations,
these modifications at low energy scales can be regarded as the evolution of its surface states.

We demonstrate that within each successive smaller momentum and energy region a corresponding low-energy model exists,
both the bulk Hamiltonian and BCs of which can be derived using a systematic low-energy-expansion procedure.
Each such low-energy model will be simpler than the previous one, but will still fully capture the surface states in that region.
Most importantly, within this validity range, there will be a quantitative asymptotic agreement between the surface states of the models.
This proves that low-energy models supplemented with proper BCs are perfectly suitable for the study of surface states.

We demonstrate this agreement in the vicinity of the quadratic node of $j=\f32$ states
between the surface states obtained from the Kane and Luttinger models,
when the latter arises as the low-energy limit of the former.
This also provides the physical insight that the mechanisms responsible for the surface states are already contained in the low-energy Luttinger model.
In the same spirit, we also derive and explore the linear-in-momentum low-energy model
in the vicinity of the Dirac points
and demonstrate the asymptotic quantitative agreement between the surface states obtained from this model and the Luttinger model.

We note that, although in this work we use the Kane model as the larger, ``more microscopic'' model from which the Luttinger model originates,
which is applicable to materials like HgTe and $\al$-Sn, the results we obtain for the surface states of the Luttinger model
may be applicable to other Luttinger semimetal materials (such as iridates) that are not necessarily described by the Kane model,
as long as the used BCs apply to them.

The rest of the paper is organized as follows.
In \secr{LM}, we present the Hamiltonian of the generalized Luttinger model in the presence of strain and BIA terms.
In \secr{bulk}, we discuss the bulk properties of the four semimetal phases.
We also derive the Hamiltonians of several low-energy linear-in-momentum models around the nodes of these phases.
In \secr{KM}, we present the relation between the Kane and Luttinger models, when the latter arises as the low-energy limit of the former.
We present a systematic low-energy expansion procedure of deriving the Luttinger model from the Kane models.
In particular, in \secr{bc}, we present the BCs for the Luttinger model, previously derived in Ref.~\ocite{KharitonovLSM},
that originate from the hard-wall BCs of the Kane model, and discuss their status in the presence of strain and BIA terms.
In \secr{scales}, we discuss in detail the hierarchy structure of the four successive semimetal phases,
their momentum and energy scales, and their low-energy models.
In \secr{method}, we outline the general semi-analytical method of calculating surface states for continuum models with BCs.
In \secr{ssL}, we present the surface states for the unperturbed Luttinger semimetal
obtained wihtin the Luttinger model, and demonstrate quantitative asymptotic agreement between them and those of the Kane model.
In \secr{ssD}, we calculate the surface states for the Dirac semimetal within the Luttinger model.
In \secr{ssDlin}, we calculate the surface states for the Dirac semimetal within the linear-in-momentum model.
In \secr{ssLN}, we calculate the surface states for the line-node semimetal.
In \secr{ssW}, we calculate the surface states for the Weyl semimetal.
In \secr{WLMKM}, we compare the surface states in the Weyl-semimetal phase calculated within the Luttinger and Kane models
for different strengths of compressive strain.
Concluding remarks are presented in \secr{conclusion}.

\section{Generalized Luttinger model for $j=\f32$ states \lbl{sec:LM}}

In this section, we present the continuum low-energy model for $j=\f32$ states in the vicinity
of the $\Ga$ point, for materials with the cubic $\Ob_h$ and tetrahedral $\Tb_d$  point groups.
The wave function
\beq
    \psih(\rb)=\lt(\ba{c} \psi_{+\f32}(\rb) \\ \psi_{+\f12}(\rb) \\ \psi_{-\f12}(\rb) \\ \psi_{-\f32}(\rb) \ea\rt)
\lbl{eq:psi}
\eeq
is the coordinate-dependent four-component spinor,
where the subscript denotes the $j_z$ projections of the $j=\f32$ angular momentum on the $z$ axis and $\rb=(x,y,z)$ is the radius vector.

We present the most general form of the Hamiltonian for $j=\f32$ states, which can be derived using the method of invariants;
in fact, all possible terms have already been found in previous works~\cite{Luttinger1956,Winkler2003}.
In this approach, for a given symmetry, determined by the spatial point group and time-reversal symmetry $\Tc_-$,
one constructs all possible basis matrix functions of momentum that are invariant under that symmetry.
The most general form of the Hamiltonian is then an arbitrary linear combination of these basis functions.
The values of the coefficients are not fixed at all by this symmetry approach.
For a model of a real material, they take on specific values.
In a more general theoretical study, these coefficients may be considered as free parameters of a family of Hamiltonians.

In addition to the spatial symmetries considered below,
we will always also assume spinful time-reversal symmetry $\Tc_-$, with $\Tc_-^2=-\um$, throughout the paper.
Although for brevity $\Tc_-$ will not always be mentioned explicitly, it should be kept in mind that the invariants
allowed in the general Hamiltonians below are also restricted by $\Tc_-$.

It is particularly instructive to consider the following symmetry hierarchy, represented as a chain
\beq
    \Ob(3) \supset \Ob_h \supset \Tb_d
\lbl{eq:symm}
\eeq
of subgroups, and to consider the Hamiltonians according to it.
This way, if $\Hh^{\Ob(3)}(\pbh)$ is the most general form of the Hamiltonian for the spherical symmetry group $\Ob(3)$,
then the most general form of the Hamiltonian for $\Ob_h$ can be presented as
\beq
    \Hh^{\Ob_h}(\pbh)
    =\Hh^{\Ob(3)}(\pbh)+\de\Hh^{\Ob_h\subset\Ob(3)}(\pbh),
\lbl{eq:HOh}
\eeq
where $\de\Hh^{\Ob_h\subset\Ob(3)}(\pbh)$ is the linear combination of all invariants of $\Ob_h$ that do not contain invariants of $\Ob(3)$.
Here,
\[
    \pbh=(\ph_x,\ph_y,\ph_z)=-i(\pd_x,\pd_y,\pd_z)
\]
is the momentum operator (throughout, we use the units in which the Planck constant $\hbar=1$).
Similarly, the most general form of the Hamiltonian for $\Tb_d$ reads
\beq
    \Hh^{\Tb_d}(\pbh)=\Hh^{\Ob_h}(\pbh)+\de\Hh^{\Tb_d\subset\Ob_h}(\pbh),
\lbl{eq:HTd}
\eeq
where $\de\Hh^{\Tb_d\subset\Ob_h}(\pbh)$ is the linear combination of all invariants of $\Tb_d$ that do not contain invariants of $\Ob_h$.
It is quite common in real materials that these additional symmetry-lowering terms $\de\Hh^{\ldots}(\pbh)$
are smaller in magnitude, while the higher-symmetry terms are dominant; in such scenario, this symmetry hierarchy becomes particularly useful.

We derive the most general forms of the Hamiltonian for these symmetries up to the cubic order in momentum;
the need for including the cubic terms is explained in \secr{bulkW}.

For full spherical symmetry $\Ob(3)$, the most general form of the Hamiltonian up to cubic order for $j=\f32$ states reads
\beq
    \Hh^{\Ob(3)}(\pbh)=\al_0\um_4\pbh^2 + \al_z\Mh_z(\pbh).
\lbl{eq:HO3}
\eeq
It is a linear combination of two invariants, $\um_4\pbh^2$ and
\beq
    \Mh_z(\pbh)=\f54\um_4\pbh^2-(\Jbh\cdot \pbh)^2,
\lbl{eq:Mz}
\eeq
with arbitrary coefficients $\al_{0,z}$.
Throughout, $\um_n$ is the unit matrix of order $n$; $\Jbh=(\Jh_x,\Jh_y,\Jh_z)$ are the $j=\f32$ angular-momentum matrices.
Because both $\Ob(3)$ and $\Ob_h$ point groups include spatial inversion $\rb\rarr-\rb$, only even powers of momentum are allowed in the Hamiltonian
within the multiplet of $j=\f32$ states of the same inversion parity.

Next, upon lowering the symmetry down to cubic, $\Ob(3)\rarr\Ob_h$, only one additional term is allowed in the Hamiltonian \eq{HOh}:
\beq
    \de\Hh^{\Ob_h\subset\Ob(3)}=\al_\square \Mh_\square(\pbh),
\lbl{eq:dHOh}
\eeq
where
\beq
    \Mh_\square(\pbh)
        =\Jh_x^2\ph_x^2+\Jh_y^2\ph_y^2+\Jh_z^2\ph_z^2
            -\f25(\Jbh\cdot \pbh)^2-\f34\um_4\pbh^2.
\lbl{eq:Msq}
\eeq
This invariant of $\Ob_h$ belong to an angular-momentum-4 irreducible representation of $\Ob(3)$,
and hence does not contain the invariants of $\Ob(3)$, which are the angular-momentum-$0$ irreducible representations;
the last two terms, invariants of $\Ob(3)$, have been introduced precisely
to remove the angular-momentum-0 component present in the combination $\Jh_x^2\ph_x^2+\Jh_y^2\ph_y^2+\Jh_z^2\ph_z^2$.
We refer to the term \eqn{HO3} and the coefficient $\al_\square$ as {\em cubic anisotropy}.

The Hamiltonians \eqssn{HOh}{HO3}{dHOh} are known as the Luttinger model~\cite{Luttinger1956}.

When lowering the symmetry further from cubic to tetrahedral, $\Ob_h\rarr\Tb_d$,
the additional, odd-in-momentum terms allowed in the Hamiltonian \eqn{HTd} are
\beq
    \de\Hh^{\Tb_d\subset\Ob_h}(\pbh)=\be_1\Mh_1(\pbh)+\sum_{i=1,2,3,4} \be_{3i}\Mh_{3i}(\pbh).
\lbl{eq:dHTd}
\eeq
There is one linear-in-momentum
\beq
    \Mh_1(\pbh)
    =\{\Jh_x,\Jh_y^2-\Jh_z^2\}\ph_x+\text{c.p.}
\lbl{eq:M1}
\eeq
and four cubic-in-momentum
\begin{align}
	\Mh_{31}(\pbh)&=\ph_x(\ph_y^2-\ph_z^2)\Jh_x+\x{c.p.}, \lbl{eq:M31}\\
	\Mh_{32}(\pbh)&=\ph_x(\ph_x^2-\ph_z^2)\Jh_x^3+\x{c.p.}, \lbl{eq:M32}\\
	\Mh_{33}(\pbh)&=\ph_x(\ph_y^2+\ph_z^2)\{\Jh_x,\Jh_y^2-\Jh_z^2\}+\x{c.p.}, \lbl{eq:M33}\\
	\Mh_{34}(\pbh)&=\ph_x^3\{\Jh_x,\Jh_y^2-\Jh_z^2\}+\x{c.p.} \lbl{eq:M34}
\end{align}
invariants of $\Tb_d$ that do not contain invariants of $\Ob_h$ (and there are no additional constant or quadratic-in-momentum terms).
Here $\{\Ah,\Bh\}=\Ah\Bh+\Bh\Ah$ is the anticommutator and ``c.p.'' denotes additional terms obtained by two possible cyclic permutations
of the $x,y,z$ indices in the presented terms.
These invariants of $\Tb_d$ that do not contain invariants of $\Ob_h$
are often referred to as {\em bulk-inversion-asymmetry (BIA)} terms~\cite{Winkler2003}.

We note that, of course, in real materials, the BIA terms cannot be tuned:
in $\Ob_h$ materials, they are absent, and in $\Tb_d$ materials, their parameters $\be_1$, $\be_{3i}$ have fixed values.
Nonetheless, theoretically, it is instructive to consider them as tunable parameters,
to determine which terms are responsible for which features in the bulk and surface-state spectra.
Also, in $\Tb_d$ materials, if the magnitude of these terms is smaller
than the energy scales of interest (or, e.g., limited by the available resolution in experiments),
it would be justified to treat such material as having $\Ob_h$ symmetry.

Finally, we also include the effect of strain~\cite{Bir1974} along the $z$ direction
(for $x$ or $y$ strain directions, the results would be equivalent by symmetry).
We take into account only the dominant, constant term
\beq
    \Hh_u=-u(\Jh_z^2-\tf54\um_4).
\lbl{eq:Hu}
\eeq
Clearly, its effect on the electron states
amounts to introducing the energy difference between the $j_z=\pm\f32$ and $j_z=\pm\f12$ pairs of states.
The strength of the strain parameter $u$ indicates the change of the lattice constant under deformation,
while the sign of $u$ determines whether the lattice is stretched (tensile strain) or compressed (compressive strain),
which for the assumed $\al_z>0$ (see below), corresponds to $u>0$ and $u<0$, respectively.

Note that for both $\Ob_h$ and $\Tb_d$ point groups, the $j=\f32$ quartet remains unsplit at the $\Ga$ point $\pb=\nv$,
i.e., it belongs to one spinful 4D irreducible representation of the respective groups. This is, however, violated by strain.

We will refer to all the Hamiltonians below, containing various combinations of the above terms, as the {\em generalized Luttinger model}.

\section{Semimetal phases, bulk spectrum \lbl{sec:bulk}}

\begin{table*}
\begin{tabular}{ c | c | c | c | c || c | c}
	& $\al_0$ & $\al_z$ & $\al_\square$ & $\be_1$ & $\be_{31}$ & $\be_{32},\be_{33},\be_{34}$ \\ \hline \hline
HgTe & $16.58\f1{2m_e}$ & $18.72\f1{2m_e}$ & $-1.6\f1{2m_e}$ & $-4.31~\x{meV}\,\x{nm}$ &
	$140.36~\x{meV}\,\x{nm}^3$ & $0$
\end{tabular}
\caption{
The values of the parameters used for the generalized Luttinger model [\eqsss{HL}{HD}{HLN}{HW}]
of HgTe, with the linear and cubic bulk-inversion-asymmetry (BIA) terms included. Here, $m_e$ is the electron mass.
The curvature $\al_{0,z,\square}$ and linear BIA $\be_1$ parameters
are calculated from the parameters of the Kane model (\tabr{KM}) via the folding procedure presented in \secr{KM}.
The exact values of the cubic BIA parameters $\be_{3i}$, $i=1,2,3,4$, of HgTe are, on the one hand, not well-documented
and, on the other hand, not important for the studied system,
since at low energies their effect reduces to just one: opening of the gap [\eq{DeW}] around the line node (\secr{bulkW}).
For these reasons, the values of $\be_{3i}$ were {\em chosen arbitrarily}.
}
\label{tab:LM}
\end{table*}

The focus of this work are the surface states of the various semimetal phases
that arise from the Hamiltonians for $j=\f32$ states presented above.
In this section, we describe the properties of their bulk band structures.

\subsection{Luttinger semimetal for $\Ob_h$ and $\Ob(3)$ \lbl{sec:bulkL}}

First, consider the unperturbed Luttinger Hamiltonian [\eqss{HOh}{HO3}{dHOh}]
\beq
    \Hh^L(\pbh)=\Hh^{\Ob_h}(\pbh)
\lbl{eq:HL}
\eeq
for $\Ob_h$ symmetry, without strain or BIA terms.
For full spherical symmetry $\Ob(3)$, when $\al_\square=0$, the bulk spectrum of $\Hh^{\Ob(3)}(\pb)$ [\eq{HO3}] can be found as follows.
For momentum
\[
    \pb=(p_x,p_y,p_z)=p\nb,\spc p=|\pb|,
\]
of any direction, characterized by the unit vector $\nb$,
the Hamiltonian $\Hh^{\Ob(3)}(\pb)$ [\eq{HO3}] has axial rotation symmetry about this direction.
Hence, it is diagonal in the basis of the $j=\f32$ states with definite projections $j_\nb$ of the angular momentum on this direction,
which are the eigenstates of the matrix $\Jh_\nb=(\Jbh\cd\nb)$ with the eigenvalues $j_\nb=+\f32,+\f12,-\f12,-\f32$.
This diagonal Hamiltonian can be immediately recognized as $\Hh^{\Ob(3)}(0,0,p_z)$ in the original basis of states with definite $j_z$.
Further, the dispersion relations are the same for the $\pm j_\nb$ states with opposite projections.
So, there is a double-degenerate (due to inversion and $\Tc_-$) band
\beq
    \eps^{L,\Ob(3)}_{|j_\nb|=\f12}(p)=\al_{+_b} p^2
\lbl{eq:eL12}
\eeq
of the $j_\nb=\pm\f12$ states and a double-degenerate band
\beq
    \eps^{L,\Ob(3)}_{|j_\nb|=\f32}(p)=\al_{-_b} p^2
\lbl{eq:eL32}
\eeq
of the $j_\nb=\pm\f32$ states, where
\beq
    \al_{\pm_b}=\al_0 \pm_b \al_z
\lbl{eq:alpm}
\eeq
are their curvatures. Throughout, we introduce the subscript $b$ for the signs $\pm_b$
to denote the upper (conduction) and lower (valence) {\em bulk} bands, to distinguish these from multiple other signs that will appear.

When these curvatures $\al_{\pm_b}$ have opposite signs, the system is a quadratic-node semimetal, referred to as the {\em Luttinger semimetal}.
For absent $\al_0=0$, the spectrum has particle-hole symmetry; hence, finite $\al_0$ describes particle-hole asymmetry.
Throughout the paper, we assume
\[
    \al_z>0
\]
to be positive and that the system is in the Luttinger semimetal regime, so that
\beq
    -\al_z<\al_0<\al_z.
\lbl{eq:LSMregime}
\eeq
In this case, the $j_\nb=\pm\f12$ states are have a particle-like quadratic dispersion with the positive curvature $\al_{+_b}>0$
and the $j_\nb=\pm\f32$ states have a hole-like quadratic dispersion with the negative curvature $\al_{-_b}<0$.
This is the case, e.g., for HgTe (\tabr{LM}) and $\al$-Sn~\cite{Madelung2004}.
The case of negative $\al_z<0$ could be related by considering the Hamiltonian $-\Hh(\pbh)$.

For cubic symmetry, the spectrum can also be found analytically and reads
\begin{widetext}
\begin{align}
    \eps_{\pm_b}^L(\pb)=\al_0\pb^2\pm_b
        \sq{(\al_z-\tf{3}{5}\al_\square)^2\pb^4
        +3\al_\square(2\al_z-\tf{1}{5}\al_\square)(p_x^2p_y^2+p_x^2p_z^2+p_y^2p_z^2)}.
\lbl{eq:eL}
\end{align}
\end{widetext}
In the presence of the cubic anisotropy $\al_\square$, the spectrum becomes anisotropic,
but so long as $\al_{\pm_b}\gtrless 0$ remain dominant over $\al_\square$,
the system remains a quadratic-node {\em Luttinger semimetal}.

Further on, by $\Hh^L(\pbh)$ [\eq{HL}] we denote the Hamiltonian of the Luttinger model 
specifically in the Luttinger semimetal regime.

\subsection{Dirac semimetal for an $\Ob_h$ system under strain \lbl{sec:bulkD}}

Consider now the $\Ob_h$-symmetric Luttinger-semimetal Hamiltonian \eqn{HL} with the added strain \eqn{Hu}:
\beq
    \Hh^D(\pbh)=\Hh^{\Ob_h}(\pbh)+\Hh_u.
\lbl{eq:HD}
\eeq
Since the $\Hh^u$ term has $\Db_{\iy h}$ spatial symmetry,
the Hamiltonian $\Hh^D(\pbh)$ has the spatial symmetry with the point group
\[
    \Db_{4h}=\Ob_h\cap\Db_{\iy h}.
\]

For $\pb=(0,0,p_z)$, there are two double-degenerate bands
\[
    \eps^D_{|j_z|=\f12}(0,0,p_z)
        =[\al_0+(\al_z-\tf{3}{5}\al_\square)]p_z^2+u,
\]
\[
    \eps^D_{|j_z|=\f32}(0,0,p_z)
        =[\al_0-(\al_z-\tf{3}{5}\al_\square)]p_z^2-u.
\]
For this momentum direction, strain shifts these bands in opposite directions.
As a result, for compressive strain with $u<0$ (and for $\al_z>0$), the two bands cross at two momenta $p_z=\pm_u p_u$, with
\beq
	p_u=\sq{\f{|u|}{\al_z-\f35\al_\square}},
\lbl{eq:pu}
\eeq
and energy
\beq
    \eps_u=\al_0 p_u^2 =\f{\al_0}{\al_z-\f35\al_\square}|u|.
\lbl{eq:epsu}
\eeq
Throughout the rest of the paper, we assume compressive strain with $u<0$.
The bulk spectrum of this Dirac semimetal can also be found analytically for any momentum and consists of two double-degenerate bands
\begin{widetext}
\beq
    \eps^D_{\pm_b}(\pb)=\al_0 p^2 \pm_b
        \sq{(\al_z-\tf{3}{5}\al_\square)^2 p^4
            +3\al_\square(2\al_z-\tf{1}{5}\al_\square)(p_x^2p_y^2+p_\p^2p_z^2)
            +(\al_z-\tf{3}{5}\al_\square)(2p_z^2-p_\p^2)u+u^2},
\lbl{eq:eD}
\eeq
\end{widetext}
where $p_\p^2=p_x^2+p_y^2$.

Expansion about these points $\pb=(0,0,\pm_u p_u)$ shows (see also \secr{bulkDlin} below)
the linear dispersion of the two double-degenerate bands around them.
Semimetals with such linear nodes in systems with inversion and time-reversal $\Tc_-$ symmetries
are called {\em Dirac semimetals} and these points are called Dirac points.
Here, the Dirac points are protected by $\Db_{4h}$ spatial symmetry.
Hence, the Luttinger semimetal (with $\al_z>0$) under compressive strain turns into a Dirac semimetal.
We introduce the subscript $u$ for the signs $\pm_u$ pertaining to the two Dirac points,
to distinguish them from multiple other signs (such as $\pm_b$).

\subsubsection{Linear model of the Dirac semimetal \lbl{sec:bulkDlin}}

For the Dirac semimetal above [\eq{HD}], we also derive the low-energy model
that describes the system in the vicinity of the Dirac points $(0,0,\pm p_u)$ [\eq{pu}], linear in the momentum deviations $\kb$ from them,
\beq
    \pb=(0,0,\pm_u p_u)+\kb.
\lbl{eq:k}
\eeq
Following the standard procedure of the $k\cd p$ expansion~\cite{Bir1974,Winkler2003}, in the vicinity of the Dirac points,
the wave function of the Luttinger model may be presented as
\begin{align}
    \psih(\rb)=e^{+_u ip_uz}\Psih_{+_u}(\rb)+e^{-_u i p_uz}\Psih_{-_u}(\rb),
\lbl{eq:psiPsiD}
\end{align}
where $\Psih_{\pm_u}(\rb)$ are the envelope functions varying over spatial scales much larger than $1/p_u$.
Joined together, they form the wave function
\beq
    \Psih(\rb)=\lt(\ba{c} \Psih_{+_u}(\rb) \\ \Psih_{-_u}(\rb)\ea\rt),
\lbl{eq:PsiD}
\eeq
and of the low-energy model.
The Hamiltonian for it has the block-diagonal structure due to translation symmetry,
\begin{align}	
	\Hch^D(\kbh)=&
	\lt(\ba{cc}
		\Hch^D_{+_u}(\kbh) & \hat{0} \\
		\hat{0} & \Hch^D_{-_u}(\kbh)\ea\rt),
\lbl{eq:HcD}
\end{align}
where
\[
    \kbh=(\kh_x,\kh_y,\kh_z)=-i(\pd_x,\pd_y,\pd_z)
\]
is the momentum operator for $\Psih(\rb)$.

The linear expansion of \eq{HD} about the Dirac points gives
\beq
    \Hch^D_{\pm_u}(\kbh)=
	\lt(\ba{cc}
		\Hch^D_{\pm_u,+_{j_z}}(\kbh) & \nm \\
		\nm & \Hch^D_{\pm_u,-_{j_z}}(\kbh) \ea\rt)
	=\um_4\eps_u
	\pm_u\lt(\ba{cccc}
		(v_0-v_z)\kh_z& -v_\p \kh_-&0&0 \\
		-v_\p\kh_+&(v_0+v_z)\kh_z&0&0 \\
		0&0&(v_0+v_z)\kh_z&v_\p\kh_- \\
		0&0&v_\p\kh_+&(v_0-v_z)\kh_z \ea\rt),
\lbl{eq:HcDu}
\eeq
where $\kh_\pm=\kh_x \pm i \kh_y$ and
\beq
    v_0=\al_0 2p_u,\spc v_z= (\al_z-\tf35\al_\square)2p_u , \spc v_\p=\tf{\sq3}2 (\al_z+\tf25\al_\square) 2p_u.
\lbl{eq:v}
\eeq
We see that the pairs of $j_z=+\f32,+\f12$ and $j_z=-\f12,-\f32$ states, to be denoted as $\pm_{j_z}$, respectively,
are decoupled in the linear Hamiltonian \eqn{HcDu}.
We also note that even in the presence of cubic anisotropy $\al_\square\neq 0$
the linear model has emergent axial rotation symmetry about the $z$ direction,
since the anisotropic terms $\Jh_x^2\ph_x^2+\Jh_y^2\ph_y^2$ in $\Hh^D(\pbh)$ [\eq{HD}] are quadratic in $\kh_{x,y}=\ph_{x,y}$ and have to be discarded,
and the only linear-in-$\kbh$ contributions in $\Mh_\square(\pbh)$ actually come only from the $\Ob(3)$-symmetric terms and give $\propto\kh_z$ terms.

For each Dirac point $\pm_u$, the bulk spectrum of $\Hch^D_{\pm_u}(\kb)$ consists of the two double-degenerate bands
\beq
	\eps^{D,\x{lin}}_{\pm_u,\pm_b}(\kb)=
	\eps_u \pm_u v_0k_z\pm_b \sq{v_\p^2 k_\p^2+v_z^2k_z^2}
\lbl{eq:eDlin}
\eeq
with the linear dispersion, where
\[
    \kb=(k_\p\cos\phi,k_\p\sin\phi,k_z).
\]
The double-degeneracy comes from the two decoupled blocks $\pm_{j_z}$ in \eq{HcDu} with the same spectrum.

The linear model is valid in the vicinity of the Dirac points when $|\kb|\ll p_u$ and $|\e-\eps_u|\ll \al_{\pm_b} p_u^2$.

\subsection{Line-node semimetal for a $\Tb_d$ system under strain \lbl{sec:bulkLN}}

\begin{figure}
\includegraphics[width=.47\linewidth]{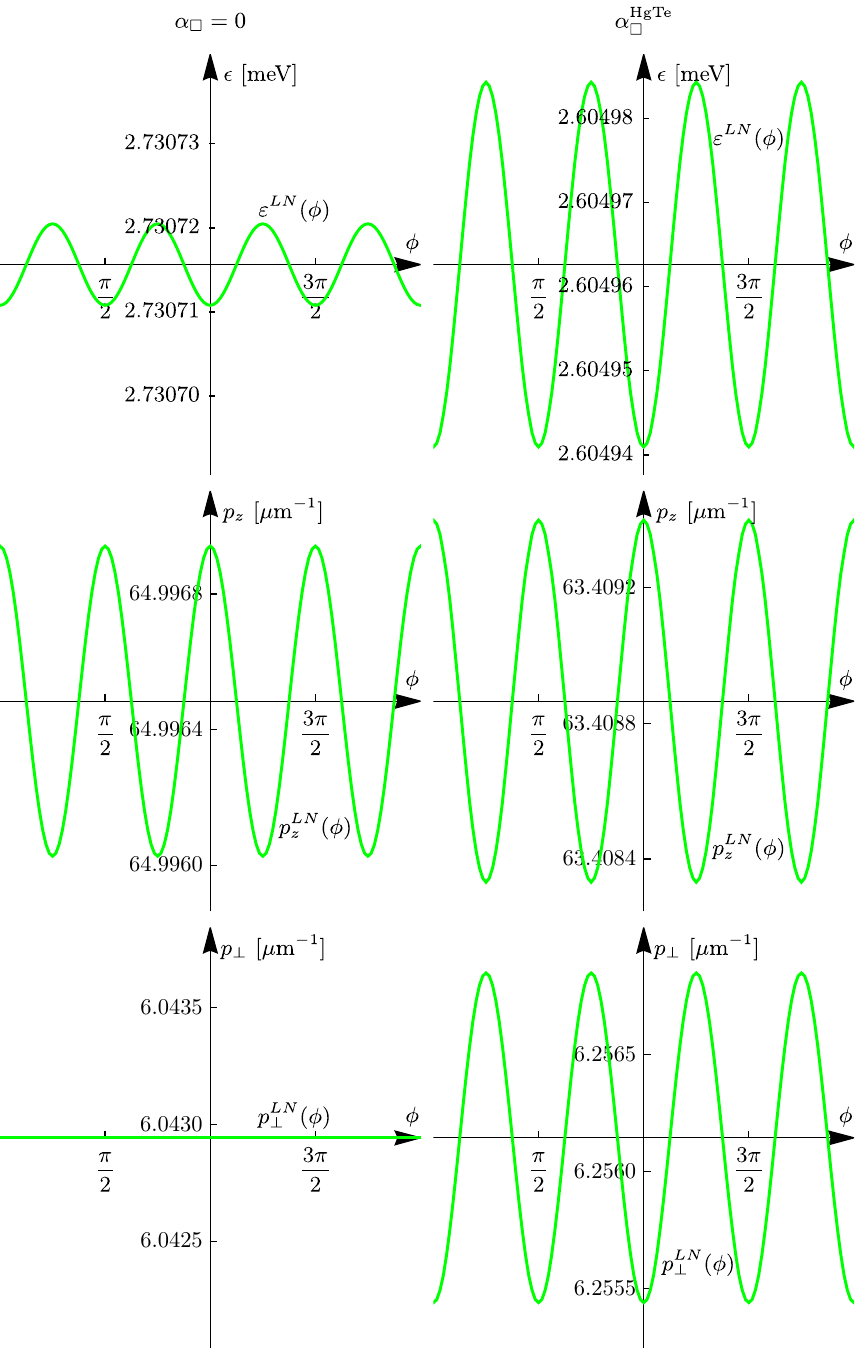}
\caption{
The dependence of the energy $\eps^{LN}(\phi)$, and momenta $p_z^{LN}(\phi)$ and $p_\p^{LN}(\phi)$ of the line node on the polar angle $\phi$
of momentum $\pb=(p_\p\cos\phi,p_\p\sin\phi,p_z)$ for the parameters of HgTe (\tabr{LM}) and the value $u=-3\x{meV}$ of compressive strain.
}
\label{fig:LN}
\end{figure}

Consider now adding strain [\eq{Hu}] to the system with $\Tb_d$ symmetry; the symmetry is lowered down to
\[
    \Db_{2d}=\Tb_d\cap\Db_{\iy h}.
\]
It is instructive to approach such system by starting with the Dirac semimetal [\eq{HD}]
with already present strain and adding the BIA terms [\eqsd{dHTd}{M34}].
Let us first include only the linear BIA term \eqn{M1},
\beq
    \Hh^{LN}(\pbh)=\Hh^{\Ob_h}(\pbh)+\Hh_u+\be_1\Mh_1(\pbh)=\Hh^D(\pbh)+\be_1\Mh_1(\pbh).
\lbl{eq:HLN}
\eeq
This generally lifts the overall double degeneracy of the bulk bands of the Dirac semimetal.
Remarkably, upon including only the linear BIA term, each four-fold degenerate Dirac point transforms into a line node of double degeneracy,
which has an approximate ring shape. This was first noticed in Ref.~\ocite{Ruan2016}.

For the analysis of the line nodes, we find the following change of basis particularly useful,
\[
    \Hh'(\pb)=\Uh_\phi^\dg\Hh(\pb)\Uh_\phi,
\]
\beq
    \Uh_\phi=\f1{\sq2}
        \lt(\ba{cccc}
            0 & -e^{-i\phi} & 0 & -e^{-i\phi} \\
            1&0&1&0\\
            e^{-i\phi}&0&-e^{-i\phi}&0\\
            0&1&0&-1
        \ea\rt),
\lbl{eq:Uphi}
\eeq
where $\phi$ is the polar angle of momentum
\beq
    \pb=(p_\p\cos\phi,p_\p\sin\phi,p_z)
\lbl{eq:pcyl}
\eeq
in the cylindrical coordinates.
The idea behind this basis is to utilize the fact that $j_z=\pm\f12$ and $j_z=\pm\f32$ are pairs of Kramers doublets.
For $\Tc_-$ and inversion symmetry in $\Db_{4h}$, the Hamiltonian is a unit matrix $\um_2$ within each Kramers doublet.
One can therefore change the bases within the Kramers doublets to modify the form of the remaining terms in the Hamiltonian,
and thereby uncover some simplifying properties.

Consider first the regime (quantified below) when the linear BIA term is much smaller than strain.
In this case, it may be included into the linear model of \secr{bulkDlin}.
To leading order, the linear BIA term \eqn{M1} may be taken as constant at the Dirac points $(0,0,\pm_u p_u)$,
\[
    \be_1\Mh_1(\pm_u\pb_u)=\pm_u m \lt(\ba{cccc} 0&0&1&0 \\ 0&0&0&-1 \\ 1&0&0&0 \\ 0&-1&0&0 \ea\rt),
\]
where
\beq
    m=\tf{\sq3}2 \be_1 2 p_u.
\lbl{eq:m}
\eeq

The Hamiltonian of the linear-in-momentum low-energy model with the linear BIA term reads
\beq	
	\Hch^{LN}(\kbh)=
	\lt(\ba{cc}
		\Hch^{LN}_{+_u}(\kbh) & \hat{0} \\
		\hat{0} & \Hch^{LN}_{-_u}(\kbh)\ea\rt)
,\spc
    \Hc^{LN}_{\pm_u}(\kbh)=\Hc^D_{\pm_u}(\kbh)+\be_1\Mh_1(\pm_u\pb_u).
\lbl{eq:HcLN}
\eeq
The BIA term couples the $\pm_{j_z}$ pairs of $j_z=+\f32,+\f12$ and $j_z=-\f12,-\f32$ states, which are decoupled in the Dirac semimetal [\eq{HcDu}].
However, utilizing the change of basis \eqn{Uphi},
\[
	\Hch'^{LN}_{\pm_u}(k_\p\cos\phi,k_\p\sin\phi,\kh_z)
	=\Uh_\phi^\dg\Hc^{LN}_{\pm_u}(k_\p\cos\phi,k_\p\sin\phi,\kh_z)\Uh_\phi,
\]
we are still able to present the Hamiltonian in the block-diagonal form
\beq
    \Hch'^{LN}_{\pm_u}(k_\p\cos\phi,k_\p\sin\phi,\kh_z)
    =\eps_u\um_4
        \pm_u\lt(\ba{cccc}
        (v_0+v_z)\kh_z & v_\p k_\p - m &0&0\\
        v_\p k_\p-m & (v_0-v_z)\kh_z &0&0\\
        0&0& (v_0+v_z)\kh_z & v_\p k_\p + m \\
        0&0& v_\p k_\p +m & (v_0-v_z)\kh_z \\
    \ea\rt),
\lbl{eq:HcLNU}
\eeq

For each Dirac point $\pm_u$, the four bands are
\beq
	\eps^{LN,\x{lin}}_{\pm_u,+_m,\pm}(\kb)=
	\eps_u \pm_u v_0 k_z\pm\sq{(v_\p k_\p-_m m)^2+(v_z k_z)^2},
\lbl{eq:eLNlin+m}
\eeq
\beq
	\eps^{LN,\x{lin}}_{\pm_u,-_m,\pm}(\kb)=
	\eps_u \pm_u v_0 k_z\pm\sq{(v_\p k_\p+_m m)^2+(v_z k_z)^2},
\lbl{eq:eLNlin-m}
\eeq
arising from the two decoupled blocks, respectively, which we label $\pm_m$.
Comparing to the double-degenerate bands \eqn{eDlin} of the Dirac semimetal, we see that, for this linear model,
the effect of the linear BIA term is to shift the dependence of the bands on $k_\p$ by $\pm_m m/v_\p$ for every $\phi$,
thereby lifting their double-degeneracy.
In other words, each $\pm_u$ Dirac point is indeed transformed into a line node, which under this approximation is a circle of radius
\beq
    p_{\be_1}=
    \f{|m|}{v_\p}=\f{|\be_1|}{\al_z+\tf25\al_\square}
\lbl{eq:pbe1}
\eeq
in the $p_z=\pm_u p_u$ plane.
This momentum scale characterizes the linear BIA term and arises when the quadratic terms of the Luttinger semimetal and the linear BIA term
are of the same magnitude (note that $p_{\be_1}$ does not contain strain).
The above linear model [\eqs{HcLN}{HcLNU}] is valid when $p_{\be_1}\ll p_u$,
which quantifies the regime when the linear BIA term is much smaller than strain.
The characteristic energy scale
\beq
	\eps_{\be_1}=\be_1 p_u 
\lbl{eq:epsbe1}
\eeq
of the linear BIA term in the regime $p_{\be_1}\ll p_u$ is set by \eq{m} and does involve strain, since the latter is dominant.
When $p_{\be_1}\sim p_u$ are comparable, $\eps_{\be_1}\sim \be_1 p_{\be_1}$ becomes strain-independent.

Further, we apply the basis change \eqn{Uphi} to the Luttinger Hamiltonian \eqn{HLN} of the line-node semimetal,
the form of $\Hh'^{LN}(\pb)=\Uh_\phi^\dg \Hh^{LN}(\pb) \Uh_\phi$ is presented in \appr{appendix}.
In the presence of cubic anisotropy ($\al_\square\neq 0$), for arbitrary $\phi$, we confirm the existence of the $\pm_u$ line nodes numerically,
with momentum and energy
\beq
	\pb=(p_\p^{LN}(\phi)\cos\phi,p_\p^{LN}(\phi)\sin\phi,\pm_u p_z^{LN}(\phi)), \spc \e=\eps^{LN}(\phi), \spc \phi\in[0,2\phi).
\lbl{eq:peLN}
\eeq
The momenta $p_\p^{LN}(\phi)$ and $p_z^{LN}(\phi)$ and energy $\eps^{LN}(\phi)$ of the line nodes have some dependence on $\phi$,
which is, however, numerically quite weak in the regime $p_{\be_1}\sim p_u$
and very weak in the regime $p_{\be_1}\ll p_u$.
Therefore, the line node is a circle to a high accuracy in the whole regime $p_{\be_1}\lesssim p_u$.
Note that in the regime $p_{\be_1}\ll p_u$, the momentum $p_\p^{LN}(\phi)\approx p_{\be_1}$ [\eq{pbe1}]
is determined by the linear BIA parameter $\be_1$ and is virtually independent of strain $u$,
while the momentum $p_z^{LN}(\phi)\approx p_u$ [\eq{pu}] and energy $\eps^{LN}(\phi)\approx \eps_u$ [\eq{epsu}]
are determined by strain and are virtually independent of the linear BIA parameter.
The same concerns the momenta $p_\p^W$ and $p_z^W$ and energy $\eps^W$ of the Weyl points below.
For the parameters of HgTe (\tabr{LM}) and the strain value $u=-3\x{meV}$ used in the subsequent calculations,
which correspond well to the regime $p_{\be_1}\ll p_u$,
the dependencies $p_\p^{LN}(\phi)$, $p_z^{LN}(\phi)$, and $\eps^{LN}(\phi)$ are presented in \figr{LN}.

Without the cubic anisotropy ($\al_\square=0$), we are also able to prove the existence of the line node analytically:
one can see explicitly that for $p_\p=p_{\be_1}|_{\al_\square=0}=\f{|\be_1|}{\al_z}$
one of the states is completely decoupled from the other three in the Hamiltonian $\Hh'^{LN}(\pb)|_{\al_\square=0}$ at any $\phi$.
Hence, the band of this state can cross with the bands of the other three states and this indeed happens for one of the latter.
Therefore, in this case, $p_\p^{LN}|_{\al_\square=0}=\f{|\be_1|}{\al_z}$ of the line node is $\phi$-independent;
however, $p_z^{LN}(\phi)|_{\al_\square=0}$ and $\eps^{LN}(\phi)|_{\al_\square=0}$ still have some weak dependence on $\phi$
(since the linear BIA term breaks axial rotation symmetry about the $z$ axis) and need to be obtained numerically.

We also notice that the Hamiltonian $\Hh'^{LN}(\pb)$ (with cubic anisotropy, $\al_\square\neq 0$)
is block-diagonal at $\phi=\pm\f\pi4$, which is due to reflection symmetry in $\Db_{2d}$.
The bulk spectrum of each block can easily be found and for one of the blocks its two bands cross at
$(p_\p,p_z)=(p_\p^{LN}(\pm\tf\pi4),p_z^{LN}(\pm\tf\pi4))$ given by
\beq
    p_\p^{LN}(\pm\tf\pi4)=p_{\be_1},\spc
    [p_z^{LN}(\pm\tf\pi4)]^2=\tf12 p_{\be_1}^2+ p_u^2,
\lbl{eq:pLN}
\eeq
for any relation between $p_u$ and $p_{\be_1}$, which signifies the node.

We use this simplification to analytically derive the next low-energy model: the linear-in-momentum model around
the $+_u$ line node at $\phi=\f\pi4$.
We expand the Hamiltonian $\Hh'^{LN}(\pb)$ of the line-node semimetal about the momentum
$(\tf1{\sq2} p_{\p\f\pi4},\tf1{\sq2} p_{\p\f\pi4},p_{z\f\pi4})$ of the line node at $\phi=\f{\pi}4$ to linear order
in momentum deviation $(q_\p,q_\phi,q_z)$ expressed in the local basis of the cylindrical coordinates,
\[
	\pb=(\tf1{\sq2}(p_{\p\f\pi4}+q_\p-q_\phi),\tf1{\sq2}(p_{\p\f\pi4}+q_\p+q_\phi),p_{z\f\pi4}+q_z).
\]
Here, we denote $p_{\p\f\pi4}=p_\p^{LN}(\f\pi4)$ and $p_{z\f\pi4}=p_z^{LN}(\f\pi4)$ for brevity,
$q_\phi$ is the momentum component along the direction $\f1{\sq2}(-1,1,0)$ tangent to the line node at $\phi=\f\pi4$,
while $q_\p$ and $q_z$ are perpendicular to it.

For the block of the two states that form the line node, we obtain
\beq
	\Hcth'^{LN}(q_\p,q_\phi,q_z;\phi=\tf\pi4)=
	\tauh_0[\eps^{LN}(\tf\pi4)+v_{0\p}q_\p+v_{0z}q_z]
		+(\tauh_x v_{x\p}+\tauh_yv_{y\p})q_\p+\tauh_z(v_{z\p}q_\p+v_{zz}q_z).
\lbl{eq:HcLN2init}
\eeq
Here, $\tauh_{0,x,y,z}$ are the unity and Pauli matrices.
The velocities $v_{\ldots}$ are determined by the parameters of the line-node Hamiltonian $\Hh^{LN}(\pb)$ [\eq{HLN}]
and are provided in \appr{appendix}.

Importantly, we note that the terms at the matrices $\tauh_x$ and $\tauh_y$ contain only $q_\p$ and no $q_z$.
Therefore, one can perform a basis change (rotation about the $z$ pseudospin axis) to transform the Hamiltonian to the form
\beq
	\Hcth''^{LN}(q_\p,q_\phi,q_z;\phi=\tf\pi4)=
	\tauh_0[\eps^{LN}(\tf\pi4)+v_{0\p}q_\p+v_{0z}q_z]
		+\tauh_x v_{x\p}' q_\p+\tauh_z(v_{z\p}q_\p+v_{zz}q_z),
\lbl{eq:HcLN2}
\eeq
where
\[
	v_{x\p}'=\sq{v_{x\p}^2+v_{y\p}^2}.
\]

The bulk spectrum of this low-energy model consists of two bands and reads
\[
    \epst^{LN}(q_\p,q_\phi,q_z;\phi=\tf\pi4)
    =\eps^{LN}(\tf\pi4)
    +v_{0\p} q_\p+v_{0z} q_z
        \pm\sq{(v_{x\p}' q_\p)^2 + (v_{z\p}q_\p+v_{zz}q_z)^2}.
\]
There is no dependence on momentum $q_\phi$ in the Hamiltonian \eqn{HcLN2} and, as a consequence, in the spectrum,
which further confirms the existence of the line node.
The line node within this model is the straight line $q_\p=q_z=0$, which is the line tangent to the exact line node.

An analogous low-energy model could be derived for every point of the line node, parameterized by $\phi$.
The dependence of the velocity parameters on $\phi$ would have to be found numerically.

We believe that this line-node semimetal phase is of accidental nature, in the sense that it is not due to any exact physical symmetry,
since the symmetry $\Db_{2d}$ is not lowered further upon including the cubic BIA terms, which lift the line-node degeneracy, see \secr{bulkW}.
We do not explore the reasons for the existence of these line nodes further here.

We also mention that, as shown in Ref.~\ocite{Ruan2016},
when strain is comparable to the linear BIA term, or weaker, or absent, $p_u\lesssim p_{\be_1}$, additional line nodes arise in the planes $p_x=\pm p_y$.
These line nodes have a clear origin and {\em are} a consequence of the exact spatial symmetries:
reflections along the directions $\f1{\sq2}(1,\pm1,0)$ contained in the $\Db_{2d}$ point group.
These line nodes arise from the crossing of the bands that belong to different (spinful)
irreducible representations of the reflection symmetry group.
In this work, we are interested in the regime $p_u\gtrsim p_{\be_1}$,
when strain is larger than the linear BIA term, although it does not have to be much larger. In this regime, these line nodes are absent.

\subsection{Weyl semimetal for a $\Tb_d$ system under strain \lbl{sec:bulkW}}

Upon adding the cubic BIA terms \eqsdn{M31}{M34} to the Hamiltonian \eqn{HLN} of the line-node semimetal, i.e., considering the Hamiltonian
\beq
    \Hh^W(\pbh)=\Hh^{\Tb_d}(\pbh)+\Hh_u,
\lbl{eq:HW}
\eeq
the degeneracy of each of the two $\pm_u$ line nodes is lifted everywhere except for the four nodal points
\[
	\pb=(\pm p_\p^W,0,\pm_u p_z^W) \mbox{ and } (0,\pm p_\p^W,\pm_u p_z^W).
\]
The resulting system is a Weyl semimetal and the nodal points are the Weyl points,
with the asymptotically linear dispersion of the bands around them.
The energy $\eps^W$ of all eight Weyl points is the same due to $\Tb_d$ and $\Tc_-$ symmetries.
For the considered hierarchy of scales (\secr{scales}), as with the line node above,
$p_\p^W\approx p^{LN}_\p(0)\approx p_{\be_1}$, $p_z^W\approx p_z^{LN}(0) \approx p_u$, and $\eps^W\approx \eps^{LN}(0)\approx \eps_u$ to a good accuracy.
The topological properties of this Weyl semimetal are discussed in detail in \secr{topoW}.

The cubic BIA terms therefore need to be taken into account to create the Weyl semimetal in this type of system;
without them, the Luttinger Hamiltonian of the most general form up to quadratic order in momentum
describes the line-node semimetal, discussed in the previous \secr{bulkLN}. This is the only reason why we include the cubic BIA terms.

The cubic BIA terms in the Luttinger model consist of four invariants, characterized by the four coefficients $\be_{3i}$, $i=1,2,3,4$.
To our knowledge, their values for materials like HgTe are not well-documented.
If the Luttinger model arises as the low-energy limit of the Kane model (see the next \secr{KM}),
the parameters of the latter, from which the cubic BIA parameters of the Luttinger model arise, are also not well-documented.

However, despite this uncertainty, we realize that there is no need to explore the whole space of the four cubic BIA parameters $\be_{3i}$.
When the cubic BIA terms are small,
their key effect reduces to opening of the gap at the line node, everywhere except for the four Weyl points.
We demonstrate this by deriving the low-energy model for the Weyl semimetal at $\phi=\f\pi4$ around the (former) line node,
which amounts to incorporating the effect of the cubic BIA terms into the low-energy model \eqn{HcLN2} for the line-node semimetal.
The Weyl-semimetal Hamiltonian $\Hh'^W(\pb)$ with the cubic BIA terms
is still exactly block-diagonal at $\phi=\f\pi4$ due to reflection symmetry in $\Db_{2d}$.
To leading order, the cubic BIA terms are simply taken at the line-node momentum and produce the momentum-independent energy terms
$\eps_\al$, $\al=0,x,y,z$, the expressions for which are provided in \appr{appendix}.
For the block of the two line-node states, we obtain
\[
	\Hcth'^W(q_\p,q_\phi,q_z;\phi=\tf\pi4)
	=\tauh_0[\eps^{LN}(\tf\pi4)+\eps_0+v_{0\p}q_\p+v_{0z}q_z]
		+\tauh_x(\eps_x+v_{x\p}q_\p)+\tauh_y(\eps_y+v_{y\p}q_\p)+\tauh_z(\eps_z+v_{z\p}q_\p+v_{zz}q_z).
\]
Performing the same basis change as in \eqs{HcLN2init}{HcLN2} to eliminate the $\tauh_y q_\p$ term, we obtain
\beq
	\Hcth''^W(q_\p,q_\phi,q_z;\phi=\tf\pi4)=\tauh_0[\eps^{LN}(\tf\pi4)+\eps_0+v_{0\p}q_\p+v_{0z}q_z]
		+\tauh_x(\eps_x'+v_{x\p}' q_\p)+\tauh_y \De^W(\tf\pi4)+\tauh_z(\eps_z+v_{z\p}q_\p+v_{zz}q_z),
\lbl{eq:HcW2}
\eeq
where
\beq
	\De^W(\tf\pi4)=\eps_y'=\f{\eps_y v_{x\p}-\eps_x v_{y\p}}{\sq{v_{x\p}^2+v_{y\p}^2}}
	=\f{\sq3 p_{\p\f\pi4}p_{z\f\pi4}(p_{z\f\pi4}^2-\tf12p_{\p\f\pi4}^2)}{\sq{ p_{\p\f\pi4}^2+4p_{z\f\pi4}^2}}
	(\be_{31}+\tf74\be_{32}+\be_{33}-\be_{34}).
\lbl{eq:DeW}
\eeq

The bulk spectrum of this model reads
\beq
    \epst^W(q_\p,q_\phi,q_z;\phi=\tf\pi4)
    =\eps^{LN}(\tf\pi4)
    + \eps_0+v_{0\p} q_\p+v_{0z} q_z
        \pm\sq{(\eps_x'+v_{x\p}' q_\p)^2 + [\De^W(\tf\pi4)]^2+(\eps_z+v_{z\p}q_\p+v_{zz}q_z)^2}.
\lbl{eq:etW2}
\eeq
The effect of the cubic BIA terms is contained in the four energy parameters $\eps_0$, $\eps_x'$, $\eps_y'=\De^W(\tf\pi4)$, and $\eps_z$,
of which $\eps_0$ is a trivial energy shift.
We notice that the term at $\tauh_y$ in \eq{HcW2} contains only the energy $\De^W(\tf\pi4)$ and no momentum dependence;
as a result, the same concern the respective square term under the square root in the spectrum \eqn{etW2}.
Hence, $\De^W(\tf\pi4)$ [\eq{DeW}] determines the minimal energy distance (the ``gap'') between the two bands at $\phi=\f\pi4$,
which is reached at the line in momentum space, where the two other squares are nullified:
\[
        \eps_x'+v_{x\p}' q_\p=0,\spc \eps_z+v_{z\p}q_\p+v_{zz}q_z=0.
\]
We see that near the line node, the cubic BIA terms \eqsdn{M31}{M34} have two effects on the spectrum of the line-node semimetal:
(i) they shift the position of the line node in both momentum and energy via $\eps_0$, $\eps_x'$, and $\eps_z$;
(ii) they open a gap, determined by $\De^W(\f\pi4)$.
The shift of the line node (which is also small due to the assumed smallness of the cubic BIA terms)
is inconsequential for the bulk or surface states, whereas the gap opening qualitatively modifies their spectrum.

We see that the gap \eqn{DeW} is determined by the linear combination of the four parameters $\be_{3i}$ of the cubic BIA terms \eqsdn{M31}{M34},
which does not even depend on the relation between $p_{\p\f\pi4}$ and $p_{z\f\pi4}$.
This way, we were able to aggregate (and thus characterize by) the key effect of the four cubic BIA terms into just this one quantity:
the gap at the line node.
The exact values of the four parameters $\be_{3i}$ are of no significance,
since their effect results in just one qualitative change of the spectrum.
Therefore, there is no need to explore their whole parameter space, as any combination will describe the general behavior.
We prove this further in \secr{bccubicBIA} by comparing the bulk and surface-state spectra for two sets of values of $\be_{3i}$.
Having established that, we use just one set of values of $\be_{3i}$, provided in \tabr{LM} and discussed at the end of \secr{LfromK},
for the main calculations for the Weyl semimetal.

This gap determines the characteristic energy scale
\beq
	\eps_{\be_3}=\De^W(\tf\pi4)
\lbl{eq:epsbe3}
\eeq
of the cubic BIA terms in the considered hierarchy of scales (\secr{scales}). The corresponding momentum scale
\beq
	p_{\be_3}=\f{\eps_{\be_3}}{v^D}
\lbl{eq:pbe3}
\eeq
is related via the typical velocity $v^D=\al_z p_u$ of the linear spectrum around the Dirac point.

One could similarly derive an analogous low-energy linear-in-momentum model for the Weyl semimetal
at every point of the (former) line node, parameterized by $\phi$.
The dependence of the energy parameters on $\phi$ would also have to be found numerically.
The most important object of such a family of models would be the dependence of the gap $\De^W(\phi)$ along the (former) line node.
One can anticipate that, for a sensibly chosen local $\phi$-dependent basis of the two eigenstates of the line node,
the real gap $\De^W(\phi)$ will switch sign at the Weyl points.
It is clear by symmetry that the extrema of $\De^W(\phi)$
are reached at $\phi=\pm\f\pi4$ and are therefore given by the derived expression $\De^W(\f\pi4)$ [\eq{DeW}].

\section{Luttinger model as the low-energy limit of the Kane model \lbl{sec:KM}}

In this section, we demonstrate the relation of the Luttinger model for the $j=\f32$ states,
presented in \secr{LM} above, to the Kane model~\cite{Kane1957,Bir1974,Winkler2003},
when the former arises as the low-energy limit of the latter. The goals of considering such relation are as follows.
(i)~To demonstrate that hybridization between the $j=\f12$ and $j=\f32$ states
can lead to a qualitative change of the low-energy band structure of the $j=\f32$ states, resulting in the creation of the Luttinger semimetal phase.
(ii)~To derive the parameters of the Luttinger model from those of the Kane model for real materials, since the latter are better researched.
The Kane model is more commonly used in the studies of a large family of semiconductor and semimetal materials~\cite{Madelung2004},
such as $\al$-Sn and HgTe.
(iii)~To derive the effective BCs for the Luttinger model from the hard-wall BCs of the Kane model, which have a clear physical interpretation.
(iv)~To establish the validity range of the Luttinger model, which will then be used in the next sections
to demonstrate the quantitative asymptotic agreement within this range of the surface states obtained from the two models.

\subsection{General Hamiltonian for the Kane model}

The Kane model includes, in addition to the quartet of the $j=\f32$ states, the doublet of the $j=\f12$ states of opposite inversion parity.
The wave function reads
\beq
    \Psih^K(\rb)
    =\lt(\ba{c} \Psih_{\f12}^K(\rb) \\ \Psih_{\f32}^K (\rb)\ea\rt)
    ,\spc
    \Psih_{\f12}^K(\rb)=\lt(\ba{c}\Psi_{\f12,+\f12}^K(\rb) \\ \Psi_{\f12,-\f12}^K(\rb)\ea\rt)
    ,\spc
    \Psih_{\f32}^K(\rb)=\lt(\ba{c}
        \Psi_{\f32,+\f32}^K(\rb) \\
        \Psi_{\f32,+\f12}^K(\rb) \\
        \Psi_{\f32,-\f12}^K(\rb) \\
        \Psi_{\f32,-\f32}^K(\rb)\ea\rt),
\lbl{eq:PsiK}
\eeq
where the labels denote the $j,j_z$ quantum numbers.

The Kane Hamiltonian has the corresponding block structure
\beq
    \Hh^{K}(\pbh)
    =\lt(\ba{cc}
    \Hh_{\f12\f12}^K(\pbh) & \Hh_{\f12\f32}^K(\pbh) \\
    \Hh_{\f32\f12}^K(\pbh) & \Hh_{\f32\f32}^K(\pbh)
    \ea\rt)
    ,\spc
    \Hh_{\f32\f12}^K(\pbh)=\Hh_{\f12\f32}^{K\dg}(\pbh),
\lbl{eq:HK}
\eeq
here the cross blocks describe the hybridization between the $j=\f32$ and $j=\f12$ states.
Its most general form can also be obtained using the method of invariants within the Hilbert space \eqn{PsiK}.
Of course, the symmetry structure of the $\Hh_{\f32\f32}^K(\pbh)$ block is the same
as that of the Luttinger Hamiltonian for the $j=\f32$ states only, presented in \secr{LM}.
Following the same symmetry hierarchy \eqn{symm},
one can present the Hamiltonians for the three spatial points groups (and time-reversal symmetry $\Tc_-$) as
\beq
    \Hh^{K,\Ob_h}(\pbh)=\Hh^{K,\Ob(3)}(\pbh)+\de\Hh^{K,\Ob_h\subset\Ob(3)}(\pbh),
\lbl{eq:HKOh}
\eeq
\beq
    \Hh^{K,\Tb_d}(\pbh)=\Hh^{K,\Ob_h}(\pbh)+\de\Hh^{K,\Tb_d\subset\Ob_h}(\pbh),
\lbl{eq:HKTd}
\eeq
with analogous meaning of the terms as in \eqs{HOh}{HTd}.

The most general form of the Kane model for full spherical symmetry $\Ob(3)$ reads
\beq
    \Hh^{K,\Ob(3)}(\pbh)
    =\lt(\ba{cc}
        (\De + \ga_{\f12}\pbh^2)\um_2 & v\Mh_v(\pbh) \\
        v\Mh_v^\dg(\pbh) &
        \ga_0\pbh^2\um_4+\ga_z\Mh_z(\pbh)
    \ea\rt).
\lbl{eq:HKO3}
\eeq
\beq
    \Mh_v(\pbh)=\tf3{\sq2}(\hat{\Tb}\cdot\pbh).
\lbl{eq:M12321}
\eeq
Here, $\e=\De,0$ are the energy levels of the $j=\f12$ and $j=\f32$ states, respectively, at zero momentum;
$\ga_{\f12}$, $\ga_{0,z}$ are the curvature parameters of the quadratic terms.
The $4\times 2$ matrices $\hat{\Tb}=(\Th_x,\Th_y,\Th_z)$ are the basis matrices for the cross block $\Hh_{\f12\f32}^K(\pbh)$
belonging to the angular-momentum-1 irreducible representation of $\Ob(3)$~\cite{Trebin1979, Winkler2003}.
Thus, for spherical $\Ob(3)$ symmetry (and for cubic $\Ob_h$ below),
there is just one invariant $\Mh_v(\pbh)$ for the hybridization between the $j=\f32$ and $j=\f12$ states;
its coefficient $v$ is real due to  time-reversal symmetry $\Tc_-$ and has the dimensionality of velocity.

Upon lowering the symmetry down to cubic, $\Ob(3)\rarr\Ob_h$, the only additional term, quadratic in momentum,
is the cubic anisotropy term for the $j=\f32$ states
\beq
    \de\Hh^{K,\Ob_h\subset\Ob(3)}(\pbh)
    =\lt(\ba{cc}
        \nm & \nm \\
        \nm &
        \ga_\square \Mh_\square(\pbh)
    \ea\rt).
\lbl{eq:dHKOh}
\eeq
Throughout, $\nm$ denote the null matrices of the respective sizes.

Upon further lowering the symmetry down to tetrahedral, $\Ob_h\rarr\Tb_d$, the additional, BIA terms up to cubic order are
\beq
    \de\Hh^{K,\Tb_d\subset\Ob_h}(\pbh)=
    \lt(\ba{cc}
        \ldots  & B_-\Mh_-(\pbh)+B_+\Mh_+(\pbh) \\
    B_-\Mh_-^\dg(\pbh)+B_+\Mh_+^\dg(\pbh)
    & \be_1^0\Mh_1(\pbh)+\sum_{i=1,2,3,4}\be_{3i}^0 \Mh_{3i}(\pbh)
    \ea\rt).
\lbl{eq:dHKTd}
\eeq
There are linear and cubic BIA terms in $\Hh_{\f32\f32}^K(\pbh)$ of the same structure.
There are two quadratic invariants in the $\Hh^K_{\f12\f32}(\pbh)$ block,
with real coefficients $B_\pm$ due to time-reversal symmetry $\Tc_-$
and the matrix functions
\beq
    \Mh_-(\pbh)=-\tf{\sq{3}}{2}
        [(\Th_{xx}-\Th_{yy})(\ph_z^2-\tf13\pbh^2)
         -\Th_{zz}(\ph_x^2-\ph_y^2)]
    ,\spc
    \Mh_+(\pbh)=\sq{3}i(\Th_x \ph_y \ph_z+\text{c.p.}),
\lbl{eq:Mpm}
\eeq
where $\Th_{xx}$, $\Th_{yy}$, $\Th_{zz}$
are some of the basis matrices for the cross block $\Hh_{\f12\f32}^K(\pbh)$
belonging to the angular-momentum-2 irreducible representation of $\Ob(3)$~\cite{Trebin1979, Winkler2003}.
(Note that these three matrices are linearly dependent: $\Th_{xx}+\Th_{yy}+\Th_{zz}=\nm$.)
There is also a cubic term in the $\Hh_{\f12\f12}^K(\pbh)$ block in \eq{dHKTd};
we do not present it here and only denote with $\ldots$,
since it does not contribute to the Luttinger Hamiltonian upon the folding procedure.

Finally, the strain term reads
\beq
    \Hh^K_u=\lt(\ba{cc} \nm & \nm \\ \nm & -u^0(\Jh_z^2-\f54\um_4) \ea\rt).
\lbl{eq:HKu}
\eeq
To the lowest, zeroth order, strain only affects the $\Hh_{\f32\f32}(\pbh)$ block.

\subsection{Hybridization effect between $j=\f32$ and $j=\f12$ states}

\begin{figure*}
\centering
\includegraphics[width=.8\linewidth]{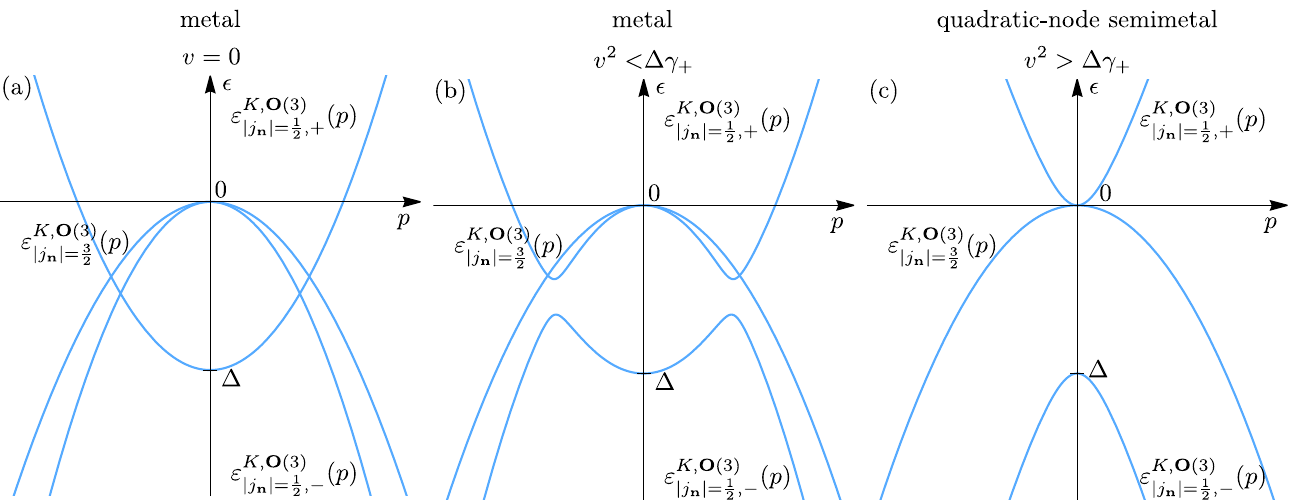}
\caption{
The effect of hybridization $v$ between the $j=\f12$ and $j=\f32$ states
of the Kane model [\eqs{HKO3}{HKO3block}] with spherical symmetry $\Ob(3)$.
The ``inverted'' regime $\De<0$ is assumed, when the $j=\f12$ level $\e=\De$ is below the $j=\f32$ level $\e=0$ at momentum $p=0$,
and the bare curvature parameters are assumed to satisfy $\ga_\f12>0$ and $\ga_\pm<0$.
The $|j_\nb|=\f12$ bands $\eps^{K,\Ob(3)}_{|j_\nb|=\f12,\pm}(p)$ [\eq{eK12}]
and the $|j_\nb|=\f32$ bands $\eps^{K,\Ob(3)}_{|j_\nb|=\f32}(p)$ [\eq{eK32}] are shown.
Due to symmetry, the $|j_\nb|=\f32$ states are unaffected by hybridization.
(a) For absent hybridization, $v=0$, the $|j_\nb|=\f12$ bands of $j=\f32$ and $j=\f12$ states cross.
(b) For weaker hybridization, $v^2<\De\ga_+$, the coupling of the $|j_\nb|=\f12$ states results in avoided crossing of the bands $\eps^{K,\Ob(3)}_{|j_\nb|=\f12,\pm}(p)$.
(c) For stronger hybridization, $v^2>\De\ga_+$, level repulsion causes the bands $\eps^{K,\Ob(3)}_{|j_\nb|=\f12,\pm}(p)$ to become monotonous,
so that the band $\eps_{|j_\nb|=\f12,+}^{K,\Ob(3)}(p)=\eps^{K,\Ob(3)}_{j=\f32,|j_\nb|=\f12}(p)$ [\eq{eK3212}]
at small $p$ switches from hole to electron character and the system becomes a Luttinger semimetal with a quadratic node.
The Luttinger-semimetal regime (c) of the Kane model is realized in materials like HgTe and $\al$-Sn and is considered in this work.
}
\label{fig:hybr}
\end{figure*}

\begin{table*}
\begin{tabular}{ c | c | c | c | c | c | c | c || c | c }
	& $\De$ & $\ga_{\f12}$ & $\ga_0$ & $\ga_z$ & $\ga_\square$ & $v^2$ & $\be_1^0$ & $B_+$ & $B_-$ \\ \hline \hline
HgTe & $-303~\x{meV}$ & $\f1{2m_e}$ & $-4.1\f1{2m_e}$ & $-1.96\f1{2m_e}$ & $-1.6\f1{2m_e}$ &
$12.53\f1{2m_e}\x{eV}$ & $-4.31~\x{meV}\,\x{nm}$ & $-75.48~\x{meV}\,\x{nm}^2$ & $0$
\end{tabular}
\caption{
The values of the parameters used for the Kane model $\Hh^K(\pb)$ [\eq{HK}] of HgTe.
Except for $B_\pm$, the parameters of HgTe are taken from Refs.~\ocite{Cardona1986,Novik2004}.
The not well-documented bare cubic BIA parameters $\be_{3i}^0=0$ of the $j=\f32$ states were assumed zero
and the parameters $B_+$ and $B_-=0$ were determined via \eq{be3} from the chosen values of the cubic BIA parameters $\be_{3i}$ (\tabr{LM})
for the Luttinger model, to have correspondence between the models, see the end of \secr{LfromK} for the explanation.
}
\label{tab:KM}
\end{table*}

Suppose one starts with a model ``larger'' than the Luttinger model, that besides the $j=\f32$ states
also includes other states, separated at $\pb=\nv$ by finite energies;
the above Kane model with additional $j=\f12$ states, separated by $\De$, is one specific example.
But one is still only interested in momentum and energy scales in the vicinity of the $j=\f32$ states at $\pb=\nv$,
so that the energy deviations are much smaller than $\De$.
In this low-energy limit, one would expect that the model that includes only the $j=\f32$ states,
the most general form of which has been presented above in \secr{LM}, would be sufficient.
And this is indeed true; however, the relation between these two models is nontrivial.

One cannot simply neglect the $j=\f12$ states in \eqs{PsiK}{HK} and consider the block $\Hh_{\f32\f32}^K(\pb)$ with its ``bare'' parameters
as the effective low-energy Hamiltonian just for the $j=\f32$ states, even in this low-energy limit, for energies and momenta close to $(\e,\pb)=(0,\nv)$.
This is because the polynomial momentum terms in $\Hh_{\f32\f32}^K(\pb)$ are themselves small compared to $\De$ for small $\pb$.
The hybridization with the $j=\f12$ states via the cross block $\Hh_{\f12\f23}^K(\pb)$ at nonzero $\pb$ generates
effective polynomial momentum terms within $j=\f32$ states (via virtual transitions to the $j=\f12$ states and back),
which, even though they are much smaller than $\De$, can be comparable to or even more dominant than these bare terms.
As a result, these hybridization effects can lead to a significant, qualitative change of the local band structure
in the vicinity of the $j=\f32$ level $(\e,\pb)=(0,\nv)$.

We first illustrate this hybridization effect for the whole bulk spectrum of the Kane model for full rotation symmetry $\Ob(3)$.
As for the Luttinger model in \secr{bulkL}, in this case, for momentum $\pb=p\nb$ of arbitrary direction,
the Hamiltonian is block-diagonal in the basis of states with definite projection $j_\nb=+\f32,+\f12,-\f12,-\f32$
on that direction and these blocks are the same for the subspaces of $\pm j_\nb$ states with opposite projections.
Hence, the $|j,j_\nb\ran=|\f32,\pm\f32\ran$ states are decoupled from the rest.
The two pairs $(|\f12,+\f12\ran,|\f32,+\f12\ran)$ and $(|\f12,-\f12\ran,|\f32,-\f12\ran)$ of the states have the Hamiltonian block
\beq
        \lt(\ba{cc}
        \De+\ga_{\f12}p^2 & v p  \\
        v p & (\ga_0+\ga_z)p^2 \ea\rt),
\lbl{eq:HKO3block}
\eeq
which can easily be diagonalized.

Altogether, the bulk spectrum of the $\Ob(3)$-symmetric Kane model consists of three double-degenerate bands
\begin{widetext}
\begin{align}
\label{eq:eK12}
    \eps_{|j_\nb|=\f12,\pm}^{K,\Ob(3)}(p)
    =&\tf\De2+\tf12(\ga_{\f12}+\ga_+)p^2
    \pm \sq{[\tf\De2+\tf12(\ga_{\f12}-\ga_+)p^2]^2+v^2p^2},\\
\label{eq:eK32}
    \eps_{j=\f32,|j_\nb|=\f32}^{K,\Ob(3)}(p)=&\ga_- p^2,
\end{align}
\end{widetext}
where we introduce
\beq
    \ga_\pm=\ga_0\pm \ga_z.
\lbl{eq:gapm}
\eeq
At small momenta, the spectrum becomes
\begin{align}
    \eps_{j=\f12,|j_\nb|=\f12}^{K,\Ob(3)}(p)
        =&    \eps_{|j_\nb|=\f12,-}^{K,\Ob(3)}(p)=
	\De + \lt(\ga_{\f12}+\f{v^2}{\De}\rt)p^2 + \Oc(p^4),
\lbl{eq:eK1212}
        \\
    \eps_{j=\f32,|j_\nb|=\f12}^{K,\Ob(3)}(p)
        =&\eps_{|j_\nb|=\f12,+}^{K,\Ob(3)}(p)=
    	\lt(\ga_+ -\f{v^2}{\De}\rt)p^2 + \Oc(p^4),
\lbl{eq:eK3212}
        \\
    \eps_{j=\f32,|j_\nb|=\f32}^{K,\Ob(3)}(p)
        =&\ga_- p^2,
\lbl{eq:eK3232}
\end{align}
where one may identify and label the two bands $\pm$ of $|j_\nb|=\f12$ states
as originating from the $j=\f12$ and $j=\f32$ states at $p=0$. This quadratic expansion as valid at $p\ll p_\De$, where
\beq
	p_\De=\sq{\f{\De}{\ga_{\f12}-\ga_++\f{2v^2}{\De}}}
\lbl{eq:pDe}
\eeq
is the momentum scale associated with the level spacing $\De$, obtained by comparing the terms under the square root in \eqn{eK12},
which thereby determines the validity range of the Luttinger model.

Examining \eqss{eK1212}{eK3212}{eK3232}, we see that the effect of hybridization at small momenta is to change the curvatures of the bands,
via the term $\f{v^2}{\De}$;
$\ga_{\f12}$, $\ga_\pm$ [\eq{gapm}] are the ``bare'' curvatures of the quadratic bands in the absence of hybridization, when $v=0$.
Due to spherical symmetry, this affects only the $|j_\nb|=\f12$ bands, while the $|j_\nb|=\f32$ bands with the curvature $\ga_-$ remain unaffected.

For many semiconductor materials, the bare curvature parameters are such that
\[
    \ga_\f12>0, \ga_\pm<0,
\]
i.e., without hybridization ($v=0$) the $j=\f12$ states would have an electron character
and both bands of the $j=\f32$ states would have a hole character. In the so-called inverted regime
\[
    \De<0,
\]
the $j=\f32$ level $\e=0$ at $\pb=\nv$ is above the $j=\f12$ level $\e=\De$.
Figure~\ref{fig:hybr} shows the effect of the hybridization $v$ between the $|j_\nb|=\f12$ states in this regime.
Without hybridization ($v=0$), the two bands \eqn{eK12} of the $|j_\nb|=\f12$ states cross.
Introducing hybridization opens up a gap between these bands.
For weaker hybridization, $v^2<\De \ga_+$, the character of the bands at small momentum remains the same,
and the $|j_\nb|=\f12$ bands are nonmonotonous at larger $p$. The system is a metal with a Fermi surface in this regime.
For stronger hybridization,
\[
    v^2>\De \ga_+,
\]
the $j=\f32$, $|j_\nb|=\f12$ band \eqn{eK3212} at small $p$ switches from hole to electron character,
and the system becomes a Luttinger semimetal with a quadratic node.
This is the regime of parameters of the Kane model realized in $\al$-Sn, HgTe, and many similar materials and considered in the rest of the work.

\subsection{Derivation of Luttinger model from Kane model via folding procedure \lbl{sec:LfromK}}

When the Luttinger model for the $j=\f32$ states arises as the low-energy limit of the Kane model for the $j=\f32$ and $j=\f12$ states,
the former can be derived from the latter via a systematic ``folding'' procedure~\cite{Winkler2003}, which we now present.
This procedure systematically excludes the remote $j=\f12$ states from the Hilbert space, while taking into account the effect of hybridization to them,
and establishes the relation between the parameters of the two model.
The procedure can be carried out to derive both the Hamiltonian and BCs.

It is technically simpler to carry out the derivation of the Hamiltonian in momentum space;
however, analogous steps can be performed in real space.
For a plane-wave wave function $\Psih^K e^{i \pb\rb}$ [\eq{PsiK}] with momentum $\pb$ and constant $\Psih^K$,
the stationary Schr\"odinger equation for the Kane model in the block form [\eq{HK}] reads
\begin{align}
    \lt(\ba{cc} \Hh_{\f12\f12}^K(\pb) & \Hh_{\f12\f32}^K(\pb) \\ \Hh_{\f32\f12}^K(\pb) & \Hh_{\f32\f32}^K(\pb)\ea\rt)
    \lt(\ba{c} \Psih_\f12^K \\ \Psih_\f32^K \ea\rt)
    =\e \lt(\ba{c}\Psih_\f12^K\\ \Psih_\f32^K \ea\rt).
\end{align}
We exclude $\Psih_\f12^K$ from the above equations to obtain the equation
\begin{align}
    \lt(\Hh_{\f32\f32}^K(\pb)
        +\Hh_{\f32\f12}^K(\pb)\f{1}{\e \um_2-\Hh_{\f12\f12}^K(\pb)}\Hh_{\f12\f32}^K(\pb)\rt)
       \Psih_\f32^K
    =\e\Psih_\f32^K
\lbl{eq:Psi32eq}
\end{align}
solely for $\Psih_{\f32}^K$.
This equation has a form of the stationary Schr\"odinger equation for $\Psih_{\f32}^K$,
where the matrix in the left-hand-side could be viewed as an effective Hamiltonian for it.
The first term $\Hh_{\f32\f32}^K(\pb)$ is the ``bare'' Hamiltonian
and the second term is the ``correction'' due to the hybridization with the $j=\f12$ states.
Physically, it can be seen as the effect of virtual transitions to the $j=\f12$ states and back.

However, this interpretation of the left-hand-side as a Hamiltonian is quantitatively rigorous only to lowest order,
when the energy and momentum in the inverse matrix are taken at the values $(\e,\pb)=(0,\nv)$ of the $j=\f32$ level.
In this limit, the $j=\f32$ part $\Psih_\f32^K$ of the wave function \eqn{PsiK} of the Kane model may be identified as the wave function [\eq{psi}]
\[
    \psih \larr \Psih_\f32^K
\]
of the low-energy Luttinger model, and the matrix in the left-hand-side as its Hamiltonian
\beq
    \Hh(\pb)+\Oc(\pb^4) \larr \Hh_{\f32\f32}^K(\pb)+\Hh_{\f32\f12}^K(\pb)\f{1}{-\Hh_{\f12\f12}^K(\nv)}\Hh_{\f12\f32}^K(\pb),
\lbl{eq:H<-K}
\eeq
so that \eq{Psi32eq} indeed takes the form of the effective stationary Schr\"odinger equation
\[
    \Hh(\pb)\psih=\e\psih.
\]

For a given symmetry, the correction part in the Luttinger Hamiltonian \eqn{H<-K}
has, of course, the same symmetry structure as the bare Hamiltonian $\Hh_{\f32\f32}^K(\pb)$,
and can be presented as a linear combination of the respective invariants.
The whole effective Luttinger Hamiltonian \eqn{H<-K} has therefore the symmetry structure presented in \secr{LM}
with the coefficients $(\al_{0,z,\square}, u, \be_1, \be_{3i})$.
And the effect of hybridization with the $j=\f12$ states in the low-energy limit can be regarded as the ``renormalization''
of the ``bare'' coefficients $(\ga_{0,z,\square}, u^0, \be_1^0, \be_{3i}^0)$ of $\Hh_{\f32\f32}^K(\pb)$.
Since the hybridization block $\Hh_{\f12\f32}^K(\pb)$ starts with the linear terms [\eq{M12321}], the generated terms are at least quadratic in $\pb$.
Therefore, there are no generated strain $\Jh_z^2-\f54\um_4$ [\eq{Hu}] and linear BIA $\Mh_1(\pb)$ [\eq{M1}] terms in \eq{H<-K}
and these parameters remain equal to their bare values:
\[
    u=u^0,
\]
\beq
    \be_1=\be_1^0.
\lbl{eq:be1be10}
\eeq
The quadratic terms in \eq{H<-K} are generated from the linear term [\eq{M12321}] in both $\Hh_{\f32\f12}^K(\pb)$ and $\Hh_{\f12\f32}^K(\pb)$.
Since these terms and $\Hh_{\f12\f12}^K(\nv)$ have $\Ob(3)$ symmetry, only the $\um_4\pb^2$ and $\Mh_z(\pb)$ invaiants of $\Ob(3)$ are generated,
and no cubic anisotropy invariant $\Mh_\square(\pb)$ of $\Ob_h$. The renormalized curvature parameters are
\beq
    \al_0=\ga_0-\f{v^2}{2\De} ,\spc \al_z=\ga_z-\f{v^2}{2\De} ,\spc \al_\square=\ga_\square.
\lbl{eq:alga}
\eeq
The corresponding curvature parameters
\[
    \al_{+_b}=\ga_+-\f{v^2}{\De},\spc
    \al_{-_b}=\ga_-
\]
of the bulk bands $\eps_{|j_\nb|=\f12}^{L,\Ob(3)}(p)=\al_{+_b} p^2 $ and $\eps_{|j_\nb|=\f32}^{L,\Ob(3)}(p)=\al_{-_b} p^2$ [\eqss{eL12}{eL32}{alpm}]
of the $\Ob(3)$-symmetric Luttinger model agree with those of \eqs{eK3212}{eK3232}, respectively, of the $\Ob(3)$-symmetric Kane model at small momenta.
In particular, the $\eps_{|j_\nb|=\f12}^{L,\Ob(3)}(p)$ band is affected by the hybridization with the $j=\f12$ states
and the $\eps_{|j_\nb|=\f32}^{L,\Ob(3)}(p)$ band is not, due to the symmetries discussed above.
In the inverted regime $\De<0$, for strong enough hybridization, the $\eps_{|j_\nb|=\f12}^{L,\Ob(3)}(p)$ band becomes electron-like, $\al_{+_b}>0$,
even if the bare one was hole-like, $\ga_+<0$, and the system is in the Luttinger semimetal regime.

Additional cubic BIA terms are generated in \eq{H<-K} from the linear term  $\Mh_v(\pb)$ [\eq{M12321}] in one hybridization block
and the quadratic term $\Mh_\pm(\pb)$ [\eq{Mpm}] in the other.
The renormalized coefficients of the cubic BIA terms of the Luttinger Hamiltonian \eqn{H<-K} read
\beq
	\be_{31}=\be_{31}^0+\sq{\f23}\f{v(B_+-\f94B_-)}{\De},\spc
	\be_{32}=\be_{32}^0+\sq{\f23}\f{vB_-}{\De},\spc
	\be_{33}=\be_{33}^0+\sq{\f23}\f{vB_-}{6\De},\spc
	\be_{34}=\be_{34}^0-\sq{\f23}\f{vB_-}{3\De}.
\lbl{eq:be3}
\eeq

To have more practical relevance, we perform calculations of the surface states of the line-node and Weyl semimetals
for the parameters of HgTe.
The parameters $\ga_{0,z,\square}$, $v$, $\De$, and $\be_1^0$
of the Kane model for HgTe are quite well-established~\cite{Cardona1986,Novik2004}, presented in \tabr{KM}.
These determine the curvature parameters $\al_{0,z,\square}$ [\eq{alga}] and the linear BIA parameter [\eq{be1be10}] of the Luttinger model,
presented in \tabr{LM}.
Note that for HgTe the renormalization correction term $\sim v^2/\De$ in \eq{alga}
is much stronger than the bare curvature values $\ga_{0,z}$; as a result,
there is quite a strong particle-hole asymmetry in the Luttinger semimetal: $\al_+ \gg |\al_-|$.

On the other hand, to our knowledge, the bare cubic BIA parameters $\be_{3i}^0$ and $B_\pm$ parameters are not well-documented.
These parameters determine according to \eq{be3} the cubic BIA parameters $\be_{3i}$ of the Luttinger model.
However, as we have demonstrated in \secr{bulkW}, within the Luttinger model, for a well-defined hierarchy of scales (\secr{scales})
the exact values of the four parameters $\be_{3i}$ are of no significance,
since the key qualitative effect of the cubic BIA terms is the opening of the gap \eqn{DeW} in the Weyl-semimetal phase at the line node,
whose magnitude is determined by the linear combination of $\be_{3i}$.
For this reason, we {\em choose} to perform the calculations within the Luttinger model for only $\be_{31}$ nonzero and $\be_{32}=\be_{33}=\be_{34}=0$,
as provided in \tabr{LM}.
In \secr{bccubicBIA}, we also explicitly demonstrate the equivalence of the spectra between this case and
the one with both $\be_{31}$ and $\be_{34}$ nonzero and $\be_{32}=\be_{33}=0$.
The chosen value of $\be_{31}$ is close to the one estimated from the unpublished DFT calculations.
To have correspondence \eqn{be3} with the parameters of the Kane model,
when the comparison between the two is performed in \secr{WLMKM}, we {\em assume} the bare cubic BIA parameters $\be_{3i}^0=0$ of the Kane model
zero and calculate $B_+$ and $B_-=0$ from the chosen values of $\be_{3i}$ of the Luttinger model.

We also observe an interesting property (although it does not seem consequential).
Inserting the expressions \eqn{be3} into \eq{DeW} for the gap at the line node due to the cubic BIA terms,
\[
	\De^W(\tf\pi4)
	=\f{\sq3 p_{\p\f\pi4}p_{z\f\pi4}(p_{z\f\pi4}^2-\tf12p_{\p\f\pi4}^2)}{\sq{ p_{\p\f\pi4}^2+4p_{z\f\pi4}^2}}
	\lt(\be_{31}^0+\tf74\be_{32}^0+\be_{33}^0-\be_{34}^0 +\sq{\f23}\f{v B_+}{\De}\rt).
\]
we note that $B_-$ drops out. This means that, at least in the regime of small cubic BIA terms, there is no effect of the $B_-$ term on the gap.

\subsection{Boundary conditions for the Luttinger semimetal model \lbl{sec:bc}}

The bound states of any continuum model can be explored once its bulk Hamiltonian has been supplemented with proper BCs.
Possible BCs of continuum models is an interesting and rather large topic on its
own~\cite{AkhiezerGlazman,ReedSimon,Berry,BerezinShubin,Bonneau,Tokatly,McCann,AkhmerovPRL,AkhmerovPRB,Ostaay,Hashimoto2016,Hashimoto2019,Ahari,
KharitonovLSM,KharitonovQAH,Seradjeh,Walter,Shtanko,KharitonovSC,Enaldiev2015,Volkov2016,Devizorova2017,KharitonovFGCM,KharitonovGBCs}.
The most general form of them is governed by the fundamental principle of quantum mechanics,
the norm conservation of the wave function, which translates to nullification of the probability current at the surface.
Such general BCs for the Luttinger model is an open problem.

In this work, we focus on just one instance of possible BCs for the Luttinger model in the semimetal regime.
Namely, we consider BCs for the Luttinger model that correspond to the well-known hard-wall BCs for the Kane model.
For $\Ob(3)$ symmetry, we have derived such BCs via the folding procedure in the previous work
and demonstrated that the Luttinger semimetal model with these BCs exhibits surface states.

For the hard-wall BCs for the Kane model, all components of the wave function vanish at the boundary;
e.g., for the sample occupying the half space $z>0$ with the boundary $z=0$, they read
\begin{align}
    \Psih^K(x,y,z=0)=\hat{0}.
\lbl{eq:bcK}
\end{align}
These BCs have a clearly physical interpretation:
the vacuum at $z<0$ can be described by the Kane model with the positive $\De'>0$ and large $\De'\gg |\De|$ level spacing, so that the system is an insulator.
Then any solutions to the stationary Schr\"odinger equation at energy $\e\sim\De$ will decay into $z<0$.
In the limit $\De'\rarr+\iy$, all components of the Kane-model wave function \eqn{PsiK} vanish already at the surface.

The corresponding previously derived BCs for the Luttinger semimetal model read
\begin{align}
    \psih(x,y,z=0)=\hat{0}.
\lbl{eq:bc}
\end{align}
These BCs apply within the validity range of the Luttinger model, at $|\pb|\ll p_\De$ and $|\e|\ll \De$.

When the Hamiltonian is modified by adding new terms, one must check whether the BCs are still valid.
Fundamentally, any BCs that nullify the probability current through the surface are valid.
Since the current contributions of all terms up to quadratic order in momentum
contain the wave function, both BCs \eqsn{bcK}{bc} remains valid when such term are added.
In particular, these BCs remain valid in the presence of the strain, linear BIA, and cubic anisotropy terms.

What concerns the cubic BIA terms, on the other hand, their current contributions generally do not vanish for the BCs \eqn{bc}.
However, as we explained above, the sole purpose of including them is to create a Weyl semimetal by lifting the accidental degeneracy of the line node.
The momentum scale $p_{\be_3}$ [\eq{pbe3}] of the cubic BIA terms in the vicinity of the line node
is assumed to be much smaller than the scales of strain $p_u$ [\eq{pu}] and linear BIA term $p_{\be_1}$ [\eq{pbe1}]
that create the very line-node structure.
While the momentum regions where cubic BIA terms would becomes comparable are outside of the validity ranges of the models.
Hence, the effect of the cubic BIA terms on the BCs may be neglected in the regime of interest,
even if the probability current of the cubic terms does not vanish exactly;
meaning that the difference between the wave function satisfying the BCs \eqn{bc} and the correctly modified BCs,
for which the current with the cubic BIA terms included would vanish, is negligible.
The corresponding necessary adjustment of the method of calculating surface states (\secr{method})
when the cubic BIA terms are included is explained in \secr{bccubicBIA}.

\section{Hierarchy of scales and low-energy models for the study of surface states\lbl{sec:scales}}

The Hamiltonians for the Kane, Luttinger, and linear models presented in the previous \secssr{LM}{bulk}{KM},
with multiple ``successive'' semimetal phases,
are a perfect framework to illustrate the following important point, which is rather general and applies to other systems with similar properties.

This system is quite special in that each consecutive perturbation (compressive strain, linear BIA term, cubic BIA terms)
does not gap out the previous semimetal phase, but creates a new type of semimetal phase
(in contrast, for tensile strain, the system would be a topological insulator, at which point significant modifications of the band structure would stop).
When a new perturbation is introduced, the most ``eventful'', significant qualitative changes of the band structure
occur around the nodes of the previous semimetal phase, in the region of the size set by the scale of the perturbation,
where they transform into the nodal fine structure of the new semimetal phase. These changes occur for both bulk and surface states.
At the same time, outside of these ``eventful'' successively smaller momentum regions,
the effects of these new perturbations remain comparatively small and do not lead to qualitative changes.
There, the behavior of both bulk and surface states remains only weakly affected,
as it is governed by more dominant terms that define the previous semimetal phase.
Importantly, since we have demonstrated that already the unperturbed Luttinger semimetal exhibits surface states,
this sequential evolution of the surfaces states starts with Luttinger semimetal and continues all the way down to the Weyl semimetal.

If there is a well-defined hierarchy of the scales of perturbations,
there is a corresponding hierarchy of successively smaller momentum and energy regions,
in which different distinct behaviors of the bulk and surface states will manifest, accompanied by crossovers between them.

Turning to the theoretical description of such a system, if one starts with a rather general model
that contains all the perturbations with a well-defined hierarchy, then there exists a corresponding hierarchy of low-energy models,
valid within the respective momentum and energy ranges around the nodes of successive semimetal phases.
Each successive low-energy model will be simpler, with less degrees of freedom (wave-function components or momentum powers in the Hamiltonian).
For multiple scales, one can talk about a chain of embedded low-energy models.
Such low-energy models can be derived by utilizing a variant of the systematic low-energy expansion procedure.
Importantly, since not only the Hamiltonians, but also the BCs can be derived this way,
the low-energy model will capture all the properties of the surface states within its validity range,
where they will be in the quantitative asymptotic agreement
with the surface states obtained from all the ``larger'' models that embed this low-energy model.

On the one hand, the largest model has the largest validity range, and all the lower-energy effects can be taken into account within it.
The main advantage of such model is that the behaviors of the surface states
at all lower scales, as well as the crossovers between them, will be captured by it.
This demonstration of different behaviors in various ranges is possible only within the largest model, which includes all these scales.
However, such model has more degrees of freedom and is more complicated for the theoretical analysis.

On the other hand, ``smaller'' low-energy models embedded in it have a narrower validity range,
but their analysis is simpler, oftentimes simple enough that the surface states can be found analytically.
Perhaps the most important aspect of employing low-energy models is that they allow one to clearly identify the mechanisms of the surface states,
by isolating the minimal ingredients needed for them and discarding other effects that turn out to be nonessential.

Which model or set of models is more preferable for the analysis is a separate question.
Our methodological goal here is to explicitly demonstrate the said relations between the models and thereby prove
that low-energy models supplemented with proper BCs are perfectly applicable for studying surface states.

Specifically for the system described in the previous \secssr{LM}{bulk}{KM}, when all perturbations are present,
we assume the following hierarchy of their energy [\eqs{epsbe1}{epsbe3}] and momentum [\eqsss{pu}{pbe1}{pbe3}{pDe}] scales:
\beq
    \eps_{\be_3} \ll \eps_{\be_1} \lesssim |u| \ll |\De| \ll \eps^*_K,
\lbl{eq:ehy}
\eeq
\beq
    p_{\be_3} \ll p_{\be_1} \lesssim p_u \ll p_\De \ll p^*_K.
\lbl{eq:phy}
\eeq
This hierarchy is also satisfied well in HgTe (for the properly chosen strain).

The Kane model is the ``largest'' considered continuum model.
It has some cutoff energy and momentum scales $(\eps^*_K,p_K^*)$, so that it is valid (i.e., is quantitatively accurate)
at energies $\e$ and momenta $\pb$ below these scales, satisfying the conditions
\[
    |\e|\ll \eps_K^*
,\spc
    |\pb|\ll p_K^*.
\]
The cutoff scales are not explicitly present in the model itself.
Outside of this range, other remote bands or higher-order momentum terms would need to be taken into account (or a lattice model could be considered)
and in relation to such larger model, the Kane model would in turn serve as a low-energy model.

The first low-energy model, embedded into the Kane model, is the Luttinger model,
applicable around the level $(\e,\pb)=(0,\nv)$ of the $j=\f32$ states.
The energy separation $\De$ between the $j=\f12$ and $j=\f32$ states
and the corresponding momentum scale $p_\De$ [\eq{pDe}] of the Kane model serve as the cutoff scales
\[
    (\eps_L^*,p_L^*)=(|\De|,p_\De)
\]
for the Luttinger model, which is valid at
\[
    |\e|\ll |\De|
,\spc
    |\pb|\ll p_\De.
\]
At this stage, the $j=\f12$ states are excluded from the Hilbert space.

Note that if one wants to focus only on the four semimetal phases,
whose nodal structures occur in the vicinity of the $j=\f32$ level $(\e,\pb)=(0,\nv)$,
the Kane model is essentially unnecessary, as it contains redundant degrees of freedom of the $j=\f12$ states.
Since the scales of all the perturbations are assumed to be much smaller than $(|\De|,p_\De)$,
all their effects can be captured within the Luttinger model
and quantitative asymptotic agreement between the bulk and surface states of the Luttinger and Kane models within this range will manifest.
We demonstrate this agreement in \secr{LLMKM} for the case of the Luttinger semimetal, without any perturbations.
In this case, no further scales are present,
\[
    p_\De\ll p_K^*
\]
is the whole sequence of scales, and there is no other low-energy model to consider.
We explicitly demonstrate how the range of the asymptotic agreement changes with $\De$.
We also demonstrate this agreement in \secr{WLMKM} for the case of the Weyl semimetal, with all the perturbations present.

When compressive strain is added and the Dirac semimetal is created,
the linear-in-momentum model \eqn{HcD} around the Dirac points $\pm_u\pb_u$ [\eq{pu}] exists,
whose cutoff scales $(\eps_D^*,p_D^*)=(|u|,p_u)$ are set by strain. The validity range of this model is
\[
    |\e-\eps_u|\ll |u|
,\spc
    |\pb\mp_u \pb_u|\ll p_u.
\]
At this stage, the number of the wave-function components (per $\pm_u$ momentum region) still remains the same,
but the order of momentum is lowered from quadratic to linear.
When strain is the only added perturbation, these are no more scales and features,
\[
    p_u\ll p_\De \ll p_K^*
\]
is the whole sequence of scales, and there is no other low-energy model to consider.
We demonstrate the asymptotic agreement between the bulk and surface states of this linear-in-momentum model and the Luttinger model in \secr{ssDlin}.

Adding further the linear BIA term, the line-node semimetal is created.
In this work, we focus on the regime $p_{\be_1} \lesssim p_u $, where strain is always present, although it does not have to be much larger;
we did not find a qualitative difference in the bulk or surface-state spectrum between the $p_{\be_1}\ll p_u$ and $p_{\be_1}\sim p_u$ regimes.
Note that while the momentum scale $p_{\be_1}$ [\eq{pbe1}] of the linear BIA terms is fixed, strain is to some extent tunable in real materials.
For $p_{\be_1} \ll p_u $, the effect of the linear BIA term can also be taken into account to leading order
in the linear model around the Dirac points [\eq{HcLN}], since the arising line node fits within its momentum validity range.

Regardless of the relation, whether $p_{\be_1}\sim p_u$ or $p_{\be_1}\ll p_u$, for the line-node semimetal, as discussed in \secr{bulkLN},
the next, $\phi$-dependent low-energy model can be derived, describing the vicinity of the line nodes,
which would be linear in the momentum deviation from the line nodes $\pm_u \pb^{LN}(\phi)$ [\eq{peLN}].
For each $\phi$, this model includes only the two degenerate states at each line node,
so the number of wave-function components is reduced from four to two. The model is valid in the regions
\[
    |\e-\eps^{LN}(\phi)|\ll \eps_{\be_1}
,\spc
    |\pb \mp_u \pb^{LN}(\phi)|\ll p_{\be_1}
\]
around the line nodes.
If $p_{\be_1} \ll p_u$, such linear model could be derived from the linear model \eqn{HcLN} around the (former) Dirac points.
If $p_{\be_1}\sim p_u$, the latter cannot be used anymore and one has to derive such model from the Luttinger model \eqn{HLN} directly,
although for arbitrary $\phi$ the coefficients would have to be obtained numerically.
For $\phi=\f\pi4$, we have derived the Hamiltonian \eqn{HcLN2} for such linear model in \secr{bulkLN}.
One could also derive the BCs for this model and calculate the surface states, which we do not do here.

Further, as discussed in \secr{bulkW},
the effect of the cubic BIA terms, which create the Weyl semimetal by gapping out each line node everywhere except at the four Weyl points,
can be included within the same $\phi$-dependent model around the (former) line nodes.
The condition $p_{\be_3} \ll p_{\be_1}$ ensures that these terms are within the validity range of the model.
We have derived the Hamiltonian \eqn{HcW2} for such linear model for arbitrary $p_{\be_1}\lesssim p_u$ and $\phi=\f\pi4$ in \secr{bulkW}.
The Weyl points should be contained in this model and manifest
as vanishing of the gap $\De^W(\phi)$ at the line node at $\phi=0,\f\pi2,\pi,\f{3\pi}2$ as it switches its sign.
The very last low-energy model, linear in 3D momentum deviations from the Weyl points, describing their vicinity,
could be obtained by expanding the Hamiltonian of this low-energy model in $\phi$.
Although we do not derive this last low-energy model, we do establish the linear scaling near the projected Weyl points in \secr{ssWy},
which proves that this is be possible.

We note that (as already pointed out in Ref.~\ocite{Ruan2016}) the linear spectrum around the Weyl points is highly anisotropic:
for the two directions perpendicular to the line node, the characteristic velocity
is that $v^D = \al_z p_u$ of the linear dispersion around the (former) Dirac node,
while for the direction along the line node, the characteristic velocity
\[
	\f{\pd_\phi\De^W(0)}{p_{\be_1}}\sim \f{\eps_{\be_3}}{p_{\be_1}}
	\sim v^D \f{\eps_{\be_3}}{\eps_{\be_1}}
\]
is determined by the variation of the gap along the line node and is therefore parametrically much smaller.
Accordingly, the momentum validity range of the linear model around the Weyl points, stemming from the energy range
\[
    |\e-\eps^W|\ll \eps_{\be_3}
\]
is also anisotropic.
For example, for the four Weyl points $(\pm p_\p^W,0,\pm p_z^W)$ in the $p_y=0$ plane, the range is
\[
    |p_x\mp p^W_\p|,|p_z\mp p^W_z|\ll p_{\be_3}
\]
for the two directions perpendicular to the (gapped out) line node and
\[
    |p_y|\ll p_{\be_1}
\]
for the one direction along the line node.

We also point out that since the energy $\eps^{LN}(\phi)$ of the line node depends on $\phi$,
there is a possibility of type-II Weyl semimetal, when the variation of $\eps^{LN}(\phi)$ exceeds the variation of $\De^W(\phi)$.
In fact, the former is {\em parametrically} larger than the latter.
However, we have seen in \figr{LN} that for $p_{\be_1} \gg p_u$ these variations are {\em numerically} very small,
which ensures that in this regime the Weyl semimetal will be of type I.
For small enough strain, $p_u \lesssim p_{\be_1}$, however, the transition from type-I to type-II Weyl semimetal should eventually occur.

\section{Semi-analytical method of calculating surface states \lbl{sec:method}}
\subsection{Outline of the method}

In this section, we outline the general semi-analytical method of calculating the surface states of a given continuum Hamiltonian supplemented with BCs.
To be specific, we demonstrate that for the Luttinger Hamiltonian $\Hh(\pbh)$ up to quadratic order, without the cubic BIA terms,
which can be any of the three versions \eqssn{HL}{HD}{HLN} of the Luttinger, Dirac, or line-node semimetals, and the BCs \eqn{bc}.
The same approach holds, however, for any other continuum Hamiltonian with proper BCs, satisfying the current nullification requirement,
in particular, for the Kane Hamiltonian \eqn{HK} with the hard-wall BCs \eqn{bcK}, and the linear models with their BCs.
The important nontrivial nuance for the Hamiltonian \eqn{HW} of the Weyl semimetal with the cubic BIA terms is explained afterwards
in \secr{bccubicBIA}.

We outline the method for the semi-infinite system that occupies the $z>0$ half-space, with the surface at $z=0$;
however, this works for any half-space with arbitrary surface orientation.
Since the BCs \eqn{bc} have translation symmetry in the $x$ and $y$ directions along the surface,
the in-plane momentum $(p_x,p_y)$ is conserved
and the wave function of the surface state can be sought in the plane-wave form
\[
	\psih(x,y,z)=\psih(p_x,p_y,z)e^{i(p_xx+p_yy)}.
\]
The problem becomes effectively one-dimensional, in which the in-plane momentum $(p_x,p_y)$ enters as a parameter:
the wave function $\psih(p_x,p_y,z)$ must satisfy the stationary Schr\"odinger equation
\beq
	[\Hh(p_x,p_y,\ph_z)-\e \um]\psih(p_x,p_y,z)=\nm,
\lbl{eq:SchEq1D}
\eeq
the BCs
\beq
	\psih(p_x,p_y,z=0)=\nm,
\lbl{eq:bc1D}
\eeq
and must decay into the bulk,
\[
	\psih(p_x,p_y,z\rarr+\iy)=\nm.
\]

The method follows directly from the theory of linear differential equations.
For a given energy $\e$, one first constructs a general solution to \eq{SchEq1D} that decays into the bulk.
Such solutions can exist only for energies $\e$ within the ``gap'' of the bulk continuum spectrum
at a given $(p_x,p_y)$. Assuming there is only one such ``gap'' at every $(p_x,p_y)$,
\beq
	\e\in(E_{z>0,-_b}(p_x,p_y),E_{z>0,+_b}(p_x,p_y)),
\lbl{eq:egap}
\eeq
where
\beq
	E_{z>0,+_b}(p_x,p_y)=\min_{p_z,\sig} \eps_{+_b,\sig}(p_x,p_y,p_z),
\lbl{eq:E+}
\eeq
\beq
    E_{z>0,-_b}(p_x,p_y)=\max_{p_z,\sig} \eps_{-_b,\sig}(p_x,p_y,p_z).
\lbl{eq:E-}
\eeq
are the boundaries of the continua of the bulk spectrum:
$E_{z>0,+_b}(p_x,p_y)$ is the minimum of all upper $+_b$ (conduction) bands (labeled with $\sig$) over $p_z$
and $E_{z>0,-_b}(p_x,p_y)$ is the maximum of all lower $-_b$ (valence) bands (labeled with $\sig$) over $p_z$.
Such general solution is a linear combination of particular solutions of the form
\beq
    \chih e^{i p_z z},
\lbl{eq:chipart}
\eeq
where the momentum $p_z$ satisfies the characteristic equation
\beq
    \det[\Hh(p_x,p_y,p_z)-\e\um]=0
\lbl{eq:pzeq}
\eeq
and $\chih$ is the corresponding nontrivial ``eigenvector'' solution to
\[
    [\Hh(p_x,p_y,p_z)-\e\um]\chih=0.
\]
For energies within the gap \eqn{egap}, all momentum solutions necessarily have nonzero imaginary parts.

First consider the case without the cubic BIA terms, when the top momentum order of the Hamiltonian $\Hh(p_x,p_y,\ph_z)$ in $\ph_z$ is quadratic.
There are $2N=8$ momentum $p_{zn}(p_x,p_y,\e)$ solutions, of which there are $N=4$ with positive and negative imaginary parts.
For the system occupying the $z>0$ half-space, only the $N=4$ momentum solutions with positive imaginary parts $\x{Im}\, p_{nz}(p_x,p_y,\e)>0$,
are kept, labeled $n=1,2,3,4$, to have decaying particular solutions \eqn{chipart}.

The general solutions at a given energy $\e$, decaying into the bulk, therefore reads
\beq
    \psih(p_x,p_y,z,\e)=\sum_{n=1}^N c_n\chih_n(p_x,p_y,\e)e^{ip_{zn}(p_x,p_y,\e)z},
\lbl{eq:psissgen}
\eeq
where $c_n$ are arbitrary coefficients.
[It is assumed here that that eigenvectors $\chih_n(p_x,p_y,\e)$ are linearly independent.
In the case of degenerate momentum solutions $p_{zn}(p_x,p_y,\e)$, the number of eigenvectors may sometimes be less than the multiplicity.
In that case, enough particular solutions still exists, but their coordinate dependence differs from that of \eq{chipart}.
The adaptation to this case is also straightforward and follows from the theory of differential equations.]

Inserting this form into the BCs \eqn{bc1D}, the problem reduces to solving the system of linear homogeneous equations
for the coefficients $c_n$, which can be presented in the matrix equation
\beq
    \Xh(p_x,p_y,\e)\ch=\nm,
\lbl{eq:eqc}
\eeq
where
\[
    \Xh(p_x,p_y,\e)=
        \lt(\ba{ccc}
        \chih_1(p_x,p_y,\e) &
        \ldots &
        \chih_N(p_x,p_y,\e)
        \ea\rt)
\]
is a matrix whose columns are the eigenvectors of the particular solutions and
\[
    \ch=\lt(\ba{c} c_1\\\ldots\\c_N \ea\rt)
\]
is the vector of the coefficients.
Equation~(\ref{eq:eqc}) has nontrivial solutions for $\ch$ only if the matrix $\Xh(p_x,p_y,\e)$ is degenerate, i.e.,
\beq
    \det\Xh(p_x,p_y,\e)=0.
\lbl{eq:eqe}
\eeq
The solutions to this equation give the energies $\e=\Ec_{z>0}(p_x,p_y)$ of the surface states
and the corresponding nontrivial solutions $\ch$ to \eq{eqc} give their wave functions according to \eq{psissgen}.

Depending on the complexity of the model, for some models, the whole procedure can be carried out analytically
(such as the Luttinger model for the Luttinger semimetal and various linear-in-momentum models).
For some more complicated models, the momentum solutions $p_{zn}(p_x,p_y,\e)$ and eigenvectors $\chih_n(p_x,p_y,\e)$
can be found analytically, but \eqs{eqc}{eqe} cannot be solved analytically.
For even more complicated models, the momentum solutions and eigenvectors cannot be found analytically.
In latter two cases, the respective parts of the algorithm have to be performed numerically.
However, this is still a very resource-efficient task
(especially compared to the common finite-size numerical calculations of the surface states):
the computational complexity is determined by the number of degrees of freedom of the continuum model,
given by the product of the number of wave-function components and momentum order of the Hamiltonian.
This is why we refer to this method as {\em semi-analytical},
in contrast to the common finite-size numerical calculations of the surface states.
Another important advantage of this approach is that the calculations are performed for a true half-infinite system
and there is no limit on the energy and momentum resolution (in contrast to the finite-size calculations).
This is particularly important for the fine features of surface-state structure,
such as in the vicinity of the nodes or when the surface-state bands are close to the bulk-band boundaries, as we will see below.

\subsection{Inclusion of the cubic BIA terms \lbl{sec:bccubicBIA}}

\begin{figure*}
\centering
\includegraphics[width=.50\linewidth]{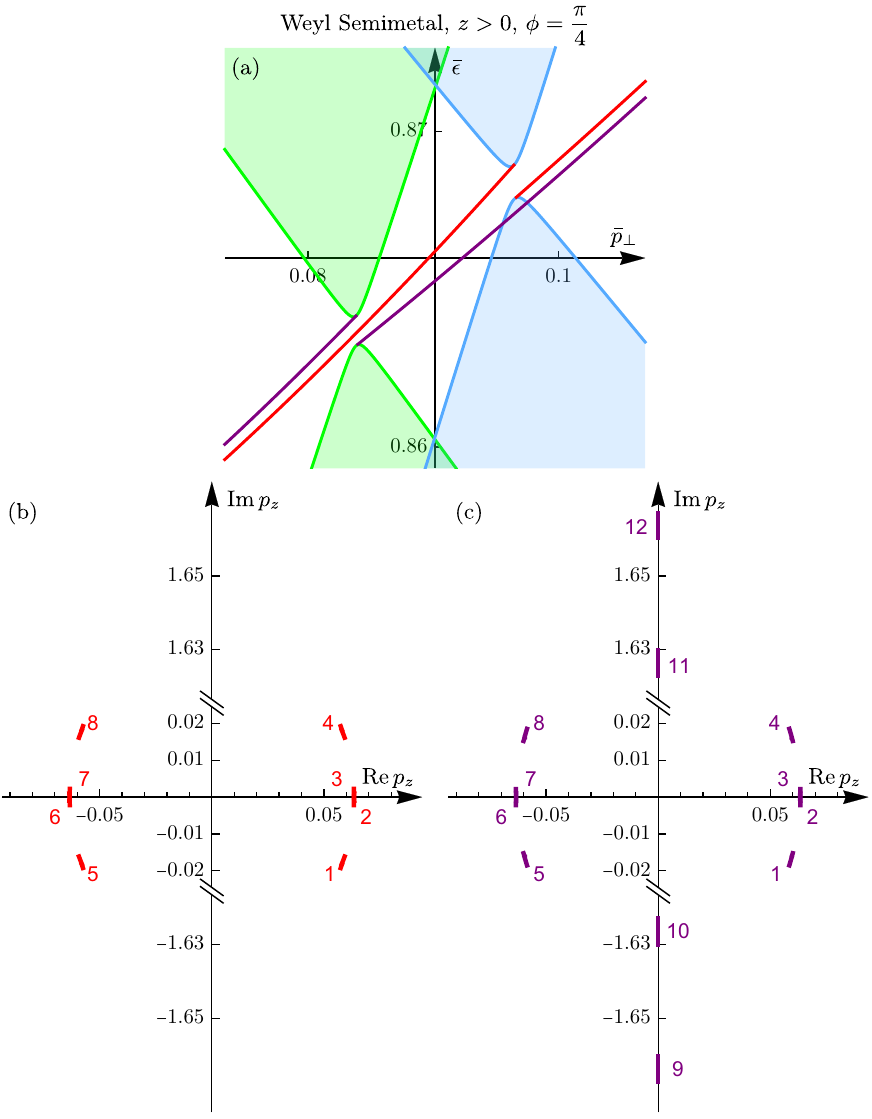}
\caption{
(a) The surface-state spectrum $\Ec_{z>0}^W(p_x=p_\p\cos\phi,p_y=p_\p\sin\phi)$ of the Weyl semimetal for the $z>0$ sample
for the surface momentum $(p_x,p_y)=p_\p(\cos\phi,\sin\phi)$ along the $\phi=\f\pi4$ path
in the vicinity of the (former) line node of the line-node semimetal
calculated within the Luttinger model for two sets of values of the cubic BIA parameters [\eqsd{M31}{M34}]:
(i) $(\be_{31},\be_{32},\be_{33},\be_{34})=(\be_{31}^*,0,0,0)$ (red surface-state band and blue bulk bands),
where $\be_{31}^*$ is the value in \tabr{LM} and the other parameters, including the cubic-power term $\propto \be_{34} \ph_z^3$, are absent,
as used in all subsequent calculations in \secr{ssW} for the Weyl semimetal;
(ii) $(\be_{31},\be_{32},\be_{33},\be_{34}=(2\be_{31}^*,0,0,\be_{31}^*)$ (purple surface-state band and green bulk bands),
with the cubic-power term present.
The second set is chosen this way, so that the low-energy gap $\De^W(\f\pi4)$ [\eq{DeW}] at the line node is the same as for the first set.
(b) and (c) show the paths of the complex momentum solutions
$p_{zn}(p_x=p_\p\cos\phi,p_y=p_\p\sin\phi,\e=\Ec_{z>0}^W(p_\p\cos\phi,p_\p\sin\phi))$ to the characteristic equation \eqn{pzeq}
for the surface states as $p_\p$ spans the range of (a).
This calculation both emphasizes the issue of the cubic-power terms explained in \secr{bccubicBIA}
and demonstrates explicitly the point made in \secr{bulkW}
that the spectrum for different values of the cubic BIA parameters is qualitatively the same, differing only by a shift in momentum and energy.
}
\lbl{fig:ssWzcubicBIA}
\end{figure*}

The above general semi-analytical method of calculating the surface states works for any continuum Hamiltonian with any well-defined BCs
that satisfy the current-nullification requirement.
As already mentioned in \secr{bc}, for the Luttinger-model Hamiltonian $\Hh^W(\pbh)$ [\eq{HW}] of the Weyl semimetal phase,
with the cubic BIA terms \eqsdn{M31}{M34} included, the BCs \eqn{bc} no longer exactly satisfy the current-nullification requirement,
albeit the deviations are parametrically small.
More precisely, this happens for the cubic terms of the momentum component perpendicular to the surface in question,
such as $\ph_z^3$ for the $z=0$ surface.
For the Hamiltonian \eqn{HW} we consider, such terms arise only from the $\be_{34}$ cubic BIA terms, while the other $\be_{31,32,33}$ terms,
although cubic in all momentum components, contain only lower powers of $\ph_z$.
Related to this deviation from the exact current nullification in the BCs \eqn{bc},
an attempt to straightforwardly apply the above method to \eqs{HW}{bc} leads to nontrivial issues that need to be resolved.

First, while these extra higher-order terms provide
the desired fine-structure (small, but essential) qualitative modifications of the bulk spectrum at lower scales of interest
(like opening of the minigap along the line node of the line-node semimetal, to transform it to a Weyl semimetal),
they can also lead to undesired qualitative changes of the bulk spectrum at large momenta, where they inevitably become dominant.
Namely, the gap $E_{z>0,+_b}(\pb_\p)-E_{z>0,-_b}(\pb_\p)$
of the bulk spectrum at a given surface momentum $\pb_\p$ as a function of momentum $p_z$ perpendicular to the surface may disappear completely,
if the bulk bands $\eps_{\pm_b}(p_x,p_y,p_z)$ at larger momenta $p_z$ cross the original gap at smaller momenta.
(The simplest example of this scenario would be adding a cubic term $\be_3\ph_z^3$ to the 1D Hamiltonian $\ph_z^2$ for a one-component wave function:
while there were no bulk states for it in the region $\e<0$, there are bulk states at all energies for the Hamiltonian $\ph^2+\be_3\ph^3$.)
In this case, the surface-state problem, if approached rigorously, is rendered meaningless.

Second, if the same BCs \eqn{bc} are applied, the above method still fails even if the gap remains.
The number of linearly independent BCs nullifying the current exactly
must always be half the number of degrees of freedom (order of momentum times the number of wave-function components);
this number is also the number of linearly independent particular solutions in the general decaying solution
(in the gapped energy region); this equality enables the surface-state solutions.
When cubic terms are included, one would have more decaying particular solutions than the constraints the BCs \eqn{bc} provide,
and the system of equations for the coefficients would be underdetermined.
To resolve the problem in the latter case, new BCs would need to be derived that would
satisfy the current-nullification constraint for the Hamiltonian with the cubic terms exactly.

Both of these issues are resolved as follows.
We realize that there is no goal to solve the problem with the cubic terms exactly:
they are only meant to provide the desired small modifications of the spectrum within the validity range of the model without them,
and they are never meant to be used in the large-momentum regions where they become dominant.
Hence, we may include their effect approximately in a controlled way, exploiting the parametric separation of scales.

When cubic terms are included, there will be $3N=12$ linearly independent particular solutions.
Of these solutions, $2N=8$ will have the previous momentum scale of interest $p\sim p_u, p_{\be_1}$
and they will essentially be all the previous solutions slightly modified by the presence of the cubic BIA terms
(upon sending the $\be_{3i}$ coefficients to zero, these solutions would by continuity recover the previous solutions).
The other $N=4$ ``new'' momentum solutions will have a parametrically large scale $p\sim \al_z/\be_{34}$ due to the small $\be_{34}$,
which is outside of the applicability range of the model;
upon sending $\be_{34}$ to zero, these solutions would approach infinity.
(If the bulk bands at large $p_z$ momenta cross the original gaps at smaller momenta, some of these latter $N$ solutions will be real,
which, however, does not matter.)

The systematic controlled resolution of the arising issues is to simply {\em discard} all latter large-momentum $N$ solutions
and leave the BCs unmodified, as per the discussion in \secr{bc}.
The remaining former $2N$ solutions will include all the desired fine-structure modifications of the spectrum.
The rest of the procedure of calculating the surface states remains well-defined.
The number $N$ of the free coefficients $c_n$ in the general decaying solution \eqn{psissgen}
will still match the number $N$ of BCs \eqn{bc} and the system \eqn{eqc} is well-defined.
The energies at which the system becomes degenerate [\eq{eqe}] will still provide the solutions for the surface states.

We illustrate this point in \figr{ssWzcubicBIA}, where we present the surface-state spectrum $\Ec_{z>0}^W(p_x=p_\p\cos\phi,p_y=p_\p\sin\phi)$
of the Weyl semimetal for the $z>0$ sample along the $\phi=\f\pi4$ path
(see \secr{surfaces} for the explanation of notation and \secr{ssWz} for the detailed presentation of this case)
in the vicinity of the (former) line node of the line-node semimetal,
calculated within the Luttinger model for two sets of values of the cubic BIA parameters:
(i) $(\be_{31},\be_{32},\be_{33},\be_{34})=(\be_{31}^*,0,0,0)$ (red surface-state band and blue bulk bands),
where $\be_{31}^*$ is the value in \tabr{LM} and the other parameters, including the cubic-power term $\propto \be_{34} \ph_z^3$, are absent,
as used in all subsequent calculations in \secr{ssW} for the Weyl semimetal;
(ii) $(\be_{31},\be_{32},\be_{33},\be_{34}=(2\be_{31}^*,0,0,\be_{31}^*)$ (purple surface-state band and green bulk bands),
with the cubic-power term present.
The second set is chosen this way, so that the low-energy gap $\De^W(\f\pi4)$ [\eq{DeW}] at the line node is the same as for the first set.

In \figr{ssWzcubicBIA}(b) and (c), we plot the paths of the complex momentum solutions
$p_{zn}(p_x=p_\p\cos\phi,p_y=p_\p\sin\phi,\e=\Ec_{z>0}^W(p_\p\cos\phi,p_\p\sin\phi))$ to the characteristic equation \eqn{pzeq}
for the surface states at $\phi=\f\pi4$ as $p_\p$ spans the range of \figr{ssWzcubicBIA}(a).
We observe full confirmation of the above explanation:
there are $2N=8$ complex momentum solutions for the case (i) without the cubic-power term in \figr{ssWzcubicBIA}(b)
and $3N=12$ solutions for the case (ii) with the cubic-power term in \figr{ssWzcubicBIA}(c).
In the latter case, there is clearly a group of 8 low-energy solutions,
labeled ``1-8'', that are in full correspondence with all 8 solutions of the case (i).
But, in addition to that, there are 4 high-energy solutions, labeled ``9-12'', with the absolute value
much larger (about 100 times) than the typical magnitude of the low-energy momentum solutions.
According to the above prescription, the latter 4 solutions were simply dropped in the calculation for the case (ii) and BCs \eqn{bc} were still used.

We see that indeed, as anticipated in \secr{bulkW},
the difference in the bulk and surface-state spectra for the two sets of cubic BIA parameters
amounts to the shift of the whole band structure in momentum and energy;
other than that, the spectra are essentially identical.
Having explicitly proven that,
in all subsequent calculations for the Weyl-semimetal phase (\secr{ssW}),
we use the set (i) $(\be_{31},\be_{32},\be_{33},\be_{34})=(\be_{31}^*,0,0,0)$ of the cubic BIA parameters with only $\be_{31}$ present.

\section{Surface states of the Luttinger Semimetal \lbl{sec:ssL}}

\subsection{Surface states from the Luttinger model}

\begin{figure*}
\centering
\includegraphics[width=\linewidth]{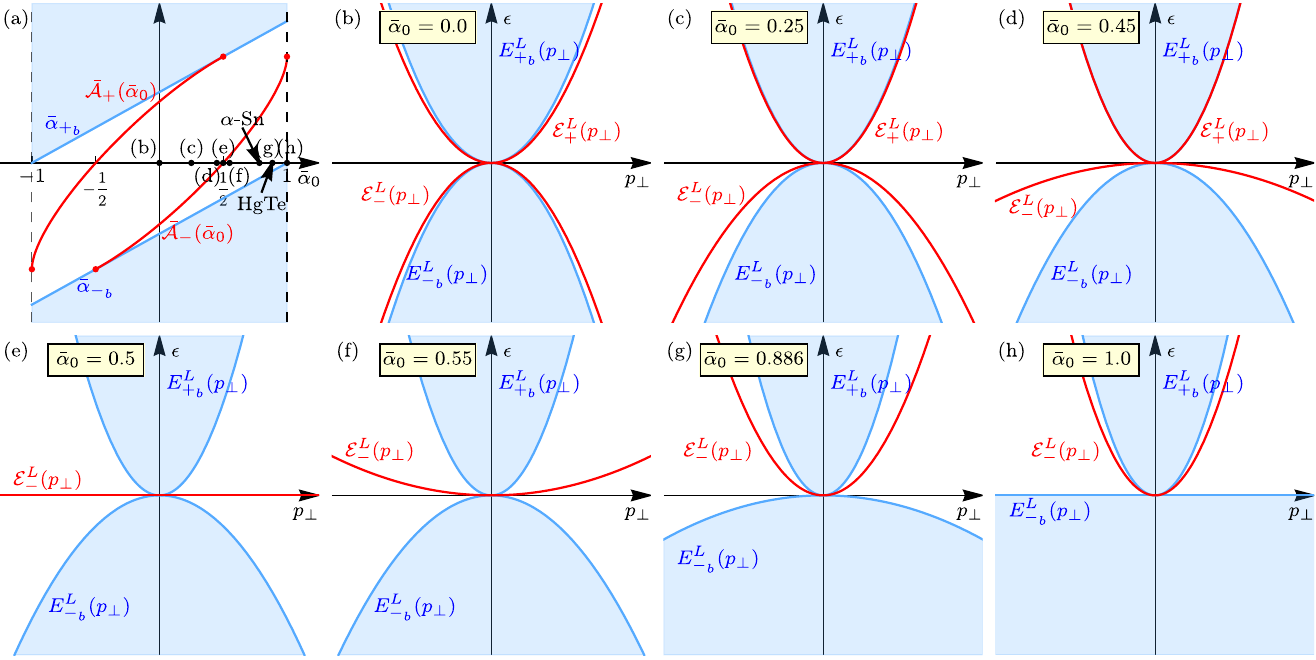}
\caption{
The surface-state spectrum of the unperturbed Luttinger semimetal with $\Ob(3)$ symmetry calculated analytically using the method of \secr{method}
within the Luttinger model with the Hamiltonian $\Hh^L(\pbh)$ [\eqs{HL}{HO3}] and boundary conditions \eqn{bc}.
The surface-state spectrum $\Ec_\pm^L(p_\p)$ [\eq{EcL}] is fully characterized by one dimensionless parameter $\alr_0=\al_0/\al_z$,
which determines the degree of particle-hole asymmetry (the label ``$z>0$'' is dropped for brevity).
In the Luttinger-semimetal regime, $-1<\alr_0<1$.
(a) The ``phase diagram'' of the surface states,
showing the dimensionless curvatures $\Acr_\pm(\alr_0)$ [\eq{Acr}, red] of the surface-state bands as functions of $\alr_0$.
In the ranges $-1<\alr_0<\tf12$, $-\tf12<\alr_0<\tf12$, and $\tf12<\alr_0<1$,
there is one band $\Ec_+^L(p_\p)$, two bands $\Ec_\pm^L(p_\p)$, and one band $\Ec_-^L(p_\p)$, respectively.
The values $\alr_0=0.886$ for HgTe (\tabr{LM}) and $\alr_0=0.784$ for $\al$-Sn
(calculated from $\al_0=18.62\f1{2m_e}$ and $\al_z=23.76\f1{2m_e}$ taken from Ref.~\ocite{Madelung2004}) are indicated.
(b)-(h) The surface-state spectrum (red) for various values of $\alr_0$ within $0\leq \alr_0\leq 1$;
the range $-1\leq \alr_0 \leq 0$ is analogous by symmetry.
(b) $\alr_0=0$, the case of particle-hole symmetry.
(c) and (d) The range $0<\alr_0<\tf12$ of two surface-state bands $\Ec^L_\pm(p_\p)\gtrless0$ with particle- and hole-like characters, respectively,
exemplified with $\alr_0=0.25$ and $0.45$.
(e) The borderline case $\alr_0=\tf12=0.5$ between the regimes $-\tf12<\alr_0<\tf12$ and $\tf12<\alr_0<1$:
the band $\Ec_+^L(p_\p)$ merges altogether with the continuum of the upper bulk band
and the remaining one band $\Ec_-^L(p_\p)\equiv0$ is flat, transitioning from hole- to particle-like character.
(f) and (g) The range $\tf12<\alr_0<1$ of one surface-state band $\Ec^L_-(p_\p)>0$ with particle-like character,
exemplified with $\alr_0=0.55$ and the HgTe value $\alr_0=0.886$.
(h) The end value $\alr_0=1-0$ of the Luttinger-semimetal regime.
The shaded blue areas depict the continua of bulk states, bordered by the bulk-band boundaries $E_{\pm_b}^L(p_\p)$ [\eqs{EL+}{EL-}, blue]
in (b)-(h) and their dimensionless curvatures $\alr_{\pm_b}=1{\pm_b}\alr_0$ in (a).
}
\label{fig:ssL}
\end{figure*}

In Ref.~\ocite{KharitonovLSM}, we have demonstrated that the Luttinger model in the Luttinger-semimetal regime \eqn{LSMregime},
with the Hamiltonian $\Hh^L(\pbh)$ [\eq{HL}] and BCs \eqn{bc}, exhibits surface states.
We also explained their existence in terms of approximate chiral symmetry, by relating this model
to the model of a 2D chiral-symmetric quadratic-node semimetal with the winding number 2.
We reproduce this result here in more detail,
as it serves as the starting point of the subsequent evolution of these surface states in the presence of perturbations.

For spherical symmetry $\Ob(3)$ ($\al_\square=0$), we found the surface states analytically.
Both the bulk and surface-state band structures are fully characterized by one dimensionless parameter
\beq
	\alr_0=\f{\al_0}{\al_z},
\lbl{eq:alr0}
\eeq
which controls the degree of particle-hole asymmetry. In the semimetal regime, $-1<\alr_0<1$ [\eq{LSMregime}].
Depending on the subrange of $\alr_0$, the surface-state spectrum consists of either two or one nondegenerate bands
\beq
    \Ec_{z>0,\pm}^{L}(p_\p;\al_0,\al_z)
    =\Acr_\pm(\alr_0) \al_z p_\p^2,
\lbl{eq:EcL}
\eeq
characterized by the dimensionless curvatures
\beq
    \ba{ll}
    \Acr_+(\alr_0)<0, &\spc -1<\alr_0<-\tf12,\\
    \Acr_\pm(\alr_0)=\tf{\sq{3}}{2}(\sq{3}\alr_0 \pm \sq{1-\alr_0^2}) \gtrless 0, &\spc -\tf12<\alr_0<+\tf12,\\
    \Acr_-(\alr_0)>0, &\spc +\tf12<\alr_0<+1,
    \ea
\lbl{eq:Acr}
\eeq
of the quadratic spectrum.
Here, we present the spectrum for $z\gtrless0$ samples,
where $p_\p=\sq{p_x^2+p_y^2}$ is the absolute value of the 2D momentum $(p_x,p_y)$ along the surface.
Clearly, for $\Ob(3)$ symmetry, an analogous form holds for any other surface orientation.
The spectrum has axial rotation symmetry about the direction perpendicular to the surface.

The surface-state bands \eqn{EcL} lie between the boundaries
\beq
	E_{z>0,+_b}^L(p_\p)=\min_{p_z}\al_{+_b}p^2=\al_{+_b} p_\p^2,
\lbl{eq:EL+}
\eeq
\beq
    E_{z>0,-_b}^L(p_\p)=\max_{p_z}\al_{-_b}p^2=\al_{-_b} p_\p^2
\lbl{eq:EL-}
\eeq
of the bulk spectrum \eqn{eL}:
\[
	\alr_{-_b}<\Acr_\pm(\alr_0)<\alr_{+_b},
\]
where $\alr_{\pm_b}=\al_{\pm_b}/\al_z=1\pm_b \alr_0$.

Both the bulk-band boundaries \eqsn{EL+}{EL-} and the surface-state bands \eqn{EcL} are quadratic $\propto p_\p^2$ in the surface momentum,
since this is the only scaling present in the Hamiltonian \eqn{HL} and BCs \eqn{bc}.
Note that the absence of a scale in the BCs \eqn{bc} we consider is just as important for this quadratic scaling of the surface-state spectrum.
As discussed in \secr{bc}, more general BCs for a quadratic Hamiltonian may involve both the wave function $\psih(x,y,0)$
and its derivatives $\pd\psih(x,y,0)$,
in which case the relative coefficients between them have the physical dimension of length, which would then enter the surface-state spectrum.

The surface-state spectrum is thus fully characterized by the dimensionless curvatures
$\Acr_\pm(\alr_0)$ [\eq{Acr}] as functions of one dimensionless parameter $\alr_0$.
Its ``phase diagram'', depicting this dependence, and notable cases and regimes are presented in \figr{ssL}.
For $-\tf12<\alr_0<\tf12$, there are two bands with the curvatures $\Acr_\pm(\alr_0)\gtrless 0$ of opposite signs
(i.e., the bands have particle-like and hole-like character).
For $\alr_0=0$ [\figr{ssL}(b)], both the bulk and two surface-state bands $\Ec^L_{z>0,\pm}(p_\p)$
with $\Acr_\pm(0)=\pm\f{\sq3}2$ are particle-hole symmetric.
Upon increasing $\alr_0>0$, the curvatures $\Acr_\pm(\alr_0)$ grow, i.e., both surface-state bands move upward, as exemplified in \figr{ssL}(c) and (d).
This continues until the bordeline case $\alr_0=\tf12$ [\figr{ssL}(e)],
when the upper band $\Ec^L_{z>0,+}(p_\p)$ merges with the bulk continuuum at $\Acr_+(\tf12)=\alr_{+_b}=\tf32$ and disappears,
while the lower band $\Ec_{z>0,-}^L(p_\p)$ becomes flat, with the zero curvature $\Acr_-(\tf12)=0$.
Since both the surface-state bands and bulk-band boundaries scale as $\propto p_\p^2$,
the merging occur altogether, for all $p_\p$ at the same time.
Upon further increasing $\alr_0$, only one band $\Acr_-(\alr_0)>0$ remains in the range $\tf12<\alr_0<1$, as exemplified in \figr{ssL}(f) and (g),
whose character has switched from hole-like to particle-like upon passing through the flat-band case at $\alr_0=\tf12$.
At $\alr_0 \rarr 1-0$ [\figr{ssL}(h)], the edge of the semimetal regime is reached,
where the negative curvature $\al_{-_b}\rarr -0$ of the valence band approaches zero.
Interestingly, in this limit, the one surface-state band $\Ec_{z>0,-}^L(p_\p)$ still remains and has an intermediate curvature value $\Acr_-(1)= \f32$.
The behavior in the negative range $-1<\alr_0< 0$ is analogous by symmetry.

For cubic symmetry $\Ob_h$, the general qualitative behavior remains, as long as the cubic anisotropy is weak.
We were able to find the surface states analytically for the $z>0$ sample for the cartesian directions:
\beq
    \Ec^{L,\Ob_h}_{z>0,\pm}(p_x,p_y=0;\al_0,\al_z,\al_\square)=\Acr^{\Ob_h}_\pm (\alr_0,\alr_\square) \al_z p_x^2
\lbl{eq:EcLOh}
\eeq
where the dimensionless curvatures
\beq
	\Acr^{\Ob_h}_\pm (\alr_0,\alr_\square)
	=\tf{\sq3}2[\sq3 \alr_0 \pm\sq{(1-\tf35\alr_\square)^2-\alr_0^2}]
\lbl{eq:AcOh}
\eeq
are fully characterized by the dimensionless parameters $\alr_0$ [\eq{alr0}] and
\beq
	\alr_\square=\f{\al_\square}{\al_z}.
\lbl{eq:alrsq}
\eeq

Cubic anisotropy (even when weak) could potentially lead to some fine-structure effects in the borderline cases,
such as $\alr_0=\pm\tf12$ of the $\Ob(3)$-symmetric case,
when one of the surface-state bands (of the $-\tf12<\alr_0<\tf12$ range) merges with the bulk bands and the other one becomes flat.
We do not explore such fine details here.
For $\alr_0$ away from such borderline cases, moderate cubic anisotropy $|\alr_\square| \ll 1$
only leads to a mild warping of the spectrum, which seems otherwise inconsequential.

\subsection{Asymptotic agreement between the Luttinger and Kane models of the Luttinger semimetal \lbl{sec:LLMKM}}

\begin{figure*}
\centering
\includegraphics[width=\linewidth]{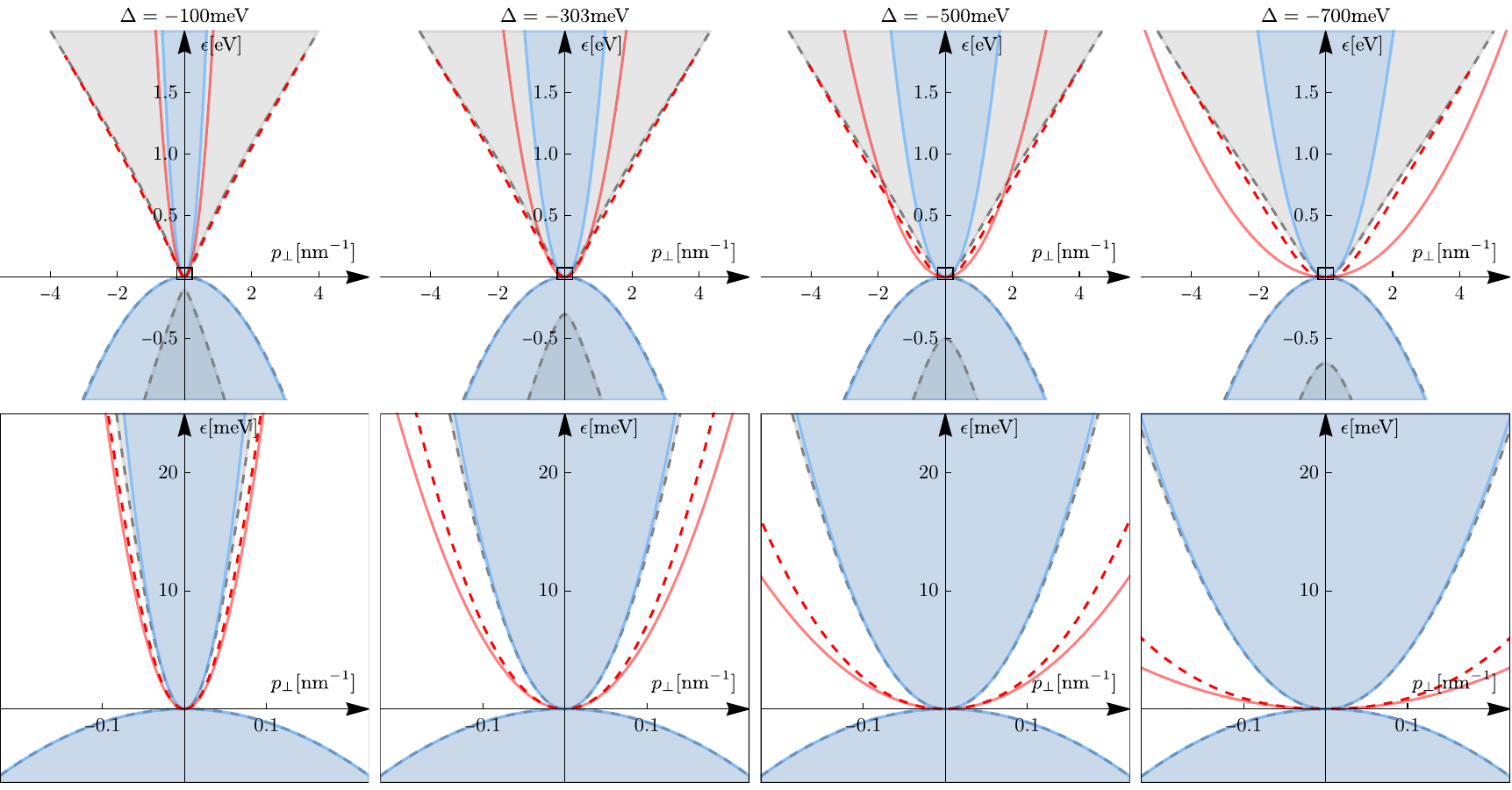}
\caption{
Comparison of the surface-state spectra of the Luttinger semimetal calculated within the Luttinger model
[$\Hh^L(\pbh)$, \eqs{HL}{bc}, lighter solid red for the surface-state bands and blue for the bulk-band boundaries]
and the Kane model [$\Hh^K(\pbh)$, \eqs{HK}{bcK}, dashed red for the surface-state bands and gray for the bulk-band boundaries]
for $\Ob(3)$ symmetry ($\al_\square=0$) for different values of the level spacing $\De$ between the $j=\f12$ and $j=\f32$ states.
The values of other parameters are those of HgTe, given in \tabsr{LM}{KM}.
There is a quantitative asymptotic agreement between the spectra as the node $(\e,p_\p)=(0,0)$ of the Luttinger semimetal is approached.
}
\label{fig:LLMKM}
\end{figure*}

In \figr{LLMKM}, we show the comparison of the bulk and surface-state spectra of the Luttinger [\eqs{HL}{bc}] and Kane [\eqs{HK}{bcK}] models
in the Luttinger-semimetal regime for the case of $\Ob(3)$ symmetry, when the former is the low-energy limit of the latter (\secr{LfromK}).
For the Luttinger model, the spectrum has been obtained analytically above [\eqs{EcL}{Acr}];
for the Kane model, the spectrum has been obtained using the semi-analytical method of \secr{method}.
As explained in \secsr{KM}{scales}, the validity range of the Luttinger model originating from the Kane model
is set by the energy spacing $\De$ between the $j=\f12$ and $j=\f32$ levels.
We present the comparison for different values of $\De$. The values of other parameters are those of HgTe, given in \tabsr{LM}{KM}.

We see that for any $\De$, there is indeed a quantitative asymptotic agreement
between the bulk and surface-state spectra of the two models as the node $(\e,p_\p)=(0,0)$ is approached.
For the Luttinger model, the spectra are exactly quadratic in $p_\p$, while for the Kane model, the spectra have the same quadratic asymptotic behavior.
Quantitative agreement occurs within the validity range $|\e|\ll |\De|$, $p_\p \ll p_\De$ [\eq{pDe}] of the Luttinger model.
We observe that the validity range does expand upon increasing $|\De|$:
for a fixed small enough $p_\p$, the difference between the surface-state bands of the Luttinger (lighter solid red) and Kane (dashed red) models
decreases upon increasing $|\De|$.

Outside of the validity range of the Luttinger model, the spectra of the Luttinger and Kane models deviate.
The main qualitative difference is that for the Luttinger model, the surface-state spectrum does not merge with the bulk-band boundaries,
since all scale quadratically, while the surface-state band of the Kane model does merge with the bulk-band boundaries.

This establishes the first relation in the hierarchy of low-energy models presented in \secr{scales}.

\section{Surfaces and symmetries \lbl{sec:surfsymm}}
\subsection{Surfaces \lbl{sec:surfaces}}

In the next \secssr{ssD}{ssLN}{ssW}, using the method of \secr{method}, we calculate the surface states
of the Dirac, line-node, and Luttinger semimetals, with strain applied along the $z$ direction [\eq{Hu}].
We do so for two characteristic surface orientations:
(i) A system occupying the half-space $y<0$, with the $y=0$ surface parallel to the strain $z$ direction;
with the $(p_x,p_z)$ surface-momentum plane,
for which the two Dirac points $\pb=(0,0,\pm_u p_u)$ are projected onto different points $(p_x,p_y)=(0,\pm_u p_u)$, \figr{e}.
(ii) A system occupying the half-space $z>0$, with the $z=0$ surface perpendicular to the strain $z$ direction;
with the $\pb_\p=(p_x,p_y)$ surface-momentum plane,
for which the two Dirac points are projected onto the same point $(p_x,p_y)=(0,0)$, \figr{e} [\figr{ssDz}].
Specifying the half-space of the system for a given surface
(e.g., whether it is $z>0$ or $z<0$ for the $z=0$ surface; equivalently, considering oriented surfaces)
is important, since the {\em chiral} properties of the surface states of the line-node and Weyl semimetals depend on that;
for the Dirac semimetal, the surface-state spectrum is the same though, due to the symmetries, discussed below.
Accordingly, the surface-state spectra will be denoted as $\Ec_{y<0}(p_x,p_z)$ and $\Ec_{z>0}(p_x,p_y)$,
with the subscripts specifying the half-space system and will be understood as {\em sets} of all the surface-state bands.
As with the Hamiltonians, the superscripts $L$, $D$, $LN$, and $W$ of the surface-state spectra will be added to denote
the Luttinger-, Dirac-, line-node-, and Weyl-semimetal phases, respectively.

\subsection{Symmetries \lbl{sec:symmetries}}

As mentioned in \secssr{bulkD}{bulkLN}{bulkW}, the bulk Dirac semimetal has $\Db_{4h}$ spatial symmetry,
and the bulk line-node and Weyl semimetals have $\Db_{2d}$ spatial symmetry.
For half-infinite samples, the symmetries are deduced as follows.

Since the projected bulk-band boundaries $E_{y<0,\pm_b}(p_x,p_z)$ and $E_{z>0,\pm_b}(p_x,p_y)$ [\eqs{E+}{E-}]
are determined as extrema of the bulk spectrum with respect to momentum component perpendicular to the surface,
their 2D symmetries are all of those of the 3D bulk system with the perpendicular component simply dropped, regardless of whether it is changed or not.
For example, if for some $(p_x,p_y)$ the extremum of the bulk bands $\eps_{\pm_b,\sig}(p_x,p_y,p_z)$ is reached at some $p_z^*$,
then, due to the $\pi$ rotation $(p_x,p_y,p_z)\rarr(p_x,-p_y,-p_z)$ symmetry in $\Db_{2d}$ of the line-node or Weyl semimetal,
the same value of the extremum is reached at $(p_x,-p_y)$, which gives the effective 2D reflection symmetry
$E_{z>0,\pm_b}(p_x,-p_y)=E_{z>0,\pm_b}(p_x,p_y)$.
Therefore, for Dirac, line-node, and Weyl semimetals,
the spatial symmetry of the bulk-band boundaries is $\Cb_{2v}$ for the $y<0$ sample and $\Cb_{4v}$ for the $z>0$ sample.

The surface-state spectrum, on the other hand, has only those symmetries of the bulk that also preserve the sample geometry.
Therefore, for the Dirac semimetal, the spatial symmetry is $\Cb_{2v}$ for the $y<0$ sample and $\Cb_{4v}$ for the $z>0$ sample;
for the line-node and Weyl semimetals, the spatial symmetry is $\Cb_2$ for the $y<0$ sample and $\Cb_{2v}$ for the $z>0$ sample.

In addition to these spatial symmetries, time-reversal symmetry $\Tc_-$ is still preserved for half-infinite samples,
since the BCs \eqn{bc} are $\Tc_-$-symmetric.

\section{Surface states of the Dirac semimetal \lbl{sec:ssD}}

\begin{figure*}
\centering
\includegraphics[width=\linewidth]{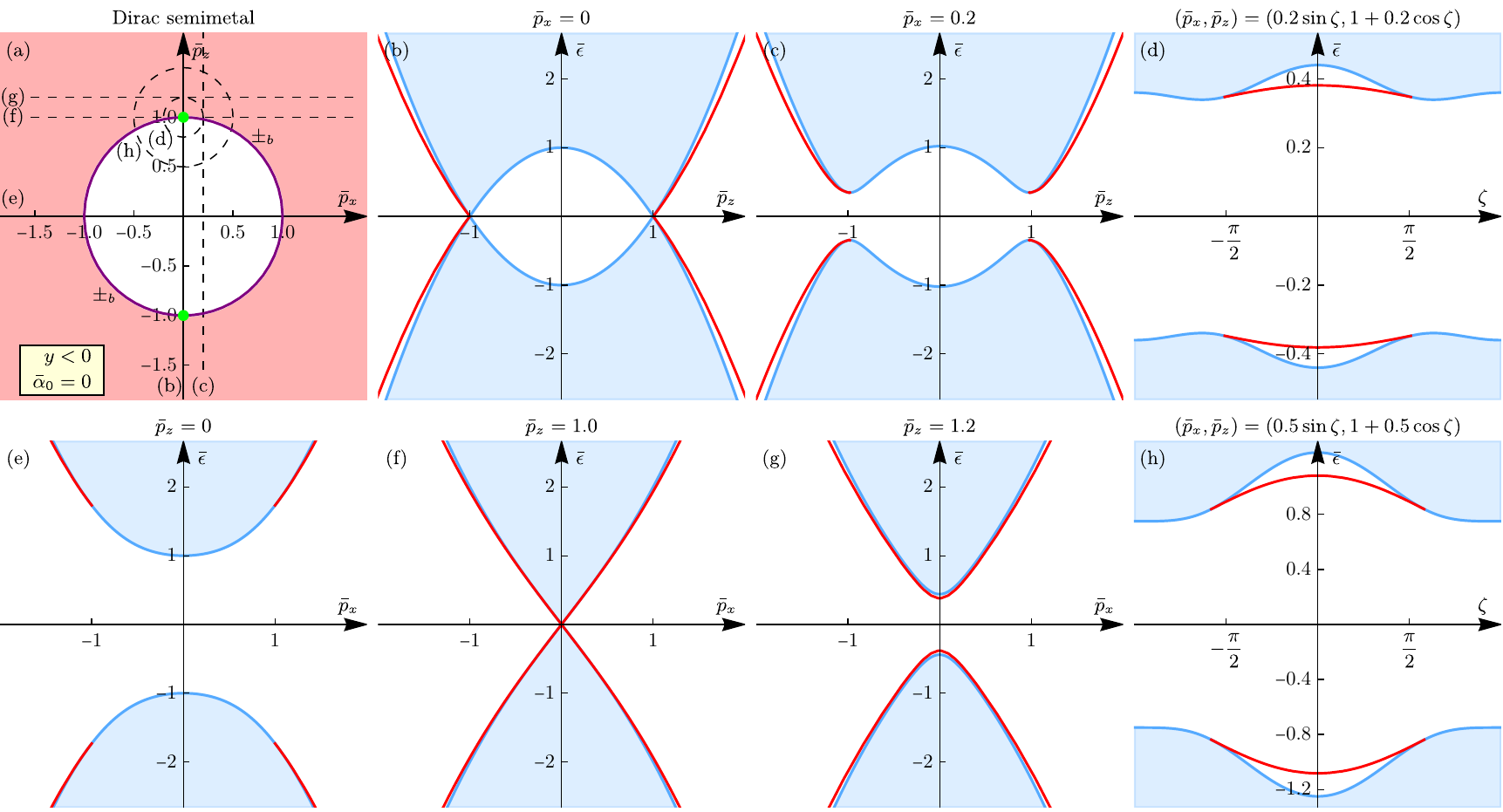}
\caption{
The surface-state spectrum of the Dirac semimetal for the $y<0$ sample
calculated using the ``semi-analytical'' method of \secr{method}
within the Luttinger model with compressive strain [$\Hh^D(\pbh)$, \eqs{HD}{bc}] and absent cubic anisotropy $\al_\square=0$.
The spectrum can be presented in the universal dimensionless form $\Ecr^D_{y<0}(\pr_x,\pr_z;\alr_0)$ [\eq{EcDyscaling}],
fully characterized by the dimensionless parameter $\alr_0=\al_0/\al_z$.
The case $\alr_0=0$ of particle-hole symmetry is presented.
(a) The plane of the dimensionless surface momentum $(\pr_x,\pr_z)$ [\eq{pr}], showing the Fermi contours (red) and
the regions occupied by the surface-state band(s)
(shaded red with different intensity for the regions with one and two bands; see, e.g., \figr{ssDy05} for the difference),
bordered by the merging contours (purple).
The projected $\pm_u$ Dirac points are shown in green.
(b), (c), (e), (f), (g) show the spectrum along straight-line paths,
(d) and (h) show the spectrum along circular paths, indicated in (a).
In (d) and (h), the vertical position of the horizontal axis is the Dirac-point energy $\epsr_{u0}$.
The surface-state bands $\Ecr^D_{y<0}(\pr_x,\pr_z;\alr_0)$
are in red and the boundaries $\Er^D_{y<0,\pm_b}(\pr_x,\pr_z;\alr_0)$ of the continuum (shaded light blue) of the bulk spectrum \eqn{eD} are in blue.
See text for more explanations.
}
\lbl{fig:ssDy0}
\end{figure*}

\begin{figure*}
\centering
\includegraphics[width=\linewidth]{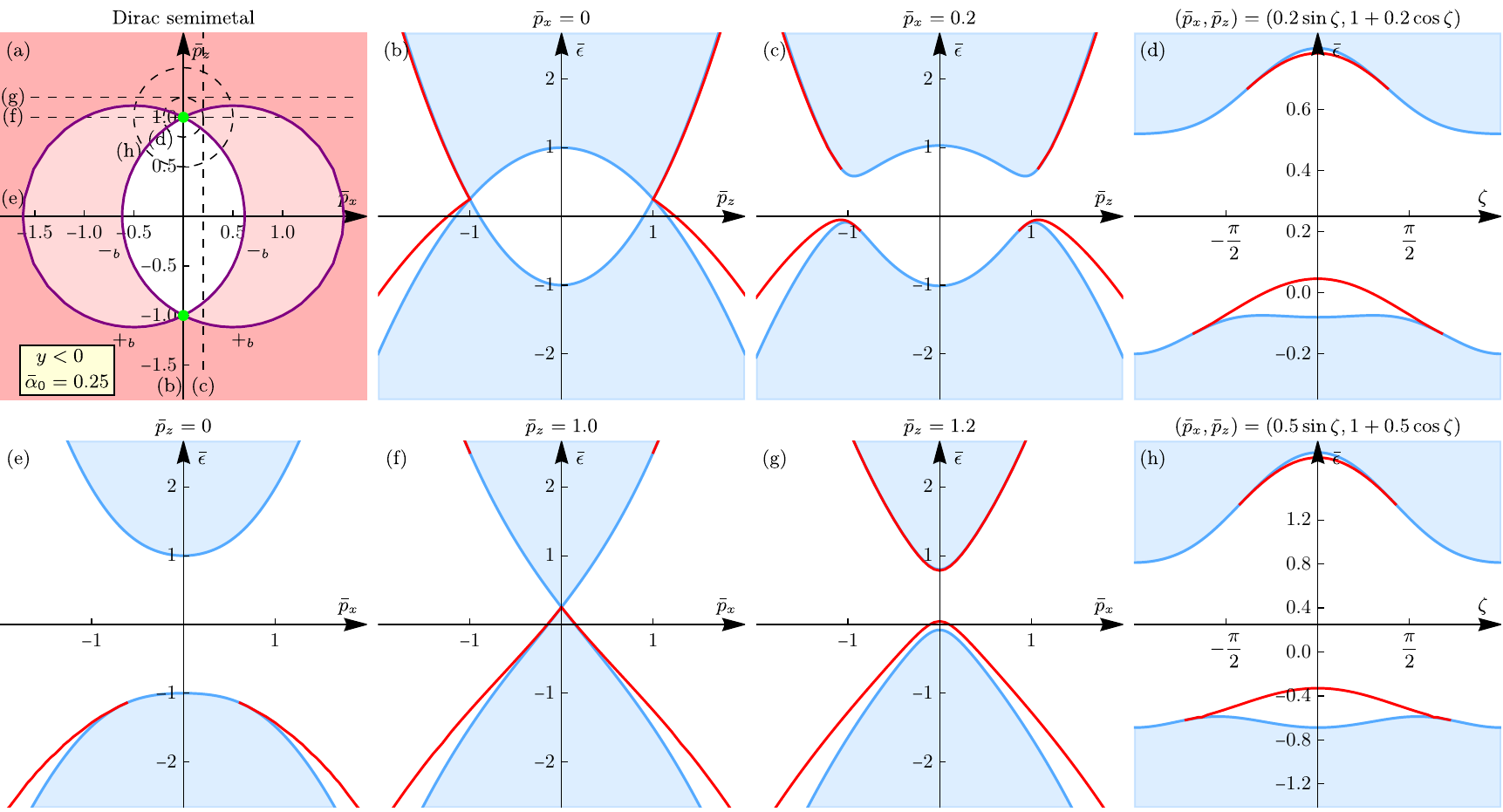}
\caption{
The surface-state spectrum $\Ecr^D_{y<0}(\pr_x,\pr_z;\alr_0)$ of the Dirac semimetal for $\alr_0=0.25$.
Other conditions same as in \figr{ssDy0}.
}
\lbl{fig:ssDy05}
\end{figure*}

\subsection{Surfaces}

In this section, we calculate the surface states of the Dirac semimetal,
which emerges when compressive strain is introduced to the Luttinger semimetal, for the Luttinger model
with the Hamiltonian $\Hh^D(\pbh)$ [\eq{HD}] and BCs \eqn{bc}.
We calculate the whole surface-state spectra $\Ec_{y<0}^D(p_x,p_z)$ and $\Ec_{z>0}^D(p_x,p_y)$
of the Dirac semimetal with the semi-analytical method of \secr{method}
for absent cubic anisotropy, $\al_\square=0$, i.e., for the $\Ob(3)$ symmetry of the Luttinger-semimetal part [\eq{HL}].
For $\al_\square=0$, due to the axial rotation symmetry of the Dirac semimetal,
the surface-state spectrum for any surface with the normal perpendicular to the $z$ axis will be the same as $\Ec^D_{y<0}(p_x,p_z)$ for the $y<0$ sample,
and the surface-state spectrum $\Ec^D_{z>0}(p_\p)$ for the $z>0$ sample will be axially symmetric and depend on $p_\p=\sq{p_x^2+p_y^2}$ only.
As in the Luttinger semimetal, including weak cubic anisotropy $|\alr_\square|\ll 1$ [\eq{alrsq}] in the Dirac semimetal
would only lead to minor inconsequential warping (away from the borderline cases $\alr_0=\pm \tf12$).

\subsection{Scaling}

\begin{figure*}
\centering
\includegraphics[width=\linewidth]{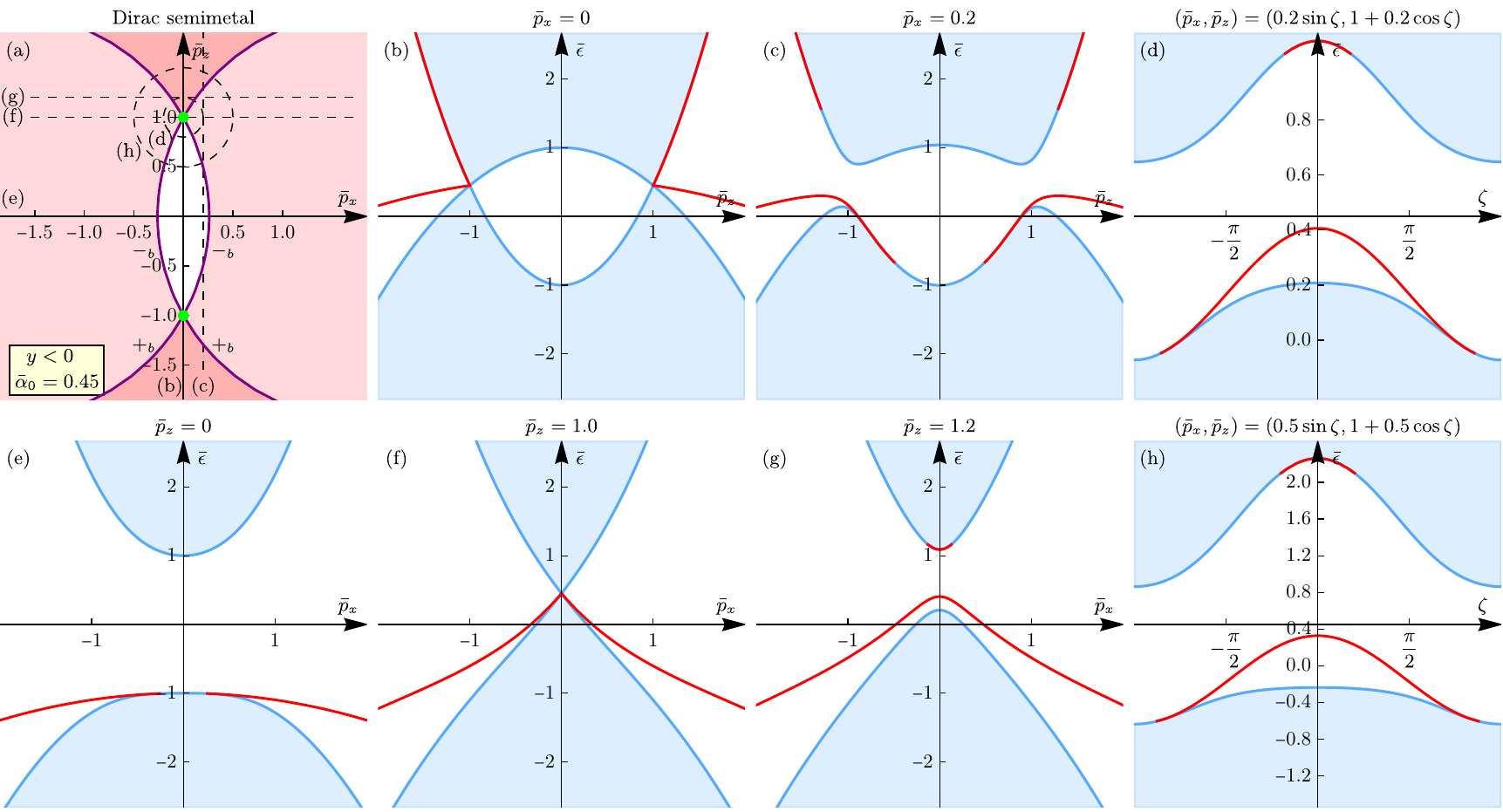}
\caption{
The surface-state spectrum $\Ecr^D_{y<0}(\pr_x,\pr_z;\alr_0)$ of the Dirac semimetal for $\alr_0=0.45$.
Other conditions same as in \figr{ssDy0}.
}
\lbl{fig:ssDy09}
\end{figure*}

\begin{figure*}
\centering
\includegraphics[width=\linewidth]{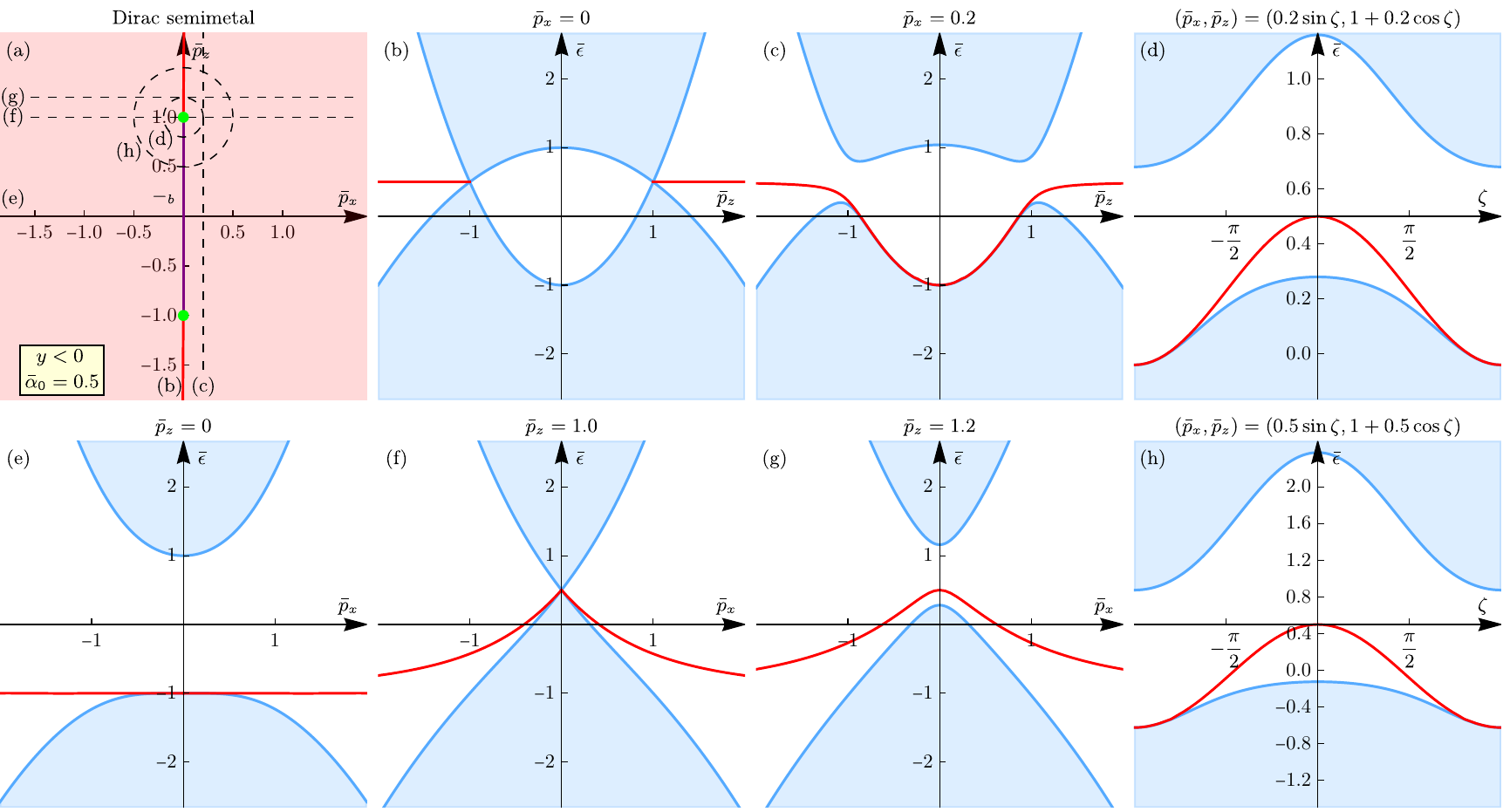}
\caption{
The surface-state spectrum $\Ecr^D_{y<0}(\pr_x,\pr_z;\alr_0)$ of the Dirac semimetal for the borderline case $\alr_0=\tf12=0.5$
between the ranges $-\tf12<\alr_0<\tf12$ and $\tf12<\alr_0<1$ of two and one surface-state bands, respectively.
Other conditions same as in \figr{ssDy0}.
}
\lbl{fig:ssDy1}
\end{figure*}

\begin{figure*}
\centering
\includegraphics[width=\linewidth]{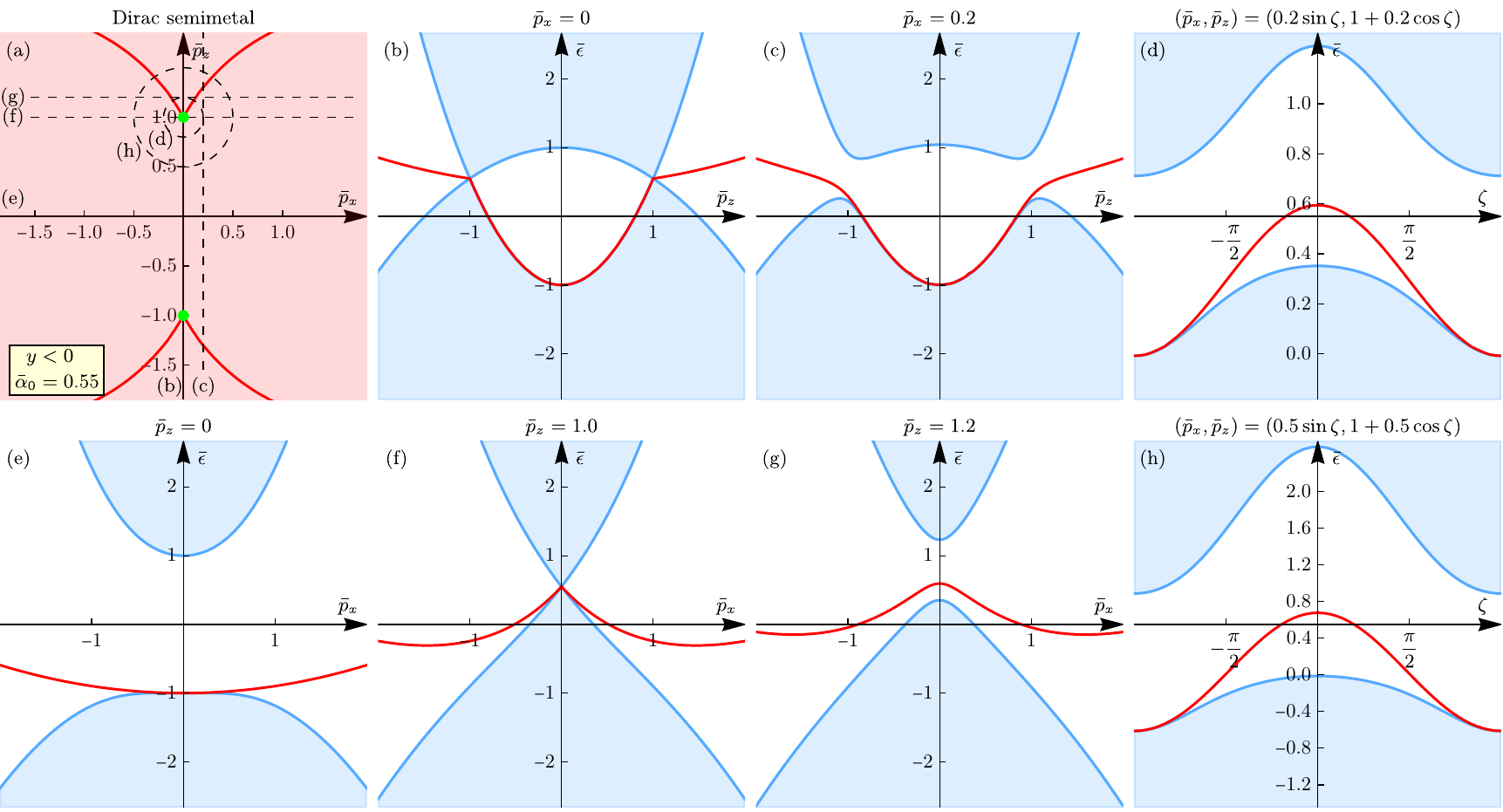}
\caption{
The surface-state spectrum $\Ecr^D_{y<0}(\pr_x,\pr_z;\alr_0)$ of the Dirac semimetal for $\alr_0=0.55$.
Other conditions same as in \figr{ssDy0}.
}
\lbl{fig:ssDy11}
\end{figure*}

\begin{figure*}
\centering
\includegraphics[width=\linewidth]{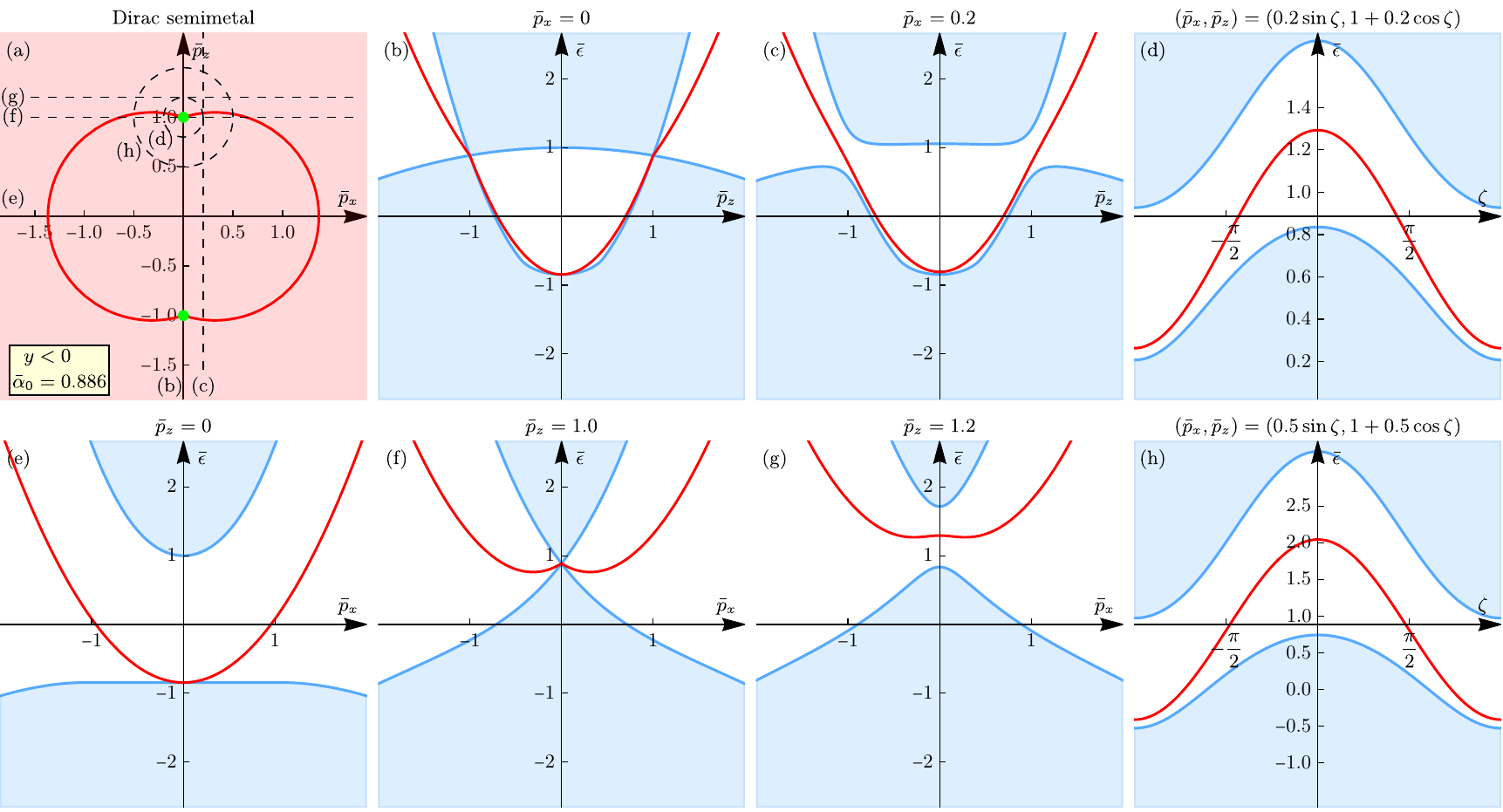}
\caption{
The surface-state spectrum $\Ecr^D_{y<0}(\pr_x,\pr_z;\alr_0)$ of the Dirac semimetal for $\alr_0=886$ of HgTe (\tabr{LM}).
Other conditions same as in \figr{ssDy0}.
}
\lbl{fig:ssDy17}
\end{figure*}

For spherical symmetry $\Ob(3)$ of the Luttinger-semimetal part [\eq{HL}],
the bulk and surface-state spectra of the Dirac semimetal described by the Luttinger model with strain [\eq{HD}]
are still fully characterized by one dimensionless parameter $\alr_0$.
The presence of strain does introduce energy $|u|$ and momentum
\beq
	p_{u0}=\sq{\f{|u|}{\al_z}}
\lbl{eq:pu0}
\eeq
scales, where the latter is the Dirac-point momentum $p_u$ [\eq{pu}] for absent cubic anisotropy $\al_\square=0$;
however, since there are no other scales, its effect is that the spectra universally depend on the relative dimensionless momentum
\beq
	\pbr=(\pr_x,\pr_y,\pr_z)=\f{\pb}{p_{u0}}.
\lbl{eq:pr}
\eeq
The surface-state spectra can always be presented in the form
\beq
    \Ec^D_{y<0}(p_x,p_z;\al_0,\al_z;u)=|u| \Ecr^D_{y<0}(\pr_x,\pr_z;\alr_0),
\lbl{eq:EcDyscaling}
\eeq
\beq
	\Ec^D_{z>0}(p_\p;\al_0,\al_z;u)=|u| \Ecr^D_{z>0}(\pr_\p;\alr_0)
\lbl{eq:EcDzscaling}
\eeq
of the universal dimensionless functions $\Ecr^D_{y<0}(\pr_x,\pr_z;\alr_0)$ and $\Ecr^D_{z>0}(\pr_\p;\alr_0)$
of the dimensionless surface momenta, fully characterized by just one dimensionless curvature parameter $\alr_0$.

We stress that this exact scaling property of the surface-state spectrum of the Dirac semimetal
originating from the Luttinger semimetal under compressive strain
is specifically due to the scaleless property of the Luttinger model:
that its Hamiltonian \eqn{HL} contains only the terms quadratic in momentum
and that the considered BCs \eqn{bc} also have no scale (see discussion in \secr{ssL}).
For models that include higher-energy scales, like the Kane model (\secr{KM}),
this is no longer exact, due to the presence of higher energy $\De$ and momentum $p_\De$ scales.
However, as long as the strain value $|u|$ is within the validity range of the Luttinger model, $|u|\ll |\De|$,
this scaling property remains accurate as well.

\subsection{Asymptotic relation to the Luttinger semimetal}

As the next general property, valid for any surface, at surface momentum much larger than $p_u$,
the surface-state spectra asymptotically approach those [\eq{EcL}] of the Luttinger semimetal:
\beq
    \Ecr^D_{y<0}(\pr_x,\pr_z;\alr_0)=\Acr(\alr_0) (\pr_x^2+\pr_z^2) + o(\pr_x^2+\pr_z^2),\spc \sq{\pr_x^2+\pr_z^2} \gg 1,
\lbl{eq:EcDyasymp}
\eeq
\beq
    \Ecr^D_{z>0}(\pr_x,\pr_y;\alr_0)=\Acr(\alr_0) (\pr_x^2+\pr_y^2) + o(\pr_x^2+\pr_y^2),\spc \sq{\pr_x^2+\pr_y^2} \gg 1,
\lbl{eq:EcDzasymp}
\eeq
where $\Acr(\alr_0)$ denotes the set of curvatures \eqn{Acr} of one or two surface-state bands, depending on $\alr_0$.
We exemplify this asymptotic relation for $y<0$ and $z>0$ samples, but the same, of course, holds for any surface orientation.
This is a manifestation of the general annouced property that the low-energy perturbation, in this case, strain,
affects the spectrum of both the bulk and surface states only at lower momentum,
while at larger momentum, $p\gg p_u$, where the quadratic terms of $\Hh^L(\pbh)$ in $\Hh^D(\pbh)$ [\eq{HD}] dominate over the constant strain term,
the spectrum asympotically approaches that of the Luttinger semimetal.

\subsection{Surface states for $y<0$ sample}

We describe the spectra $\Ecr^D_{y<0}(\pr_x,\pr_z;\alr_0)$ by
the 2D plots presented in \figsdr{ssDy0}{ssDy17} for the six parameter values
$\alr_0=0,0.25,0.45,0.5,0.55,0.886$, respectively.
Just like for the Luttinger semimetal,
the surface-state spectrum of the Dirac semimetal is qualitatively different in the regimes $|\alr_0|<\tf12$ and $\tf12<|\alr_0|<1$.
We plot the surface-state spectra $\Ecr^D_{y<0}(\pr_x,\pr_z;\alr_0)$ for the following values of $\alr_0$:
the particle-hole symmetric case $\alr_0=0$ (\figr{ssDy0}) and $\alr_0=0.25$ (\figr{ssDy05}) and $0.45$ (\figr{ssDy09}),
as three representative cases of the range $|\alr_0|<\tf12$;
the borderline case $\alr_0=\tf12=0.5$ (\figr{ssDy1}) between the two regimes;
the cases $\alr_0=0.55$ (\figr{ssDy11}) and $0.886$ (\figr{ssDy17})
(where the latter is the value for HgTe), as two representative cases of the range $\tf12<|\alr_0|<1$.
The cases $\alr_0=0.45$ and $\alr_0=0.55$, being close to the borderline case $\alr_0=\tf12$,
help better understand the transition between the two regimes.

In each of \figsdr{ssDy0}{ssDy17},
(a) shows the surface-momentum plane $(\pr_x,\pr_z)$, (b),(c),(e),(f), and (g) show the spectra along the straight-line paths,
and (d) and (h) show the spectra along the circles $(\pr_x,\pr_z)=(\pr_r\sin\zeta,1+\pr_r\cos\zeta)$, $\zeta\in[0,2\pi)$,
of radii $\pr_r=0.2,0.5$, respectively, centered at the projected Dirac point $(\pr_x,\pr_z)=(0,1)$.
The panel (a) with the surface-momentum plane shows the following:
regions (shaded red) represent one or two surface-state bands according to the intensity,
bounded by the merging contours (purple),
where the surface-state bands merge with the bulk-band boundaries $\Er^D_{y<0,\pm_b}(\pr_x,\pr_z;\alr_0)$, labeled accordingly with $\pm_b$.
Also, red curves show the fixed-energy contours of the surface-state spectrum in the surface-momentum $(\pr_x,\pr_z)$ plane, obtained as crossing
\[
	\Ecr^D_{y<0}(\pr_x,\pr_z;\alr_0)=\epsr_{u0}
\]
with the level [\eq{epsu}]
\[
	\epsr_{u0}=\f{\eps_u|_{\al_\square=0}}{|u|}=\alr_0
\]
of the Dirac points (for $\alr_\square=0$) in dimensionless units.
Assuming the Fermi level is at that Dirac-point energy, these fixed-energy contours are often called {\em Fermi contours}.

In the range $|\alr_0|<\tf12$, there are two surface-state bands, which, however, exist not at all momenta.
The two surface-state bands are present everywhere except for finite-size regions in-between the Dirac points,
enclosed by the merging contours with the $\pm_b$ bulk-band boundaries that pass through both Dirac points $\pr_z=\pm1$.
At $\alr_0=0$, due to particle-hole symmetry, the $\pm_b$ merging contours are the same [\figr{ssDy0}(a)];
for any other $0<|\alr_0|<\tf12$ [\figr{ssDy05}(a) and \figr{ssDy09}(a)], they differ.
At large momentum, the two surface-state bands asymptotically do recover those of the Luttinger semimetal, as expected according to \eq{EcDyasymp}.

In the range $\tf12<\alr_0<1$, on the other hand, there is only one surface-state band, which exists everywhere,
except, interestingly, for one point $(\pr_x,\pr_z)=(0,0)$, where it touches the $-_b$ bulk-band boundary.
The band also connects to both Dirac points.
In this range $\tf12<\alr_0<1$, at large momenta,
this one surface-state band also asymptotically recovers that of the Luttinger semimetal [\eq{EcDyasymp}].

The transition between the two regimes occurs continuously as follows:
upon increasing $\alr_0$ in the range $0<\alr_0<\tf12$,
the region of absent upper surface-state band expands, and eventually disappears at infinity at $\alr_0=\tf12$.
The region of absent lower surface-state band shrinks along the $\pr_x$ direction, so that at $\alr_0=\tf12$
the two parts of the merging contour at $\pr_x\gtrless 0$ collapse onto one straight-line segment at $\pr_x=0$, occupying $-1<\pr_z<1$,
connecting the projected Dirac points.
Upon increasing $\alr_0=\tf12+0$ infinitesimally, the now only surface-state band detaches from this line everywhere except
for the point $(\pr_x,\pr_z)=(0,0)$.

\begin{figure*}
\centering
\includegraphics[width=.5\linewidth]{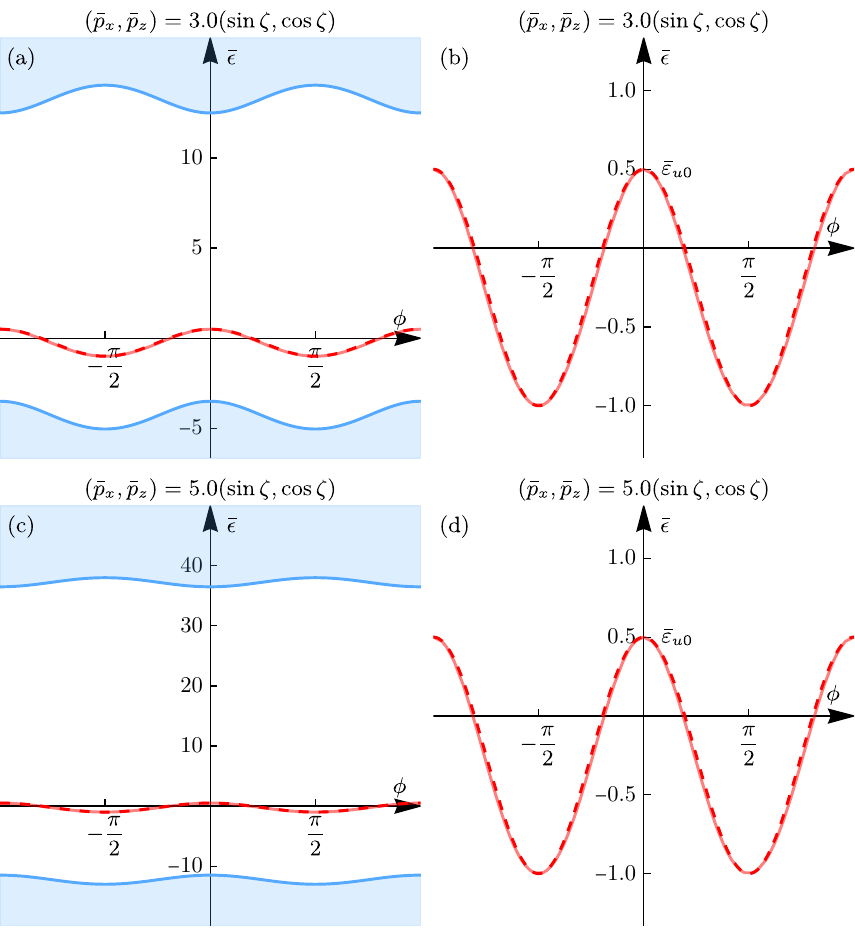}
\caption{
The surface-state spectrum $\Ecr^D_{y<0}(\pr_r\sin\zeta,\pr_r\cos\zeta;\alr_0=\tf12)$ (lighter red) of the Dirac semimetal
along the circular paths of larger radii $\pr_r=3$ [(a) and (b)] and $5$ [(c) and (d)], centered at $(\pr_x,\pr_z)=(0,0)$,
for the borderline case $\alr_0=\tf12$. Other conditions same as in \figr{ssDy0}.
The spectrum fits very well to the trial function \eqn{EcDfit} (dashed red), suggesting a possible analytical solution.
}
\lbl{fig:ssDylargepr}
\end{figure*}

The borderline case $\alr_0=\tf12$ that continuously connects these two regimes is itself particularly interesting.
For the Luttinger semimetal obtained from the Luttinger model,
one surface-state band merges with the bulk continuum, while the curvature $\Acr_-(\alr_0=\tf12)=0$ of the other quadratic band
vanishes and it becomes completely flat [\figr{ssL}(e)].
For the Dirac semimetal obtained from the Luttinger model, we believe that, based on the general asymptotic argument, one can only conclude that
the surface-state energy $\Ecr_{y<0}^D(\pr_x,\pr_z;\alr_0=\tf12)=o(\pr_x^2+\pr_z^2)$ grows slower than a quadratic power in this limit,
as written in \eqs{EcDyasymp}{EcDzasymp}.
The actual calculation shows that the limit $\Ecr_{y<0}^D(\pr_x,\pr_z;\alr_0=\tf12)$ at large $\sq{\pr_x^2+\pr_z^2}$
is finite, but depends on the direction of $(\pr_x,\pr_z)$.
In \figr{ssDylargepr}, we plot the surface-state spectrum $\Ecr_{y<0}^D(\pr_r\sin\zeta,\pr_r\cos\zeta;\alr_0=\tf12)$
along the circles of relatively large radii $\pr_r=3$ and $5$ centered at $(\pr_x,\pr_z)=(0,0)$.
It appears to fit very accurately the trial function
\beq
	\tf12+\tf34(\cos2\zeta-1).
\lbl{eq:EcDfit}
\eeq
This suggests that this could be the exact next-order term of the large-momentum expansion
[the leading term of which is the Luttinger-model result \eqs{EcDyasymp}{EcDzasymp}],
which could perhaps be demonstrated analytically, e.g., via a variant of perturbation theory.
We also observe that the spectrum along the directions $\Ecr^D_{y<0}(0,\pr_z;\alr_0=\tf12)\equiv \tf12 =\epsr_{u0}$ (for $|\pr_z|>1$) [\figr{ssDy1}(b)]
and $\Ecr^D_{y<0}(\pr_x,0;\alr_0=\tf12)\equiv-1$ [\figr{ssDy1}(e)] is exactly flat.

Regarding the Fermi contours [panel (a) of \figsdr{ssDy0}{ssDy17}], in the whole range $-\tf12<\alr_0<\tf12$,
the two surface-state bands never cross the Dirac-point energy level, and hence, there are no such Fermi contours.
In the range $\tf12<\alr_0<1$, the one surface-state band does cross the Dirac-point energy level and therefore, there is a Fermi contour.
There are two symmetric smooth parts of the contour in the $\pr_x\gtrless 0$ regions, each connected to both Dirac points.
We discuss these properties below in \secr{Ddiscussion}, in relation to previous works.

The qualitatively different situations in the regimes $|\alr_0|<\tf12$ and $\tf12<|\alr_0|<1$ is clearly manifested in the spectra along circular paths
around the projected Dirac point [panels (d) and (h) of \figsdr{ssDy0}{ssDy17} and \figr{ssDylin}],
which further confirm the general properties of the surface-state bands (in particular, their merging behavior).
For $|\alr_0|<\tf12$, there are two surface-state bands, both occupy finite segments of the circle and merge with respective bulk-band boundaries.
For $\tf12<|\alr_0|<1$, there is one surface-state band, which exists on the whole circle and has no merging points.
The transition between these regimes occurs at $\alr_0=\tf12$ as follows: one band shrinks and disappears as its two merging points meet;
the other band extends and detaches from the bulk-band boundary as its two merging points meet.

Due to the asymptotic linear scaling of the surface-state spectrum in the vicinity of the Dirac points,
for small enough $\pr_r$, the surface-state spectrum
\[	
	\Ecr^D_{y<0}(\pr_r\sin\zeta,1+\pr_r\cos\zeta;\alr_0)
	=\epsr_{u0}+\Vcr^D_{y<0}(\zeta) \pr_r+o(\pr_r), \spc \pr_r\rarr 0,
\]
is fully characterized by its velocity $\Vcr^D_{y<0}(\zeta)$ as a function of the angle $\zeta$.
Therefore, the surface-state spectrum along one circular path of small enough radii, being proportional to these velocities,
is already sufficient to fully characterize the whole local surface-state spectrum as a function of 2D surface momentum $(\pr_x,\pr_z)$.
In particular, it captures the merging behavior with the bulk bands
and the topological properties in the vicinity of the node, as will be demonstrated in detail in \secr{ssW}.

We see that the behaviors of the surface-state spectrum (controlled by a single dimensionless parameter $\alr_0$)
at large momenta [\eqs{EcDyasymp}{EcDzasymp}], where it asymptotically approaches that of the Luttinger semimetal,
and in the vicinity of the Dirac points are related in a consistent way.
There is a qualitatively different behavior in the regimes $|\alr_0|<\tf12$ and $\tf12<|\alr_0|<1$ in both limits of momentum.

\subsection{Surface states for $z>0$ sample}

\begin{figure*}
\centering
\includegraphics[width=\linewidth]{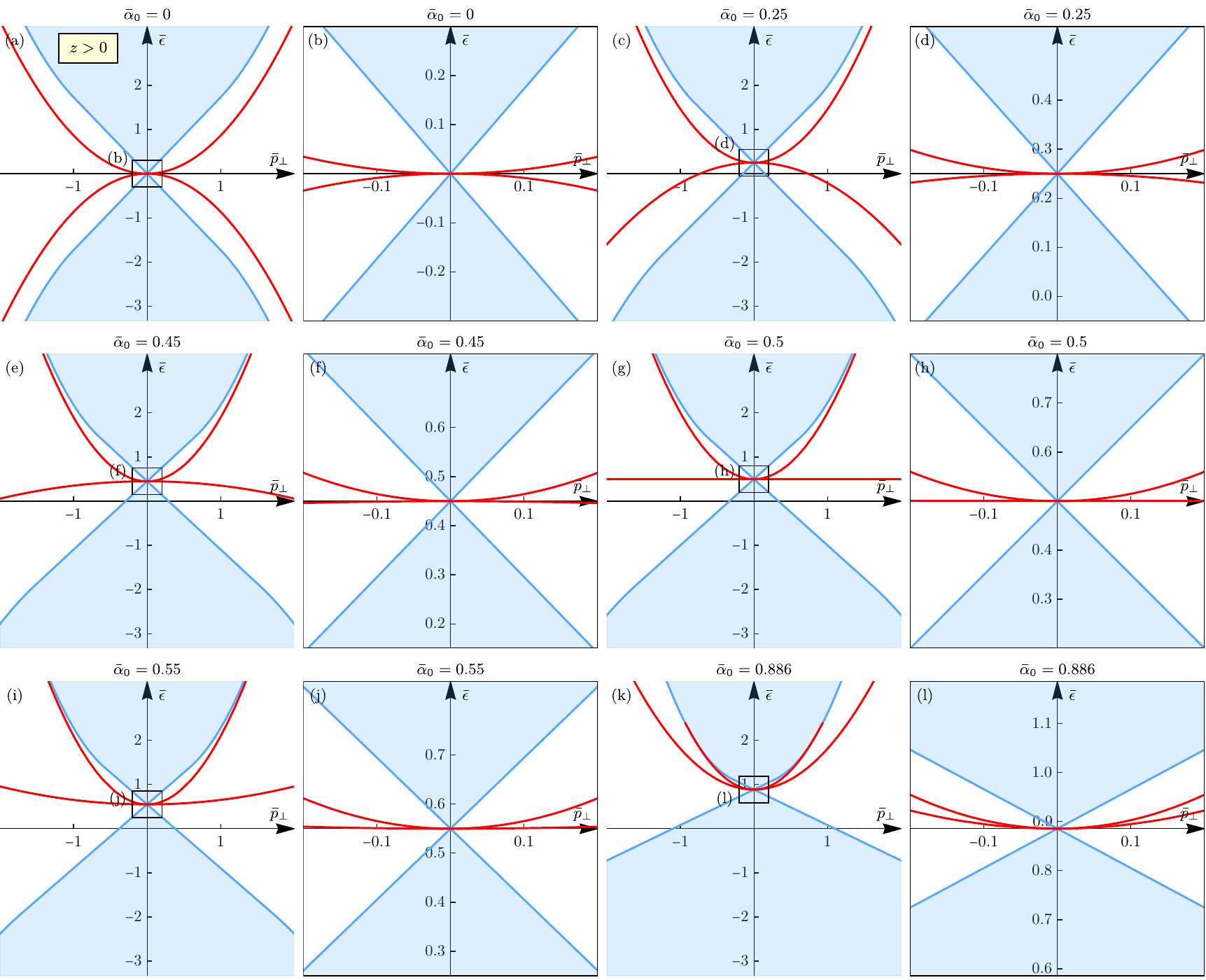}
\caption{
The surface-state spectrum $\Ecr^D_{z>0}(\pr_\p;\alr_0)$ [\eq{EcDzscaling}]
of the Dirac semimetal for the $z>0$ sample for the six cases $\alr_0=0,0.25,0.45,0.5,0.55,0.886$.
Other conditions same as in \figr{ssDy0} (including absent cubic anisotropy $\al_\square=0$).
The two Dirac points are projected onto one point $\pr_\p=0$.
In the vicinity of $\pr_\p=0$, the surface-state bands $\Ecr^D_{z>0}(\pr_\p;\alr_0)$ (always two at smaller $p_\p$)
have a quadratic asymptotic behavior, while the bulk-band boundaries $\Er^D_{z>0,\pm_b}(\pr_\p;\alr_0)$ have a linear one.
}
\lbl{fig:ssDz}
\end{figure*}

In \figr{ssDz}, we present the surface states $\Ecr^D_{z>0}(\pr_\p;\alr_0)$ of the Dirac semimetal for the $z>0$ sample
for the six cases $\alr_0=0,0.25,0.45,0.5,0.55,0.886$.
For the $\Ob(3)$ symmetry of the Luttinger-semimetal part ($\al_\square=0$), the surface-state spectrum has axial symmetry
and depends only on the absolute value $\pr_\p$ (including moderate cubic anisotropy $\al_\square$ would only lead to some warping of the spectrum).
The bulk-band boundaries $\Er^D_{z>0,\pm_b}(\pr_\p;\alr_0)$
are asymptotically linear in the vicinity of the projected double Dirac point at $\pr_\p=0$.
We find that in the whole Luttinger-semimetal regime $-1<\alr_0<1$, the structure of the surface-state spectrum near $\pr_\p=0$ is the same:
there are two surface-state bands in the vicinity of the projected Dirac point with quadratic asymptotic behavior.
Since at large momentum the behavior has to agree with that of the Luttinger semimetal [\eq{EcDzasymp}],
the behavior is different at intermediate momenta $\pr_\p\sim 1$ in the two regimes.
For $|\alr_0|<\tf12$, the two surface-state bands originating from the Dirac point evolve into the two surface-state bands
of the Luttinger semimetal at large momenta.
For $\tf12<|\alr_0|<1$, one of the bands originating from the Dirac point merges with the bulk-band boundary at intermediate momentum $\pr_\p\sim 1$
[as seen in \figr{ssDz}(k)],
while the other band evolves into the one surface-state band of the Luttinger semimetal at large momenta.
In the borderline case $\alr_0=\f12=0.5$, one of the bands is exactly flat [\figr{ssDz}(g),(h)].

\subsection{Discussion \lbl{sec:Ddiscussion}}

We now discuss the above-established properties of the surface states of the Dirac semimetal
that originates from the Luttinger semimetal under compressive strain, calculated within the Luttinger model,
in relation to the previous works.

Surface states of Dirac semimetals are often compared to those of Weyl semimetals.
The topological properties of Weyl semimetals are directly tied to those of quantum anomalous Hall systems
and their surface states obey quantum Hall topology via bulk-boundary correspondence.
We discuss the topological properties of the Dirac, line-node, and Weyl semimetal in more detail in \secr{ssW}, after we introduce the necessary concepts.
In the vicinity of a single projected Weyl point, the surface-state band is guaranteed to exist,
since the band is chiral along the path enclosing the Weyl point and its chirality is equal to the Chern number $\pm1$ of the Weyl point,
as has been recently shown for the most general linear-in-momentum model~\cite{KharitonovFGCM}.
Hence, the Fermi contour of the surface-state band (its fixed-energy cut, called so assuming the Fermi energy is at the Weyl-point level)
is also always guaranteed to exist in the vicinity of a single projected Weyl point.

A Dirac point has the structure of two related Weyl points with opposite Chern numbers $\pm1$ and has the net zero $+1-1=0$ Chern number.
Based on this, one can conclude that surface states in Dirac semimetals are not enforced to be present, at least not by quantum Hall topology,
since a Dirac semimetal is a topologically trivial system in terms of quantum Hall topology.
Nonetheless, in various theoretical models~\cite{Armitage2017},
surface-state bands are typically present in the vicinity of Dirac points.
Here, we also see that surface-state band(s) are present generically [at least for the used BCs \eqn{bc}]
in the whole Luttinger-semimetal regime $-1<\alr_0<1$.

By analogy with the surface-state Fermi contours in Weyl semimetals, which {\em are} guaranteed to exist,
in Refs.~\ocite{Kargarian2016,Kargarian2017}, a particular attention was paid to
the presence or absence of the surface-state Fermi contours in Dirac semimetals, especially in the vicinity of Dirac points.
In one~\cite{Kargarian2016} presented four-band lattice model of a Dirac semimetal,
Fermi contours existed, but were not connected to the Dirac points;
in the other~\cite{Kargarian2017} eight-band model, Fermi contours were completely absent.

Here, we present a model (Luttinger model with strain) with the minimal number of bands (four, double-degenerate valence and conduction bands)
that does not have surface states at the Dirac-point energy at all (for the $y<0$ sample, for the strain along the $z$ direction),
neither connected to nor disconnected from the Dirac points, in the substantial range $-\tf12<\alr_0<\tf12$ (half) of the parameter space $-1<\alr_0<1$.
We stress that, nonetheless, two surface-state bands do exist in this regime and they are also connected to both Dirac points;
they just never cross the Dirac-point energy level.
(Therefore, one should always make it clear whether one is talking about the presence or absence of the surface-state bands altogether,
or just whether they cross the Fermi level or not.)
We also fully characterize the evolution with $\alr_0$ between the regimes of two surface-state bands with absent Fermi contours ($-\tf12<\alr_0<\tf12$)
and one surface-state band with a present Fermi contour ($\tf12<|\alr_0|<1$).
The circular-path plots (\figr{ssDylin}) make it particularly clear how the Fermi contours disappear:
the band simply moves from crossing to not crossing the Dirac-point energy level;
in the borderline case $\alr_0=\tf12$ between the two regimes, the extremum of the band is exactly at that level.
We see that, at least in this model, this evolution also coincides with the other surface-state band disappearing
and the band in question detaching from the bulk-band boundary.

Therefore, the absence of the surface-state Fermi contours connected to the Dirac points
(which, again, does not at all rule out the presence of the surface-state bands)
does not seem to be rare at all and could be expected in more systems and models
(especially the behavior in the vicinity of the Dirac points seems to be a more generic low-energy feature than the model from which it was been derived).

\section{Surface states of the Dirac semimetal within the linear-in-momentum model \lbl{sec:ssDlin}}

In this section, we calculate the surface states of the Dirac semimetal in the vicinity of the Dirac points within the linear-in-momentum model.
The Hamiltonian \eqn{HcD} for the latter has already been derived in \secr{bulkDlin}.
Here, we first derive the BCs for the linear-in-momentum model and then analytically calculate the surface-state spectrum.
The main goal is to demonstrate the expected asymptotic agreement
in the vicinity of the Dirac points with the surface states calculated from the Luttinger  model in \secr{ssD}.

\subsection{Boundary conditions for the linear-in-momentum model of the Dirac semimetal \lbl{sec:bcD}}

\subsubsection{Systematic procedure}

We want to derive the BCs for the wave function $\Psih(\rb)$ [\eq{PsiD}] of the linear-in-momentum low-energy model [\eq{HcD}] of the Dirac semimetal
from the BCs \eqn{bc} of the Luttinger model.
This is another instance of the general systematic low-energy-expansion
procedure~\cite{KharitonovLSM,Samokhin,KharitonovSC,KharitonovQAH,AkhmerovPRB,KharitonovGBCs}
of deriving BCs from the underlying ``microscopic'' models, which for the system in question is as follows.

For any surface, consider the stationary Schr\"odinger equation
\beq
    \Hh^D(\pbh)\psih(\rb)=\eps_u\psih(\rb)
\lbl{eq:HDscheq}
\eeq
for the Luttinger Hamiltonian of the Dirac semimetal at the Dirac-point energy $\e=\eps_u$ [\eq{epsu}].
There are eight particular solutions with real momenta $\pb=(0,0,\pm_u p_u)$, the general linear combinations
of which reads
\beq
    \psih^\x{bulk}(z)=\Psih_{+_u} e^{+_u i p_u z}+ \Psih_{-_u} e^{-_u i p_u z},
\lbl{eq:psibulk}
\eeq
where $\Psih_{\pm_u}$ are arbitrary constant four-component vectors.
These vectors represent the low-energy wave function $\Psih(\rb)$ [\eq{PsiD}].

For a given surface, we look for a general solution to \eq{HDscheq} that does not grow exponentially into the bulk.
Such general solution is a linear combination of particular solutions with real momenta for the directions along the surface,
but with generally complex momenta for the directions perpendicular to the surface.
Among these particular solutions, there will necessarily be the above plane-wave bulk solutions \eqn{psibulk}
with real momenta, representing the low-energy wave function.
However, for some surfaces, there will be additional particular solutions, exponentially decaying into the bulk
over spatial scales much shorter that the large spatial scales of interest considered in the low-energy model.
Applying the BCs \eqn{bc} of the original model to such general solution and excluding the coefficients of the latter decaying solutions,
we obtain the relations for the constants $\Psih_{\pm_u}$, which represent the sought BCs for the low-energy wave function.

Below we apply this procedure specifically to the $z>0$ and $y<0$ samples.

\subsubsection{$z>0$ sample}

For the $z>0$ sample, the Dirac points $(0,0,\pm_u p_u)$ are projected onto the same momentum $(p_x,p_y)=(0,0)$ in the surface-momentum plane.
We look for the general solution $\psih(z)$ to
\[
    \Hh^D(0,0,\ph_z)\psih(z)=\eps_u\psih(z)
\]
that does no grow exponentially into the bulk, as $z\rarr +\iy$.
The Hamiltonian $\Hh^D(0,0,\ph_z)$ is diagonal and we immediately find that the bulk plane-wave solutions [\eq{psibulk}]
at the nodal points $p_z=\pm_u p_u$ are the only particular solutions. So, the general solution reads
\[
    \psih(z)=\psih^\x{bulk}(z)=\Psih_{+_u} e^{+i p_u z}+ \Psih_{-_u} e^{-i p_u z}.
\]
Applying the BCs \eqn{bc}, we obtain
\[
    \psih(z=0)=\Psih_{+_u} + \Psih_{-_u} =\nm.
\]
To leading order, these relations hold for present coordinate dependence $\Psih(\rb)$ of the low-energy wave function.
Therefore, upon the substitution $\Psih_{\pm_u}\rarr \Psih_{\pm_u}(x,y,z=0)$, these represent the BCs
\beq
    \Psih_{+_u}(x,y,z=0) + \Psih_{-_u}(x,y,z=0)=\nm
\lbl{eq:bcPsiz}
\eeq
for the low-energy wave function of the linear model of the Dirac semimetal for $z>0$ sample.
We see that different $j_z$ components are not coupled by such BCs, while the two Dirac points $\pm_u$ for each $j_z$ component are coupled.

\subsubsection{$y<0$ sample}

For the $y<0$ sample, the Dirac points $(0,0,\pm_u p_u)$ are projected
onto the different momenta $(p_x,p_z)=(0,\pm_u p_u)$ in the surface-momentum plane.
We look for the general solution of the form
\beq
    \psih(y,z)=\psih_{+_u}(y) e^{+_u i p_u z}+\psih_{-_u}(y) e^{-_u i p_u z}
\lbl{eq:psiyz}
\eeq
to the Schr\"odinger equation at $\e=\eps_u$ that does not grow exponentially into the bulk, as $y\rarr -\iy$.
Due to the linear independence of the functions $e^{\pm_u i p_u z}$, we have the decoupled equations
\beq
    \Hh^D(0,\ph_y,\pm_u p_u)\psih_{\pm_u}(y)=\eps_u\psih_{\pm_u}(y)
\lbl{eq:HDyuscheq}
\eeq
for the functions $\psih_{\pm_u}(y)$.

In this case, the situation is technically more complicated. In the Hamiltonians $\Hh^D(0,\ph_y,\pm_u p_u)$ with nonzero $\ph_y$,
i.e., for coordinate-dependent $\psih_{\pm_u}(y)$, all four wave-function components with definite $j_z$ are coupled.
For each Dirac point $\pm_u$, we find that there are six momentum $p_y$ solutions to the characteristic equation
\[
    \det[\Hh^D(0,p_y,\pm_u p_u)-\eps_u \um_4]=0,
\]
for which the particular solutions are not growing exponentially into the bulk:
there is a four-fold-degenerate solution $p_y=0$, representing the bulk low-energy wave function,
and a double-degenerate solution

\[
	p_y=-i\kappa, \spc
    \kappa
    =2 p_u \f{\f{\sq3}2(\al_z+\tf25\al_\square)}{\sq{(\al_z-\f35\al_\square)^2-\al_0^2}}= 2p_u \f{v_\p}{\sq{v_z^2-v_0^2}}
\]
for which the particular solutions decay into the bulk.
The corresponding general solution to \eq{HDyuscheq} not growing into the bulk reads
\beq
	\psih_{\pm_u}(y)=
	\lt(\ba{c} \Psi_{\pm_u,+\f32} \\ \Psi_{\pm_u,+\f12}\\ \Psi_{\pm_u,-\f12}\\ \Psi_{\pm_u,-\f32}\ea\rt)
	+\lt[
	b_{\pm_u,1} \lt(\ba{c} \pm_u \sq{v_z^2-v_0^2}\\ v_0+\tf12v_z\\ 0\\-\f{\sq3}2v_z\ea\rt)+
	b_{\pm_u,2} \lt(\ba{c} v_0-\tf12v_z\\ \mp_u \sq{v_z^2-v_0^2}\\-\f{\sq3}2v_z \\ 0\ea\rt)
	\rt]e^{\kappa y}.
\lbl{eq:psiyu}
\eeq
Here, $\Psi_{\pm_u,j_z}$, $j_z=+\f32,+\f12,-\f12,-\f32$, and $b_{\pm_u,1}$, $b_{\pm_u,2}$ are the six constant free coefficients.
Applying the BCs \eqn{bc} to the general solution \eqn{psiyz}, we have
\[
    \psih(y=0,z)=\nm.
\]
These BCs have to be satisfied at every point $(x,z)$ on the $y=0$ surface.
Due to the linear independence of $e^{\pm_u i p_u z}$,
the BCs have to be satisfied independently for the functions \eqn{psiyu}:
\[
    \psih_{\pm_u}(y=0)=\nm.
\]
For each Dirac point $\pm_u$, this gives four linear homogeneous relations between the six coefficients.
Excluding $b_{\pm_u,1}$ and $b_{\pm_u,2}$ from these relations, we arrive at the two relations
\beqar
	\tf{\sq3}2v_z\Psi_{\pm_u,+\f32}+(v_0-\tf12v_z)\Psi_{\pm_u,-\f12} \pm_u\sq{v_z^2-v_0^2}\Psi_{\pm_u,-\f32}=0,\\
	\tf{\sq3}2v_z\Psi_{\pm_u,+\f12}+(v_0+\tf12v_z)\Psi_{\pm_u,-\f32} \mp_u\sq{v_z^2-v_0^2}\Psi_{\pm_u,-\f12}=0
\lbl{eq:bcPsiy}
\eeqar
for the four coefficients. These become the BCs for the low-energy wave function upon the identification
\[
	\Psih_{\pm_u} \rarr \Psih_{\pm_u}(x,y=0,z)
\]
of the coefficients with the wave-function components at the surface. It is also instructive for further calculations of the surface states
to express these BCs in terms of the velocities $v_{0,z}$ [\eq{v}], since these will combine naturally with those in the linear Hamiltonian [\eq{HcD}].

\subsubsection{Summary}

We summarize how the BCs for the wave function $\Psih(\rb)$ of the linear model of the Dirac semimetal originate for the $z>0$ and $y<0$ samples.
This wave function has eight components and there are four BCs for any surface,
since its Hamiltonian $\Hch^D(\kbh)$ [\eq{HcD}] is linear in momentum $\kbh$.

The eight plane-wave solutions (four for each Dirac point) are present for any surface.
For the $z>0$ sample, these were the only relevant particular solutions to the stationary Schr\"odinger equation
for the Luttinger Hamiltonian at the Dirac-point energy $\e=\eps_u$.
All of these solutions have the same (absent) coordinate dependence on $(x,y)$ along the $z=0$ surface;
as a result, the four BCs for the wave function $\psih(\rb)$ of the Luttinger model translate directly to four BCs \eqn{bcPsiz}
for the eight components of the low-energy wave function $\Psih(\rb)$.

One the other hand, for the $y<0$ sample, in addition to the eight bulk plane-wave solutions,
there are extra four (two for each Dirac point) decaying solutions; so, there are twelve free coefficients (six for each Dirac point).
Here, the solutions for the two nodes have different dependencies $e^{\pm_u i p_u z}$ along the $y=0$ surface;
as a result, the four functional BCs for the Luttinger-model wave function $\psih(\rb)$, due to their translation symmetry,
reduce to eight relations for the twelve free coefficients.
Excluding the four coefficients of the decaying solutions, we again obtain four BCs \eqn{bcPsiy}
for the eight components of the low-energy wave function $\Psih(\rb)$.

\subsection{Analytical calculation of the surface states \lbl{sec:ssDlin}}

We now calculate the surface states analytically in the vicinity of the Dirac points from the linear model of the Dirac semimetal derived above.

\subsubsection{$y<0$ sample \lbl{sec:ssDliny}}

For energies
\[
	\e\in (E_{y<0,\pm_u,-_b}^{D,\x{lin}}(k_x,k_z), E_{y<0,\pm_u,+_b}^{D,\x{lin}}(k_x,k_z))
\]
within the gap between the bulk-band boundaries
\beq
	E_{y<0,\pm_u,\pm_b}^{D,\x{lin}}(k_x,k_z)=\eps_u \pm_u v_0 k_z \pm_b \sq{v_\p^2 k_x^2+v_z^2k_z^2},
\lbl{eq:EDylin}
\eeq
the characteristic equation
\beq
    \det [\Hch^D_{\pm_u,\pm_{j_z}}(\kb)-\e \um_2]=0
\lbl{eq:HcDcheq}
\eeq
with respect to $k_y$ for each $\pm_u$, $\pm_{j_z}$ block has one solution (the same for $\pm_{j_z}$)
\beq
	k_{y,\pm_u}(\e,k_x,k_z)=- i \sq{k_x^2+\f{v_z^2k_z^2-[\pm_u v_0k_z-(\e-\eps_u)]^2}{v_\p^2}}
\eeq
that corresponds to a particular solution to the stationary Schr\"odinger equation decaying into the bulk, as $y\rarr-\iy$.
For each $\pm_u$,
\[
		\Psih^{+_u}(y;\e,k_x,k_z)=\lt(\ba{c} \Psih_{+_u}^{+_u}(y;\e,k_x,k_z) \\ \nm_4\ea\rt)
,\spc
		\Psih^{-_u}(y;\e,k_x,k_z)=\lt(\ba{c} \nm_4 \\ \Psih_{-_u}^{-_u}(y;\e,k_x,k_z) \ea\rt)
\]
the general solution is their linear combination
\[
	\Psih_{\pm_u}^{\pm_u}(y;\e,k_x,k_z)=
		\lt[
	c_{\pm_u,+_{j_z}} \lt(\ba{c} \chih^{\pm_u,+_{j_z}}(\e,k_x,k_z) \\ \nm_2 \ea\rt)
	+
	c_{\pm_u,-_{j_z}} \lt(\ba{c} \nm_2 \\ \chih^{\pm_u,-_{j_z}}(\e,k_x,k_z) \ea\rt)\rt]
        e^{i k_{y,\pm_u}(\e,k_x,k_z)y},
\]
where
\beq
	\chih^{\pm_u,+_{j_z}}(\e,k_x,k_z)=\lt(\ba{c}
		v_\p k_x -\sq{v_\p^2 k_x^2+ v_z^2 k_z^2-[v_0 k_z-(\e-\eps_u)]^2}\\
		(v_0-v_z)k_z \mp_u (\e-\eps_u)
	\ea\rt),
\eeq
\beq
	\chih^{\pm_u,-_{j_z}}(\e,k_x,k_z)=\lt(\ba{c}
		-v_\p k_x +\sq{v_\p^2 k_x^2+ v_z^2 k_z^2-[v_0 k_z-(\e-\eps_u)]^2}\\
		(v_0+ v_z)k_z \mp_u(\e-\eps_u)
	\ea\rt)
\eeq
are the nontrivial solutions to
\[
	[\Hch^D_{\pm_u,\pm_{j_z}}(k_x,k_{y,\pm_u}(\e,k_x,k_z),k_z)-\e\um_2]\chih=\nm_2
\]
and $c_{\pm_u,\pm_{j_z}}$ are arbitrary coefficients.
Regarding notation, here and below, the superscript labels of the wave functions denote the quantum numbers,
while the subscripts are reserved for their components. In $\nm_4$ and $\nm_2$, the sizes of the null vectors are indicated for clarity.

Inserting these wave functions into the BCs~\eqn{bcPsiy} yields the equation for the energy $\e$ of the surface states.
Solving it, we obtain the surface-state spectrum for the linear model of the Dirac semimetal
\beq
	\Ec^{D,\x{lin}}_{y<0,\pm_u,\pm}(k_x,k_z)=\eps_u+\tf{\sq3}2[\pm_u \sq3 v_0 k_z \pm \sq{(v_z^2-v_0^2)(\tf{4v_\p^2}{3v_z^2}k_x^2 + k_z^2)}].
\lbl{eq:EcDylin}
\eeq
Due to the linear scaling, the surface-state spectrum \eqn{EcDylin}, as well as the bulk-band boundaries \eqn{EDylin},
are linear in the absolute value $\sq{k_x^2+k_z^2}$ of the surface momentum relative to the projected Dirac points
and are fully characterized by their dependence on the polar angle $\zeta$,
\[
	(k_x,k_z)=\sq{k_x^2+k_z^2}(\sin\zeta,\cos\zeta).
\]
The surface-state spectrum \eqn{EcDylin} is shown in \figr{ssDylin},
along with that calculated within the Luttinger model, for comparison; their quantitative asymptotic agreement is discussed below in \secr{Dasymptagree}.
The properties do fully agree with those obtained from Luttinger model.
The solution is fully characterized by the ratio
\[
    \f{v_0}{v_z}=\f{\al_0}{\al_z-\f35\al_\square}.
\]
In the Luttinger-semimetal regime from which this Dirac semimetal originates, $-1<v_0/v_z<1$.
Because of the linear scaling, the merging contours in $(k_x,k_z)$ are straight half-lines with specific merging angles $\zeta_m$.
The latter can be obtained by equating the surface-state energy \eqn{EcDylin} to the bulk-band boundaries \eqn{EDylin}, which gives
\[
	\cos\zeta_m=\pm \f{2\f{v_0}{v_z}\f{v_\p}{v_z}}{\sq{1-4\f{v_0^2}{v_z^2}(1-\f{v_\p^2}{v_z^2})}}.
\]

Depending on $v_0/v_z$, there are two regimes, with one and two surface-state bands \eqn{EcDylin}.
The borderline case can be found be demanding that the merging angle(s) $\zeta_m$ equal $0$ or $\pi$, which is when one
of the two bands as functions of $\zeta$ shrinks and disappears
and the other one detaches from the bulk-band boundaries.
This happens at
\[
	\f{v_0}{v_z}=\pm\tf12.
\]
For $-\tf12<v_0/v_z<\tf12$, both surface-state bands \eqn{EcDylin} are present.
For $-1<v_0/v_z<-\tf12$, only the $+$ band is present; for $\tf12<v_0/v_z<1$, only the $-$ band is present.

We also observe an interesting relation: the surface-state spectrum \eqn{EcDylin} of the linear-in-momentum model of the Dirac semimetal
can be expressed (at least for $k_x=0$) in terms of the curvatures \eqn{AcOh}
of the surface-state spectrum of the Luttinger model of the Luttinger semimetal (also for finite cubic anisotropy):
\[
	\Ec^{D,\x{lin}}_{y<0,+_u,\pm}(k_x=0,k_z)
	=\eps_u + \Acr_{\pm}^{\Ob_h}(\alr_0,\alr_\square) \al_z\cd 2p_u k_z, \spc k_z>0,
\]
and similar for the $-_u$ Dirac point or $k_z<0$.

\begin{figure*}
\centering
\includegraphics[width=\linewidth]{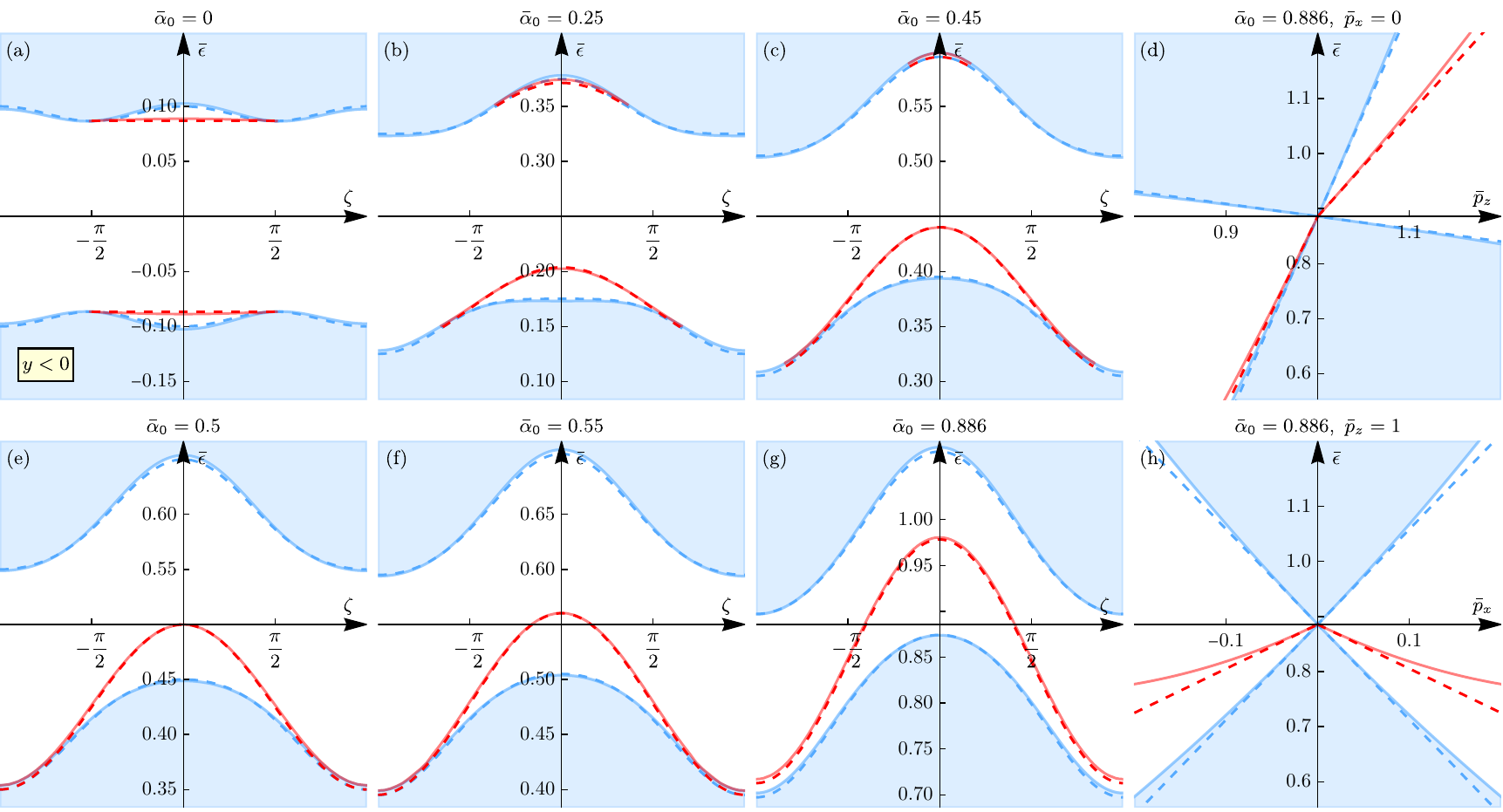}
\caption{
The surface-state spectrum $\Ec^{D,\x{lin}}_{y<0}(k_x,k_z)$ [\eq{EcDylin}] of the Dirac semimetal in the vicinity of the $+_u$ projected Dirac point
for the $y<0$ sample,
calculated analytically within the linear-in-momentum model (dashed red)
with the Hamiltonian $\Hch^D(\kbh)$ [\eq{HcD}] and BCs \eqn{bcPsiy}.
The spectrum is presented in the dimensionless momentum and energy units [\eqs{pr}{EcDyscaling}]
and for absent cubic anisotropy $\al_\square=0$.
and the comparison with spectrum calculated within the Luttinger model (solid lighter red).
The bulk-band boundaries of the linear-in-momentum and Luttinger models are in dashed blue and solid lighter blue, respectively.
(a),(b),(c),(e),(f),(g) are spectra along the circular paths of radius $\pr_r=0.05$;
(d) and (h) are spectra along straight-line paths passing through the projected Dirac point.
Quantitative asymptotic agreement between the two models as the projected Dirac point is approached is observed.
}
\lbl{fig:ssDylin}
\end{figure*}

\subsubsection{$z>0$ sample \lbl{sec:ssDlinz}}

\begin{figure*}
\centering
\includegraphics[width=.5\linewidth]{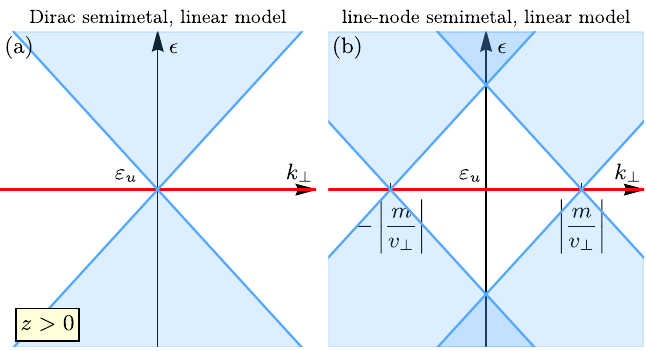}
\caption{
(a) The flat surface-state spectrum $\Ec^{D,\x{lin}}_{z>0}(k_\p)\equiv \eps_u$ [\eq{EcDlinz}] of the Dirac semimetal
for the $z>0$ sample, calculated within the linear-in-momentum model with the Hamiltonian $\Hch^D(\kbh)$ [\eq{HcD}] and BCs \eqn{bcPsiz}.
This flat spectrum and the quadratic asymptotic behavior of the surface-state bands (\figr{ssDz}) calculated within the Luttinger model
as the projected double Dirac point is approached constitute quantitative asymptotic agreement between the two models, see \secr{Dasymptagree}.
(b) The flat surface-state spectrum $\Ec^{LN,\x{lin}}_{z>0}(k_\p)\equiv \eps_u$ [\eq{EcLNlinz}] of the line-node semimetal
for the $z>0$ sample, calculated within the linear-in-momentum model with the Hamiltonian $\Hch^{LN}(\kbh)$ [\eq{HcLN}] and BCs \eqn{bcPsiz}.
}
\lbl{fig:ssDzlin}
\end{figure*}

For the $z>0$ sample, the two Dirac points are projected onto the same surface momentum $(k_x,k_y)=(0,0)$,
as a result, the $\pm_u$ parts $\Psih_{\pm_u,\pm_{j_z}}(\rb)$ are coupled by the BCs \eqn{bcPsiz}.
However, the $\pm_{j_z}$ parts are decoupled in both the Hamiltonian and BCs.

For the energies
\[
	\e\in (E_{z>0,-_b}^{D,\x{lin}}(k_\p), E_{z>0,+_b}^{D,\x{lin}}(k_\p))
\]
within the gap between the bulk-band boundaries
\[
	E^{D,\x{lin}}_{z>0,\pm_b}(k_\p)=\eps_u \pm_b v_\p k_\p \sq{1-\tf{v_0^2}{v_z^2}},
\]
the characteristic equation \eqn{HcDcheq}, with respect to $k_z$ for each $\pm_u,\pm_{j_z}$ block has one solution (the same for $\pm_{j_z}$)
\beq
	k_{z,\pm_u}(\e,k_\p)=\f{\mp_u v_0(\e-\eps_u) + i \sq{k_\p^2v_\p^2(v_z^2-v_0^2)-v_z^2(\e-\eps_u)^2}}{v_z^2-v_0^2}
\eeq
that corresponds to a particular solution to the stationary Schr\"odinger equation decaying into the bulk, as $z\rarr+\iy$.
The general solution for each $\pm_{j_z}$ is their linear combination
\begin{widetext}
\beq
	\Psih^{+_{j_z}}(z;\e,k_\p,\phi)
		=c_{+_u,+_{j_z}}\lt(\ba{c} \chih^{+_u,+_{j_z}}(\e,k_\p,\phi) \\ \nm_2\\ \nm_4 \ea\rt) e^{i k_{z,+_u}(\e,k_\p)z}
		+c_{-_u,+_{j_z}}\lt(\ba{c} \nm_4 \\ \chih^{-_u,+_{j_z}}(\e,k_\p,\phi)\\ \nm_2 \ea\rt) e^{i k_{z,-_u}(\e,k_\p)z},
\eeq
\end{widetext}
\begin{widetext}
\beq
	\Psih^{-_{j_z}}(z;\e,k_\p,\phi)
		=c_{+_u,-_{j_z}}\lt(\ba{c} \nm_2\\ \chih^{+_u,-_{j_z}}(\e,k_\p,\phi) \\ \nm_4 \ea\rt) e^{i k_{z,+_u}(\e,k_\p)z}
		+c_{-_u,-_{j_z}}\lt(\ba{c} \nm_4\\ \nm_2 \\ \chih^{-_u,-_{j_z}}(\e,k_\p,\phi) \ea\rt) e^{i k_{z,-_u}(\e,k_\p)z},
\eeq
\end{widetext}
where
\[
	\chih^{\pm_u,+_{j_z}}(\e,k_\p,\phi)=
	\lt(\ba{c}
	\mp_u v_\p k_\p e^{-i\phi}\\
	\e-\eps_u \mp_u (v_0-v_z)k_{z,\pm_u}(\e,k_\p)
	\ea\rt),
\]
\[
	\chih^{\pm_u,-_{j_z}}(\e,k_\p,\phi)=
	\lt(\ba{c}
	\pm_u v_\p k_\p e^{-i\phi}\\
	\e-\eps_u \mp_u (v_0+v_z)k_{z,\pm_u}(\e,k_\p)
	\ea\rt)
\]
are the nontrivial solutions to
\[
	[\Hch^D_{\pm_u,\pm_{j_z}}(\kb_\p,k_{z,\pm_u}(\e,k_\p))-\e\um_2]\chih=\nm_2
\]
and $c_{\pm_u,\pm_{j_z}}$ are arbitrary coefficients.

Applying the BCs \eqn{bcPsiz}, we obtain the homogeneous system of equations [akin to \eq{eqc}] for these coefficients,
which has nontrivial solutions when its determinant is zero [akin to \eq{eqe}], which yields the equation
\beq
	 \chi^{+_u,+_{j_z}}_{+\f32}(\e,k_\p,\phi)\chi^{-_u,+_{j_z}}_{+\f12}(\e,k_\p,\phi)
	-\chi^{-_u,+_{j_z}}_{+\f32}(\e,k_\p,\phi)\chi^{+_u,+_{j_z}}_{+\f12}(\e,k_\p,\phi)=0
\label{eq:eqeDlinz}
\eeq
for the energy $\e$ of the surface states for the $+_{j_z}$ block.
For every surface momentum $(k_x,k_y)$, the only solution to this equation is a nondegenerate $\e=\eps_u$.
Similar situation holds for the $-_{j_z}$ block.
Hence, the surface-state spectrum (\figr{ssDzlin}) for $z>0$ sample
obtained from the linear-in-momentum model of the Dirac semimetal is flat and double-degenerate:
\beq
    \Ec^{D,\x{lin}}_{z>0,\pm_{j_z}}(k_\p) \equiv\eps_u.
\lbl{eq:EcDlinz}
\eeq
The momentum solutions of the surface states equal
\[
    k_{z,\pm_u}(\e=\eps_u,k_\p)=i\ka, \spc \ka=\f{v_\p k_\p}{\sq{v_z^2-v_0^2}},
\]
the coefficients are $c_{+_u,\pm_{j_z}}=c_{-_u,\pm_{j_z}}=1$,
and the wave functions of the surface states reads
\beq
	\Psih^{+_{j_z}}(z;\e=\eps_u,k_\p,\phi)
		=\lt(\ba{c} \chih^{+_u,+_{j_z}}(\e=\eps_u,k_\p,\phi) \\ \nm_2\\ \chih^{-_u,+_{j_z}}(\e=\eps_u,k_\p,\phi)\\ \nm_2 \ea\rt) e^{-\ka z},
\spc	\chih^{\pm_u,+_{j_z}}(\e=\eps_u,k_\p,\phi)=\lt(\ba{c} \mp_u v_\p k_\p e^{-i\phi} \\ \mp_u (v_0-v_z)i \ka \ea\rt),
\eeq
\beq
	\Psih^{-_{j_z}}(z;\e=\eps_u,k_\p,\phi)
		=\lt(\ba{c} \nm_2\\ \chih^{+_u,-_{j_z}}(\e=\eps_u,k_\p,\phi) \\ \nm_2 \\ \chih^{-_u,-_{j_z}}(\e=\eps_u,k_\p,\phi) \ea\rt)e^{-\ka z},
\spc	\chih^{\pm_u,-_{j_z}}(\e=\eps_u,k_\p,\phi)=\lt(\ba{c} \pm_u v_\p k_\p e^{-i\phi} \\ \mp_u (v_0+v_z)i \ka \ea\rt).
\eeq

\subsection{Emergent chiral symmetry of the linear-in-momentum model}

We find a symmetry explanation for the obtained flat behavior of the surface-state spectrum \eqn{EcDlinz}
of the linear model of the Dirac semimetal for the $z>0$ sample.
We notice that both the Hamiltonian \eqn{HcD} and BCs \eqn{bcPsiz} possess chiral symmetry.
Namely, we notice that the Hamiltonian \eqn{HcD} changes its sign relative to the Dirac-point energy $\eps_u$ under the unitary transformation
\beq
    \Sh [\Hch^D(\kbh)-\eps_u\um_8] \Sh^\dg=-[\Hch^D(\kbh)-\eps_u\um_8],
\eeq
\[
    \Sh=\tauh_x\otimes \um_4,
\]
since the blocks $\Hch_{\pm_u}^D(\kbh)-\eps_u\um_8$ differ only by the sign. This means that the Hamiltonian satisfies {\em chiral symmetry}
and $\Sh$ is its operator~\cite{Chiu2016}.
The operation $\Sh$ interchanges $\pm_u$ components of the wave function:
\[
	\Sh\Psih(\rb)=\lt(\ba{c} \Psih_{-_u}(\rb) \\ \Psi_{+_u}(\rb) \ea\rt).
\]
This transformed wave function also satisfies the BCs \eqn{bcPsiz} for the $z>0$ sample,
which means that they also satisfy chiral symmetry~\cite{KharitonovLSM,KharitonovQAH,KharitonovFGCM}.
Hence, the whole $z>0$ system {\em with the boundary} has chiral symmetry.
As a result, its spectrum, including the surface states, satisfies chiral symmetry.
Regarding the double degeneracy of the flat surface-state band,
for these Hamiltonian and BCs it is guaranteed by the decoupling of the $\pm_{j_z}$ parts in both of them.
Moreover, we notice that both $\pm_{j_z}$ surface-state solutions have the same eigenvalue $-1$ of the chiral symmetry operator:
\[
	\Sh\Psih^{\pm_{j_z}}(z;\e=\eps_u,k_\p,\phi)=-\Psih^{\pm_{j_z}}(z;\e=\eps_u,k_\p,\phi).
\]
This means that the double degeneracy is also protected by chiral symmetry,
meaning that double degeneracy will persist if the Hamiltonian or BCs are modified
(for example, the decoupling of $\pm_{j_z}$ blocks could be broken) in a way that still preserves their chiral symmetry.

For the $y<0$ sample, the BCs \eqn{bcPsiy} of the linear-in-momentum model do not obey chiral symmetry and therefore,
neither does the surface-state spectrum.

We note that this is an emergent symmetry of specifically the linear-in-momentum model for the $z>0$ sample that arises in this low-energy limit.
In the Luttinger model with strain, there is no such chiral symmetry.
And indeed the surface-state spectrum is not flat;
its quadratic dispersion at small $p_\p$ is, however, in accord with the flat behavior in the linear model.

\subsection{Asymptotic agreement between the linear-in-momentum and Luttinger models of the Dirac semimetal \lbl{sec:Dasymptagree}}

Comparing the surface-state spectra of the Dirac semimetal in the vicinity of the Dirac points, $|\pb\mp_u\pb_u|\ll p_u$,
calculated within the linear-in-momentum and Luttinger models,
\figr{ssDylin} for the $y<0$ sample and \figsr{ssDz}{ssDzlin} for the $z>0$ sample (flat versus quadratic behavior),
we observe a clear quantitative asymptotic agrement, as the projected Dirac points are approached.
This establishes the next relation in the hierarchy of low-energy models presented in \secr{scales}.
Just as the Luttinger model is sufficient to fully capture the asymptotic behavior of the surface states of the Kane model
in the vicinity of the quadratic node in the Luttinger-semimetal phase (\secr{LLMKM}),
the linear-in-momentum model is sufficient to fully capture the asymptotic behavior of the surface states of the Luttinger model
in the vicinity of the Dirac nodes in the Dirac-semimetal phase.

\section{Surface states of the line-node semimetal \lbl{sec:ssLN}}

In this section, we calculate and explore within the Luttinger model the surface states of the line-node semimetal
that arises in the Dirac semimetal upon introducing the linear BIA term;
the Dirac semimetal itself, studied in the previous \secr{ssD}, arises from the Luttinger semimetal upon introducing compressive strain.
The bulk properties of this line-node semimetal have been studied in \secr{bulkLN}.

The Luttinger model \eqn{HLN} of the line-node semimetal is fully characterized by three dimensionless parameters:
two of the Luttinger-semimetal part, $\alr_0$, $\alr_\square$,
and the ratio $p_{\be_1}/p_u$ of the characteristic momenta of the linear BIA [\eq{pbe1}] and strain [\eq{pu}] terms.
We present the line-node (this section) and Weyl (next \secr{ssW}) semimetal cases for the parameters $\al_{0,z,\square}$ and $\be_1$ of HgTe
and the value $u=-3\x{meV}$ of the compressive strain.
Although we include the nonzero cubic anisotropy $\alpha_\square$,
it does not seem to play any notable role for the regime of other parameters we explore:
we have compared to the case of absent cubic anisotropy ($\al_\square=0$) and found only minor inessential quantitative differences.

For the chosen value of strain, 
the hierarchy presented in \secr{scales} is satisfied well.
As demonstrated in \secr{bulkLN},
upon introducing the linear BIA term into the Dirac semimetal, the Dirac points of the bulk spectrum transform into line nodes.
As we show below, the surface-state spectra of both the $y<0$ and $z>0$ samples
also undergo rather nontrivial transformations at the $p_{\be_1}$ scale around the former projected Dirac points.

\subsection{$y<0$ sample \lbl{sec:ssLNy}}

\begin{figure*}
\centering
\includegraphics[width=\linewidth]{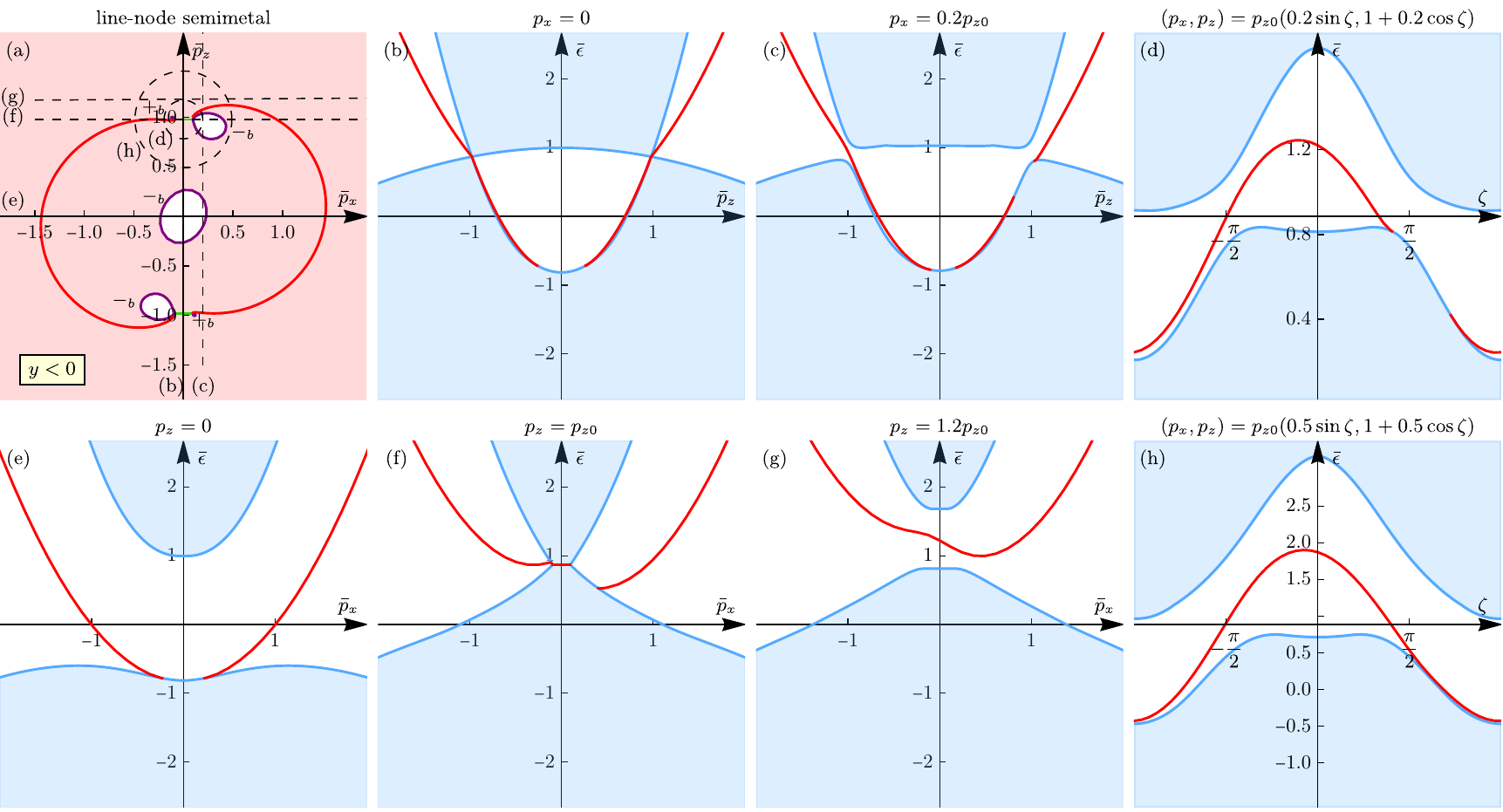}
\caption{
The surface-state spectrum $\Ec^{LN}_{y<0}(p_x,p_z)$ of the line-node semimetal for the $y<0$ sample,
calculated within the Luttinger model with compressive strain and linear BIA term [\eqs{HLN}{bc}]
for the parameters of HgTe (\tabr{LM}) and the strain value $u=-3~\x{meV}$.
The structure of the figure is the same as in \figr{ssDy17} for the Dirac semimetal.
Compared to the surface-state spectrum of the Dirac semimetal,
the key changes occur in the vicinity of the (former) projected Dirac points, which transform into projected line nodes (green).
The spectrum is presented in the dimensionless momentum and energy units [\eqs{pr}{EcDyscaling}] of the Dirac-semimetal case
for better comparison.
}
\lbl{fig:ssLNy}
\end{figure*}

\begin{figure*}
\centering
\includegraphics[width=\linewidth]{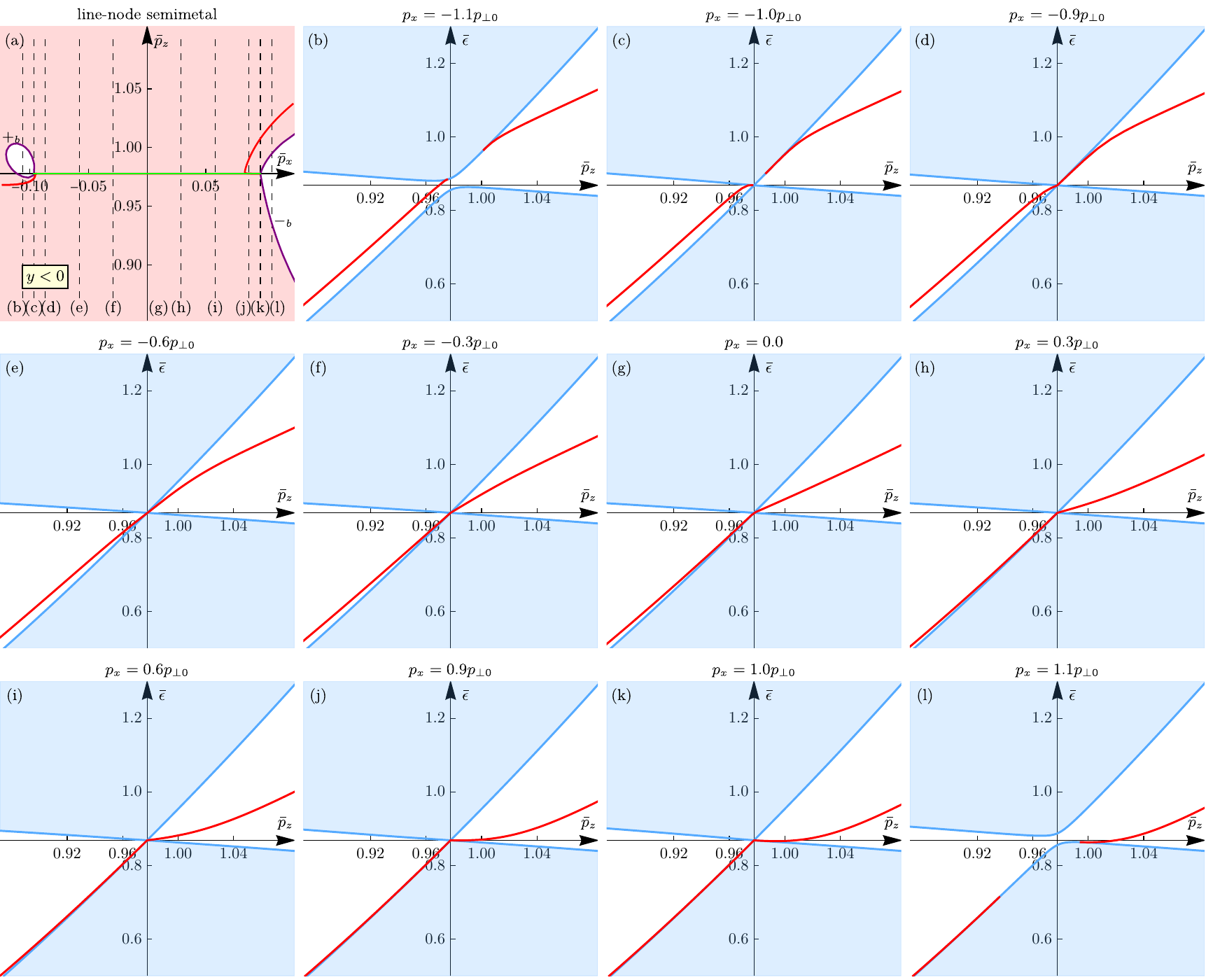}
\caption{
The surface-state spectrum $\Ec^{LN}_{y<0}(p_x,p_z)$ (red) of the line-node semimetal
along $p_x=\x{const}$ straight-line paths
in the vicinity $p_z\approx p_z^{LN}(0)=p_{z0}$ of the projected $+_u$ line node (additional characterization of \figr{ssLNy}).
For $-p_{\p0}<p_x<p_{\p0}$, $p_{\p0}=p_\p^{LN}(0)$, the path crosses the projected line node;
the bulk-band boundaries $E^{LN}_{y<0,\pm_b}(p_x,p_z)$ (blue) along this path have a linear node and the surface-state band
$\Ec^{LN}_{y<0}(p_x,p_z)$ also has a linear asymptotic behavior on both sides.
}
\lbl{fig:ssLNypx}
\end{figure*}

\begin{figure*}
\centering
\includegraphics[width=.8\linewidth]{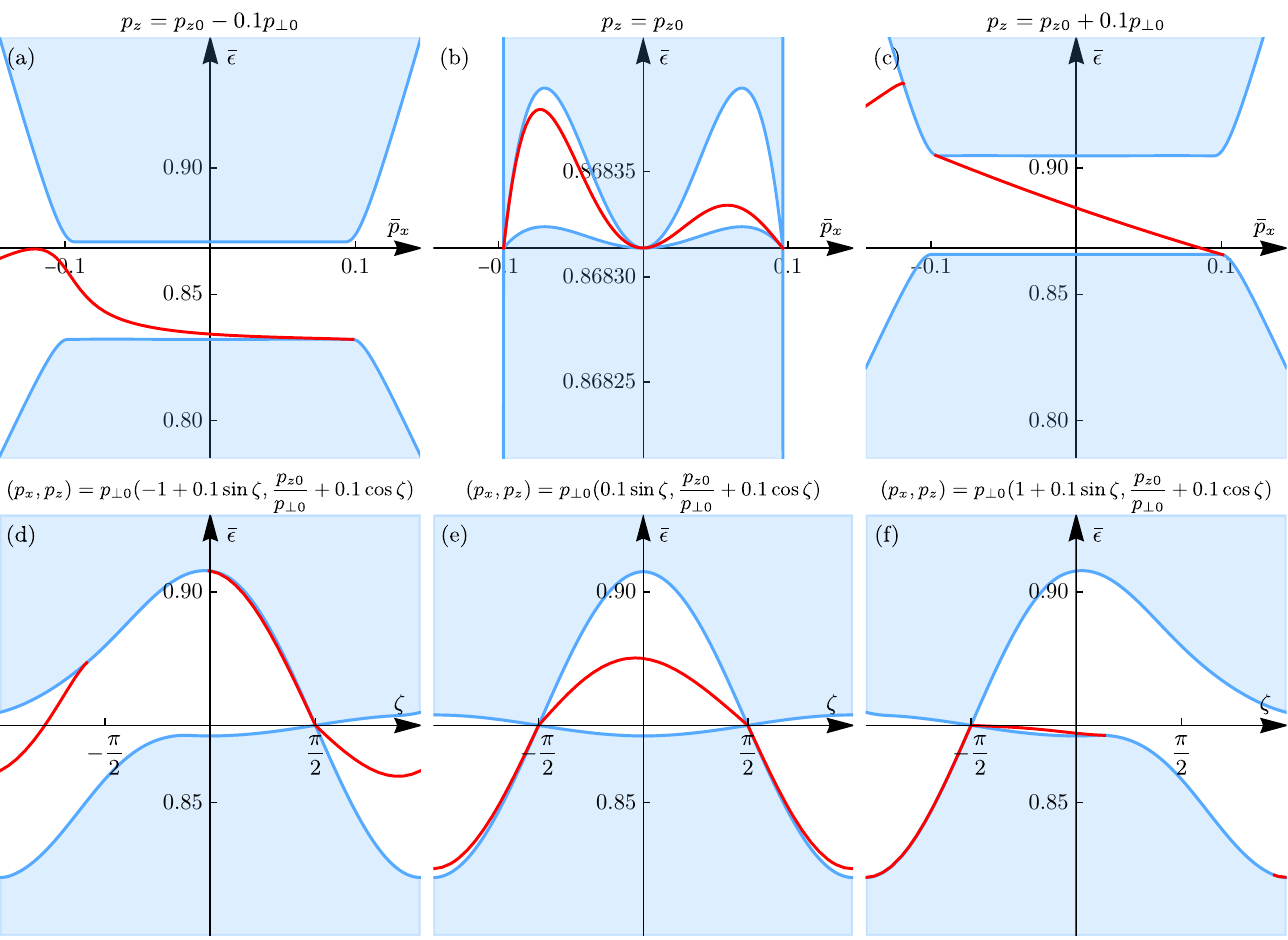}
\caption{
Additional characterization of
the surface-spectrum $\Ec_{y<0}^{LN}(p_x,p_z)$ of the line-node semimetal (\figr{ssLNy}) in the vicinity of the projected $+_u$ line node:
(a),(b),(c) along $p_z=\x{const}$ straight-line paths and (d),(e),(f) along circular paths.
The latter are useful for comparison with those around the projected Weyl points of the Weyl semimetal, presented in \figssr{ssWypr-}{ssWypr0}{ssWypr+}.
The vertical position of the horizontal axis is the energy $\eps^{LN}(0)$ of the line node.
}
\lbl{fig:ssLNy+}
\end{figure*}

\subsubsection{Bulk-band boundaries and projected line node}

First, let us discuss the bulk-band boundaries
\beq
    E_{y<0,+_b}^{LN}(p_x,p_z)=\min_{p_y,\sig} \eps_{+_b,\sig}^{LN}(p_x,p_y,p_z),
\lbl{eq:ELNy+}
\eeq
\beq
    E_{y<0,-_b}^{LN}(p_x,p_z)=\max_{p_y,\sig} \eps_{-_b,\sig}^{LN}(p_x,p_y,p_z),
\lbl{eq:ELNy-}
\eeq
and the projection of the line node onto the surface-momentum plane $(p_x,p_z)$, which are somewhat nontrivial.

Because of the symmetric orientation of the $y=0$ surface relative to the bulk symmetry axes, the boundaries $E^{LN}_{y<0,\pm_b}(p_x,p_z)$
of the conduction and valence bulk bands meet at the curve
\beq
	(p_x,p_z)=(p_\p^{LN}(\phi)\cos\phi,p_z^{LN}(\phi)), \spc \phi\in[0,\pi],
\lbl{eq:pLNy}
\eeq
in the surface momentum plane ($\phi$ becomes merely a parameter here), which is the projection of the 3D line \eqn{peLN},
and at energy $\e=\eps^{LN}(\phi)$:
\beq
	E_{y<0,+_b}^{LN}(p_\p^{LN}(\phi)\cos\phi,\pm_u p_z^{LN}(\phi))=
	E_{y<0,-_b}^{LN}(p_\p^{LN}(\phi)\cos\phi,\pm_u p_z^{LN}(\phi))
	=\eps^{LN}(\phi),\spc \phi\in[0,\pi]
\lbl{eq:ELNy=}
\eeq
The two line nodes $\pm_u$ have nonoverlapping projections, but for each of them,
the two points at $\pm\phi$, $\phi\in(0,\pi)$, of the 3D line node are projected onto one point.

As explained in \secr{bulkLN}, the line node has an approximately circular shape,
but does have a weak dependence of the momenta $p_\p^{LN}(\phi)$ and $p_z^{LN}(\phi)$ and energy $\eps^{LN}(\phi)$ on the polar angle $\phi$.
Because of the dependence $\eps^{LN}(\phi)$, there is no single energy
at which the region in the surface-momentum plane $(p_x,p_z)$ occupied by the bulk states
would be equal to the projection \eqn{pLNy} of the line node.
Rather, at any energy, the region occupied by the bulk states is solid.
In \figr{ssLNy}(a) and \figr{ssLNypx}(a), we plot the bulk and surface-state spectrum at the fixed energy $\e=\eps^{LN}(0)$.
Since the dependencies $\eps^{LN}(\phi)$ and $p_x^{LN}(\phi)$ are weak,
the region of the bulk states has only a narrow width along the $p_z$ direction.
Since $\eps^{LN}(\phi)\geq\eps^{LN}(0)$, the crossing with this level occurs for $-_b$ bulk states,
the equation
\[
	E_{y<0,-_b}^{LN}(p_x,p_z)=\eps^{LN}(0).
\]
defines the boundary of that region.
Because the $p_z^{LN}(\phi)$ and $\eps^{LN}(\phi)$ dependencies are numerically really weak,
these differences cannot be resolved in the scale of \figr{ssLNy}(a) and \figr{ssLNypx}(a)
and both the projected line node and the fixed-energy region appear as a straight-line segment.

\subsubsection{Surface-state spectrum}

We describe the surface-state spectrum $\Ec^{LN}_{y<0}(p_x,p_z)$ of the line-node semimetal for the $y<0$ sample
by the 2D plots presented in \figsdr{ssLNy}{ssLNy+}.
\figr{ssLNy} has the same structure as \figsdr{ssDy0}{ssDy17} for the Dirac semimetal.
\figsr{ssLNypx}{ssLNy+} further characterize the surface-state spectrum in the vicinity of the projected line node.
\figr{ssLNypx} shows the spectrum along the $p_x=\x{const}$ paths and \figr{ssLNy+} -- along some additional paths.

In accord with \secr{symmetries}, as the Dirac semimetal is transformed into the line-node semimetal upon introducing the linear BIA term,
the $\Cb_{2v}$ spatial symmetry of the bulk-band boundaries remains,
while the spatial symmetry of the surface-state spectrum is lowered from $\Cb_{2v}$ to $\Cb_2$.

For $p_{\be_1}\ll p_u$, the key nontrivial changes occur in the vicinity of the former Dirac points, which transform into the line nodes.
Away, the surface-state spectrum is qualitatively similar to that of the Dirac semimetal for the most part,
compare \figr{ssLNy} to \figr{ssDy17},
except for the region around the touching point at $(p_x,p_z)=(0,0)$ in the Dirac semimetal for $\tf12<|\alr_0|<1$,
due to the proximity of the surface-state band to the bulk-band boundary of the Dirac semimetal.

As in the Dirac semimetal in the regime $\tf12<|\alr_0|<1$,
there is still only one surface-state band in the line-node semimetal at $\alr_0=0.886$ of HgTe.
However, unlike in the Dirac semimetal, there appear patches of absent surface states.
The touching point $(p_x,p_z)=(0,0)$ in the Dirac semimetal
grows into a finite-size region of absent surface states of size $\sim p_{\be_1}$ in the line-node semimetal.
In the vicinity of the $+_u$ projected line node, two patches of absent surface states emerge, both scaling as $\sim p_{\be_1}$,
with merging contours connected to the two ends of the projected line node:
a smaller patch around $(p_x,p_z)=(-p_{\p0},p_{z0})$ and a larger one around $(p_x,p_z)=(p_{\p0},p_{z0})$;
we denote $p_{\p0}=p_\p^{LN}(0)$ and $p_{z0}=p_z^{LN}(0)$ for brevity.
The smaller patch is visible in \figr{ssLNypx}(a), while in \figr{ssLNy}(a) it is not well-resolved.
Other than these regions, the surface-state band merges with the bulk-band boundaries at the projected line node
from both sides $p_z \gtrless p_z^{LN}(\phi(p_x))$, as clearly seen in \figr{ssLNypx} [$\phi(p_x)$ is the inverse of $p_x=p_\p^{LN}(\phi)\cos\phi$].
The spectrum around the $-_u$ projected line node is the same by symmetry.

The weak dependence of the momentum $p^{LN}_z(\phi)$
of the line node is manifested in the spectrum along the $p_z=p_{z0}$ path, shown in \figr{ssLNy+}(b):
the ``gaps'' in the bulk spectrum are only due to $p_z^{LN}(\phi(p_x))$ depending on $p_x$.

In \figr{ssLNypx}, the spectrum along a series of straight-line paths with $p_x=\x{const}$ is shown.
Except for the close proximity to the ends $p_x=\pm p_{\p0}$ of the projected line node,
the surface-state band $\Ec_{y<0}^{LN}(p_x,p_z)$ at fixed $p_x\in(-p_{\p0},+p_{\p0})$
is asymptotically linear in $p_z$ as it approaches the projected line node from both sides.
Due to the variations of $p_z^{LN}(\phi(p_x))$ being much weaker than the $0.1p_{\p0}$ scale, the velocities
\[
	\pd_{p_z}\Ec_{y<0}^{LN}(p_x,p_z\rarr p_z^{LN}(\phi(p_x))\pm 0)
\]
can be effectively deduced from the straight-line paths at fixed $p_z=p_{z0}-0.1p_{\p0}$ and $p_{z0}+0.1p_{\p0}$,
shown in \figr{ssLNy+}(a) and (c) [the level of the horizontal axis in \figr{ssLNy+} is $\eps^{LN}(0)$].
On the $p_z^{LN}(\phi(p_x))<p_z$ side of larger $p_z$,
around the right end of the projected line node, the velocity changes from negative to positive as $p_x$ decreases from $p_{\p0}$;
the velocity depends quite linearly on $p_x$.
The termination point of the Fermi contour, defined by $\Ec^{LN}_{y<0}(p_x,p_z)=\eps^{LN}(0)$,
on the $p_z^{LN}(\phi(p_x))<p_z$ side [\figr{ssLNypx}(a)]
is also in accord with the switch of the sign of the velocity $\pd_{p_z}\Ec_{y<0}^{LN}(p_x,p_z)$ with $p_x$.
On the $p_z<p_z^{LN}(\phi(p_x))$ side of smaller $p_z$, on the other hand,
the velocity stays negative and close to that of the lower bulk-band boundary in the whole range of $p_x$.

In \figr{ssLNy+}(d),(e),(f), we also plot the spectrum along the
circular paths of radius $p_r=0.1 p_{\p0}$ centered at $(p_x,p_z)=(- p_{\p0},p_{z0})$, $(0,p_{z0})$, and $(+p_{\p0},p_{z0})$.
Such circles cross the projected line node.
These circular paths will become particularly insightful for the Weyl semimetal, where their centers turn into projected Weyl points;
we will discuss the comparison of these spectra between the line-node and Weyl semimetals in \secr{ssWy}.

\subsection{$z>0$ sample \lbl{sec:ssLNz}}

\begin{figure*}
\centering
\includegraphics[width=\linewidth]{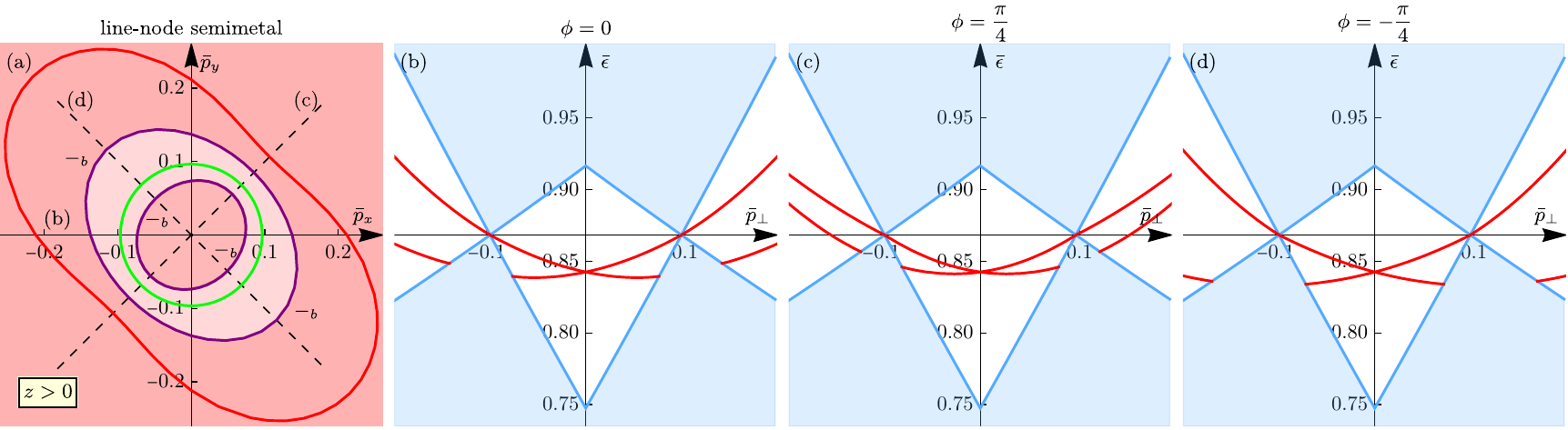}
\caption{
The surface-state spectrum $\Ec^{LN}_{z>0}(p_x,p_y)$ of the line-node semimetal for the $z>0$ sample.
Other conditions same as in \figr{ssLNy}.
(a) Surface-momentum plane.
(b),(c),(d) The spectrum along the $\phi=0,+\f\pi4,-\f\pi4$ paths, respectively.
}
\lbl{fig:ssLNz}
\end{figure*}

\subsubsection{Bulk-band boundaries and projected line node}

For the $z>0$ sample, the two line nodes $\pm_u$ [\eq{peLN}] project onto the same curve
\[
	(p_x,p_y)=(p_\p(\phi)\cos\phi,p_\p(\phi)\sin\phi), \spc \phi\in[0,2\pi),
\]
in the surface-momentum plane.
The bulk-band boundaries
\beq
    E_{z>0,+_b}^{LN}(p_x,p_y)=\min_{p_z,\sig} \eps_{+_b,\sig}^{LN}(p_x,p_y,p_z),
\lbl{eq:ELNz+}
\eeq
\beq
    E_{z>0,-_b}^{LN}(p_x,p_y)=\max_{p_z,\sig} \eps_{-_b,\sig}^{LN}(p_x,p_y,p_z).
\lbl{eq:ELNz-}
\eeq
meet at this projected line node at the line-node energy:
\beq
	E_{z>0,+_b}^{LN}(p_\p^{LN}(\phi)\cos\phi,p_\p^{LN}(\phi)\sin\phi)=
	E_{z>0,-_b}^{LN}(p_\p^{LN}(\phi)\cos\phi,p_\p^{LN}(\phi)\sin\phi)
	=\eps^{LN}(\phi),\spc \phi\in[0,2\pi).
\lbl{eq:ELNz=}
\eeq
Because of the energy dependence $\eps^{LN}(\phi)$ of the line node,
at a fixed energy $\eps^{LN}(0)$, there is also a finite-width region of bulk states, whose contours are defined by
\[
	E_{z>0,-_b}^{LN}(p_x,p_y)=\eps^{LN}(0).
\]
Due to the weak numerical dependence of the line-node energy, these deviations from the circle are not visible in the scale of \figr{ssLNz}.

\subsubsection{Surface-state spectrum}

The surface-state spectrum $\Ec^{LN}_{z>0}(p_x,p_y)$ of the line-node semimetal for the $z>0$ sample is presented in \figr{ssLNz}.
Again, compared to the Dirac semimetal,
the key changes occur at the scale $p_\p\sim p_{\be_1}$ set by the linear BIA term.
In the Dirac semimetal, there are two surface-state bands in the vicinity of the projected double Dirac point, both connected to the latter.
In the line-node semimetal, the two surface-state bands are degenerate at $p_\p=0$, due to time reversal symmetry $\Tc_-$.
For a given $\phi$, as a function of $p_\p$, the upper surface-state band passes through the projected line node.
The lower surface-state band merges with the lower bulk-band boundary and then reappears in the outer region,
and we notice that it does so virtually unhybridized with the bulk bands, as if passing them through.
In fact, it visually appears that the main qualitative effect of linear BIA term is to shift
the two surface-state bands of the Dirac semimetal in opposite directions along $p_\p$;
we demonstrate below that in the linear-in-momentum model this is a precise mathematical property.

While for every $\phi$ the dependence of the spectrum on $p_\p$ is qualitatively the same,
there is some quantitative dependence on $\phi$ as well, i.e., axial rotation symmetry is broken,
which is predominantly due to the linear BIA term rather than cubic anisotropy.
We plot the spectrum for fixed $\phi=0,\pm\f{\pi}4$ as a function of $p_\p$ (the spectrum for $\phi=\f\pi2$ is the same as for $\phi=0$).
The spectrum is most stretched along the $\phi=-\f{\pi}4$ direction and most squeezed along the $\phi=\f{\pi}4$ direction,
as also clearly evident in the in the Fermi contour (red) of the lower surface-state band in the outer region.

\subsubsection{Analytical calculation for the linear-in-momentum model}

We also find the surface-state spectrum analytically for the linear-in-momentum model \eqn{HcLN},
which includes the linear BIA term to leading order. This provides some additional insights.
We have seen that in the basis \eqn{Uphi} the Hamiltonian of the model is decoupled [\eq{HcLNU}] into two $2\tm 2$ blocks, labeled $\pm_m$.
The BCs \eqn{bcPsiz} for the $z=0$ surface have been derived in \secr{bcD}.
Clearly, this form of BCs holds upon any same change of basis within the $\pm_u$ parts of the wave function.
Repeating the steps of \secr{ssDlinz}, we obtain that the double-degenerate flat-band solution
\beq
    \Ec^{D,\x{lin}}_{z>0,\pm_m}(k_\p) \equiv \eps_u,
\lbl{eq:EcLNlinz}
\eeq
remains also in the presence of the leading BIA term. Indeed, its effect solely amounts to shifting the spectra
of the two decoupled blocks in momentum $k_\p$ by $\pm m/v_\p$, compare \eqs{eLNlin+m}{eLNlin-m} to \eq{eDlin}.
Chiral symmetry still holds in the presence of the BIA term and still provides an explanation for this behavior.
The surface-state spectrum is presented in \figr{ssDylin}(b). The bulk-band boundaries [\eqs{eLNlin+m}{eLNlin-m}] are
\[
	E^{LN,\x{lin}}_{z>0,\pm_b}(k_\p)=\eps_u \pm_b \sq{1-\tf{v_0^2}{v_z^2}}|v_\p k_\p - m|.
\]

Another conclusion that can be drawn is that all the deviations (lifting of the degeneracy and nonzero dispersion)
of the surface-state bands (\figr{ssLNz}) from this double-degenerate flat-band behavior
are due to the next-order terms of the Luttinger model that have been neglected in the linear-in-momentum model.
	
\section{Bulk topology and bulk-boundary correspondence of a Weyl semimetal}

In \secr{ssW}, we calculate and analyze the surface states of the Weyl semimetal.
One of our main goals in that section is exploring the manifestation of bulk topology in the behavior of the surface states;
such relation is known as {\em bulk-boundary correspondence}~\cite{Chiu2016}.
In this section, we remind the basic topological properties of Weyl semimetals
and formulate a type of bulk-boundary correspondence
that directly relates the characteristics of the surface-state spectrum to the Weyl points,
which we will actively use in \secr{ssW}.

\subsection{Bulk topology of a Weyl semimetal \lbl{sec:topoW}}

The topological properties of 3D Weyl semimetals are directly tied to those of a 2D quantum anomalous Hall (QAH) systems
(Chern insulators)~\cite{Armitage2017}.
In the general topological classification scheme~\cite{Ryu2010,Chiu2016},
both systems belong to class A with no assumed symmetries in their respective dimensions.
Any surface $\Sig$ in the 3D momentum space that is closed or extends to infinity may be regarded as an effective 2D QAH bulk system.
The Chern number
\beq
	C(\Sig)=\f1{4\pi}\int_\Sig \dx^2 \sigb_\pb \, \Bb(\pb)
\lbl{eq:Cdef}
\eeq
of this system is given by the flux of the wave-function-basis-independent Berry curvature vector field $\Bb(\pb)$ through that surface
($\sigb_\pb$ is the normal area vector at point $\pb\in\Sig$ on the surface).

Weyl points are singularities of the Berry curvature,
at which the orientation of the field depends on the direction from which the Weyl point is approached.
The Chern number
\beq
	C^W(\pb_0)=C(\Sig_{\pb_0})
\lbl{eq:CWdef}
\eeq
of a Weyl point at $\pb_0$ is defined as the Chern number over a surface $\Sig_{\pb_0}$ enclosing only that point,
e.g., a sphere of a sufficiently small radius centered at the Weyl point.
Weyl points can be seen as sources and drains of the Berry curvature field, depending on the signs of their Chern numbers.
Note that, as a flux integral, the sign of the Chern number depends on the assigned orientation of the surface.
A unified convention must be chosen for the orientations of the surfaces.

Since the Berry curvature $\Bb(\pb)$ is a rotorless vector field, $[\n_\pb\tm\Bb(\pb)]=\nv$,
the Chern numbers of different surfaces obey conservations laws, which follow from the Gauss theorem.
The difference of the Chern numbers of the two surfaces $\Sig_1$ and $\Sig_2$ not crossing any Weyl poins
is given by the sum of the Chern numbers of all Weyl points enclosed in the region between the two surfaces,
\beq
	C(\Sig_1)-C(\Sig_2)=\sum_{\pb_\al \x{between $\Sig_1$ and $\Sig_2$}} C^W(\pb_\al).
\lbl{eq:dC}
\eeq
In particular, the Chern number of all surfaces that can be continuously deformed to one another without crossing any of the Weyl points is the same.
Sets of such surfaces can be regarded as topological phases of effective 2D QAH systems.
The formula \eqn{dC} can be interpreted as the change as the Chern numbers across a topological phase transition realized
as the surface $\Sig$ is continuously deformed from $\Sig_1$ and $\Sig_2$.
A surface $\Sig$ passing through at least one Weyl point can be seen as a topological phase transition,
where the effective 2D QAH system becomes a gapless nodal semimetal.

For the considered generalized Luttinger model,
for the Dirac semimetal, the Chern number of any surface not crossing the Dirac points is zero by cancellation,
because a Dirac node consists of two Weyl nodes with opposite Chern numbers.
Upon introducing linear and cubic BIA terms,
by continuity, the Chern number remains zero for any surface $\Sig$
that is not crossed by the created line node or Weyl points during this process.
Knowing this and the Chern numbers of the Weyl points, one can then easily determine the Chern numbers of any other surface using \eq{dC}.
The Chern numbers $C^W(\pb_\al)$ of the Weyl points can be calculated using the expansion of the Hamiltonian about the Weyl points.
We choose the sign convention so that
the Chern numbers of the four Weyl points of the two groups $\pb=(\pm p_\p^W,0,\pm p_z^W)$ and $\pb=(0,\pm p_\p^W,\pm p_z^W)$
equal $-1$ and $+1$, respectively.
This corresponds to the convention as in Ref.~\ocite{KharitonovFGCM}, with the outward sphere orientation
and the flux is of the local pseudospin of the lower (filled) bulk band.

Among possible surfaces $\Sig$ in 3D momentum space, it is common to consider
families of parallel planes with the same common orientation and different positions.
For the considered Weyl semimetal, for the $p_x=\x{const}$ family of planes $\Sig_{p_x}$,
there are four ``topological phases'' separated by the planes $p_x=-p^W_\p,0,+p^W_\p$ of three ``phase transitions''.
The sum of the Chern numbers of the two Weyl points in the $p_x=-p^W_\p$ plane equals $-2$,
that of the four Weyl points in the plane $p_x=0$ equals $+4$,
and that of the two WPs in the $p_x=+p^W_\p$ plane equals $-2$.
According to \eq{dC}, these determine the changes of the Chern numbers of the planes as the phase-transition plane is crossed.
The Chern numbers of the $p_x=\x{const}$ planes in the phases $p_x<-p^W_\p$, $-p^W_\p<p_x<0$, $0<p_x<p^W_\p$,
and $p^W_\p<p_x$ therefore equal $C(\Sig_{p_x})=0,-2,+2,0$, respectively, for the normal of the plane pointing in the positive $p_x$ direction.
For the $p_z=\x{const}$ planes,
since the sum of the Chern numbers of the four Weyl points in the $p_z=\pm p_z^W$ planes is zero, the Chern number of any $p_z=\x{const}$ plane is zero.

These bulk topological relations are also manifested in the momentum planes of specific sample surfaces,
as the surfaces in 3D momentum space project as paths onto the 2D surface-momentum plane,
as shown in \figr{ssWy}(a) for $y<0$ sample and \figr{ssWz}(a) for $z>0$ sample and will be discussed in more detail in the respective sections.

\subsection{Generalized bulk-boundary correspondence of a Weyl semimetal \lbl{sec:topoWgen}}

\subsubsection{Bulk-boundary correspondence for paths in the surface-momentum plane \lbl{sec:bbcW}}

In Ref.~\ocite{KharitonovFGCM}, the formulation of the bulk-boundary correspondence for Weyl semimetals
was presented that is more general than commonly considered.
Also, as its special case, a version of the bulk-boundary correspondence follows that fully characterizes the vicinity of the Weyl point:
the chirality of the surface-state spectrum along a path enclosing the projected Weyl point is equal to the Chern number of the Weyl point.
In Ref.~\ocite{KharitonovFGCM}, it was presented for one Weyl point.
The generalization to any number of Weyl points is straightforward, as we now present.

To be specific, we consider the $z>0$ sample; the formulation is, of course, applicable to any surface orientation.
Consider an arbitrary path $\pb_\p^\ga(t)$, parameterized by $t$, in the surface-momentum plane $\pb_\p=(p_x,p_y)$
that is either closed or terminates with both ends at infinity $|\pb_\p|=+\iy$, and does not cross any projected Weyl points.
Each such path defines an effective 2D QAH system with the Hamiltonian $\Hh^W(\pb_\p^\ga(t),\ph_z)$
on a {\em generalized cylinder} $\Sig^\ga$ in the 3D momentum space,
obtained by passing straight $p_z$ lines (perpendicular to the surface) through this path $\ga$.
The Chern number $C(\Sig^\ga)$ [\eq{Cdef}] of such system is given by the flux through this generalized cylinder.

The surface-state spectrum $\Ec^W_{z>0}(\pb_\p^\ga(t))$ along this path becomes the ``edge-state'' spectrum as a function of $t$
of this effective 2D QAH system defined on the space $(p_z,t)$. The bulk-boundary correspondence for this system states that the signed number
\beq
	N(\Sig^\ga)=N[\Ec^W_{z>0}(\pb_\p^\ga(t))]
\lbl{eq:Nga}
\eeq
of the effective chiral edge-state bands is given by the Chern number $C(\Sig^\ga)$ of the cylinder $\Sig^\ga$:
\beq
	N(\Sig^\ga)=C(\Sig^\ga),
\lbl{eq:bbcga}
\eeq
An edge-state band is called chiral if it connects lower $-_b$ (``valence'') and upper $+_b$ (``conduction'') bulk bands.
The signed number $\pm1$ of one chiral band, also called ``chirality'',
is determined by the order of the merging points $t_{\pm_b}$ with these bulk bands on the $t$ axis.
If there are several chiral bands, then the total signed number (chirality) $N(\Sig^\ga)$ is given by the sum of their individual chiralities.

The bulk-boundary correspondence of Weyl semimetals is commonly expressed
in terms of ``conventional'' 2D QAH systems defined on infinite 2D planes in 3D momentum space.
This now becomes a special case of the above more general formulation, when the paths $\pb_\p^\ga(t)$ are straight lines,
and the generalized cylinders on which the QAH systems are defined are such planes.
For example, for a plane with the fixed $p_x=P_x=\x{const}$, to be denoted as $\Sig_{P_x}$,
the parameter $t=p_y$ is the momentum component itself.
The surface-state spectrum $\Ec^W_{z>0}(P_x,p_y)$ is interpreted in this picture as effective edge-state spectra (functions of $p_y$)
of a family a 2D QAH systems defined on the planes $\Sig_{P_x}$.
According to the bulk-boundary correspondence [\eqs{Nga}{bbcga}], the signed number
\beq
	N(\Sig_{P_x})=N[\Ec^W_{z>0}(P_x,p_y)]=C(\Sig_{P_x})
\lbl{eq:NPx}
\eeq
of the chiral edge-state bands is determined by the Chern number $C(\Sig_{P_x})$ of the plane $\Sig_{P_x}$.
The Chern numbers $C^W(\pb_\al)$ of the Weyl points determine via \eqn{dC}
the corresponding {\em changes} in the chiralities of the effective edge-state spectra upon crossing the Weyl points by the plane:
\beq
	N(\Sig_{P_{x1}})-N(\Sig_{P_{x2}})=\sum_{\pb_\al \x{ between } \Sig_{P_{x1}} \x{ and } \Sig_{P_{x2}}} C^W(\pb_\al).
\lbl{eq:dN}
\eeq

However, the above more general formulation \eqn{bbcga}
allows us to formulate, as its special case, a different type of bulk-boundary correspondence of Weyl semimetals
that is much better suited for the characterization of the surface-state spectrum in the vicinity of projected Weyl points.
Consider a closed path $\ga$ that encloses some number of projected Weyl points $\pb_\al$. The Chern number
\beq
	C(\Sig^{\ga})=\sum_{\pb_\al \x{ inside } \Sig^{\ga}} C^W(\pb_\al)
\lbl{eq:Cga=CW}
\eeq
of the 2D QAH system defined on the cylinder $\Sig^\ga$ generated by such path
is equal to the net sum of the Chern numbers [\eq{CWdef}] of the Weyl points enclosed by the cylinder $\Sig^\ga$,
since such cylinder and the set of spheres around the Weyl points can be continuously deformed into each other.
According to the bulk-boundary correspondence \eqn{bbcga}, the signed number (chirality)
\beq
	N(\Sig^\ga)=N[\Ec^W_{z>0}(\pb_\p^\ga(t))]=\sum_{\pb_\al \x{ inside } \Sig^{\ga}} C^W(\pb_\al)
\lbl{eq:Nga}
\eeq
of the effective chiral edge-state bands $\Ec^W_{z>0}(\pb_\p^\ga(t))$ along the path $\ga$
is given directly by this net sum of the Chern numbers of the enclosed Weyl points.

We will explicitly demonstrate the manifestation of both of these types of bulk-boundary correspondence
for the studied Weyl-semimetal model.

\subsubsection{Bulk-boundary correspondence for the oriented Fermi contours \lbl{sec:bbcWFermicontours}}

In fact, for a given surface, the essential information about the topological properties of the surface-state spectrum $\Ec^W_{z>0}(\pb_\p)$
and its bulk-boundary correspondence for {\em any} path $\pb_\p^\ga(t)$
can be captured in one 2D plot of the {\em oriented} fixed-energy contours
in the surface-momentum plane $\pb_\p$.
Such fixed-energy contours (also called Fermi contours, assuming the Fermi level is at the Weyl-point energy)
where the surface-state bands cross the energy $\eps^W$ level of the Weyl points, are defined by the equation
\[
	\Ec^W_{z>0}(\pb_\p)=\eps^W.
\]

The chirality $N[\Ec^W_{z>0}(\pb^\ga_\p(t))]$ of the effective edge-state spectrum $\Ec_{z>0}^W(\pb^\ga_\p(t))$ along an arbitrary path $\pb_\p^\ga(t)$
can alternatively be determined as the net signed number of its crossings of a certain energy level, such as $\eps^W$,
where the sign of the crossing is determined by the sign of the ``velocity''
\[
	\pd_t\Ec^W_{z>0}(\pb^\ga_\p(t))=\vb^W_{z>0}(\pb^\ga_\p(t))\cd \pd_t\pb_\p^\ga(t)
\]
with respect to $t$.
This sign is therefore determined by the scalar product of the path's tangent vector $\pd_t\pb_\p^\ga(t)$ and the velocity vector
\[
	\vb^W_{z>0}(\pb_\p)=\pd_{\pb_\p}\Ec^W_{z>0}(\pb_\p)
\]
at the intersection of the path and the Fermi contour.
The velocity $\vb^W_{z>0}(\pb_\p)$ is orthogonal to the Fermi contour
and which of the two options of its orientation is realized
can be specified without plotting the vector by assigning an orientation to the Fermi contour.
It makes sense to coordinate the contour orientation with the Chern numbers of the projected Weyl points
so that the points with positive and negative Chern numbers acts as effective sources and drains, respectively.

Summarizing, for any path $\ga$, one can determine the chirality $N(\Sig^\ga)$ of the effective edge state-spectrum along it
by considering its intersection points with the Fermi contour and
counting the numbers of these crossings with the signs of the velocities deduced from the local orientation of the contour.

Generally, there may be several branches of the Fermi contours connected that meet at one projected Weyl point,
in which case there is a conservation law according to the ``source and drain'' picture,
that the net number of the oriented branches has to equal the net Chern number of the projected Weyl point.
In particular, each single projected Weyl point with the Chern number $\pm1$ emanates or terminates one Fermi contour.
A single Fermi contour connecting two projected single Weyl points with the Chern numbers $\pm1$ is
typically referred to as a {\em Fermi arc} and is the simplest situation.
For $0$ or $\pm2$ net Chern numbers of the projected Weyl points, the Fermi-contour structures are more complicated, as we will see below.

\section{Surface states of the Weyl semimetal\lbl{sec:ssW}}

In this section, we calculate and explore the surface states of the Weyl semimetal
that arises in the Luttinger semimetal upon introducing compressive strain and linear and cubic BIA terms.
Without the cubic BIA terms, the system is a line-node semimetal, whose surface states were studied in the previous \secr{ssLN}.
We calculate the surface states within the Luttinger model, described by the Hamiltonian $\Hh^W(\pb)$ [\eq{HW}] and BCs \eqn{bc},
for the half-infinite $y<0$ and $z>0$ samples, using the semi-analytical method presented in \secr{method}.
We present the Weyl-semimetal case for the parameters $\al_{0,z,\square}$ and $\be_1$ of HgTe provided in \tabr{LM}
and compressive strain $u=-3\x{meV}$, the same values as for the line-node semimetal case.
The choice of the cubic BIA parameters has been explained in \secr{LfromK}.
Such values obey well the hierarchy $p_{\be_3}\ll p_{\be_1} \ll p_u$ of scales described in \secr{scales}.

\subsection{$y<0$ sample\lbl{sec:ssWy}}

\begin{figure*}
\centering
\includegraphics[width=\linewidth]{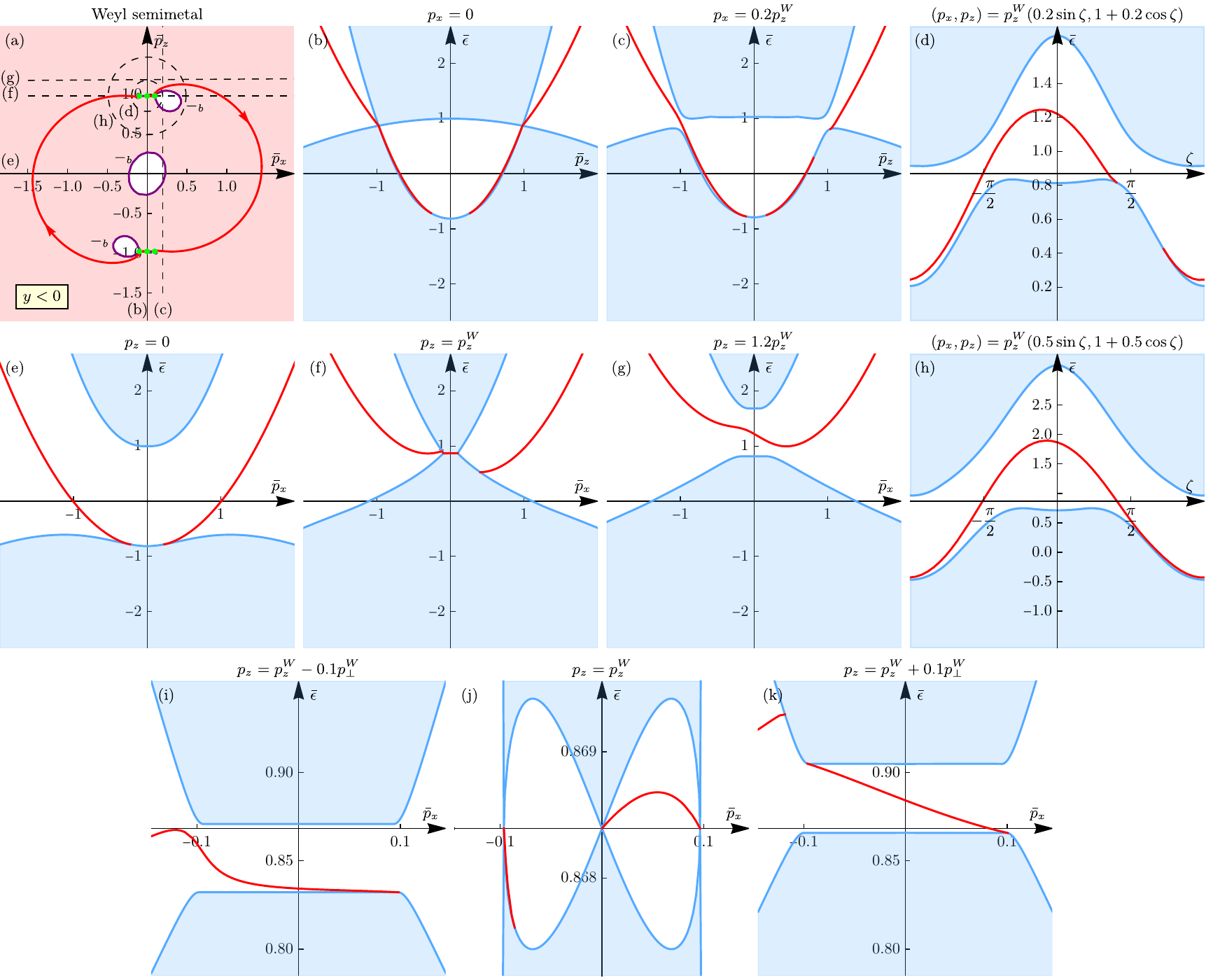}
\caption{
The surface-state spectrum $\Ec^W_{y<0}(p_x,p_z)$ of the Weyl semimetal for the $y<0$ sample,
calculated within the Luttinger model with compressive strain and linear and cubic BIA terms [\eqs{HW}{bc}]
for the parameters of HgTe (\tabr{LM}) and the strain value $u=-3~\x{meV}$.
The structure of the figure is the same as in \figr{ssDy17} for the Dirac semimetal and \figr{ssLNy} for the line-node semimetal.
The projected Weyl points are shown in green.
Away from the (former) projected line nodes, the bulk and surface-state spectra are very similar to that of the line-node semimetal (\figr{ssLNy}).
}
\lbl{fig:ssWy}
\end{figure*}

\begin{figure*}
\centering
\includegraphics[width=\linewidth]{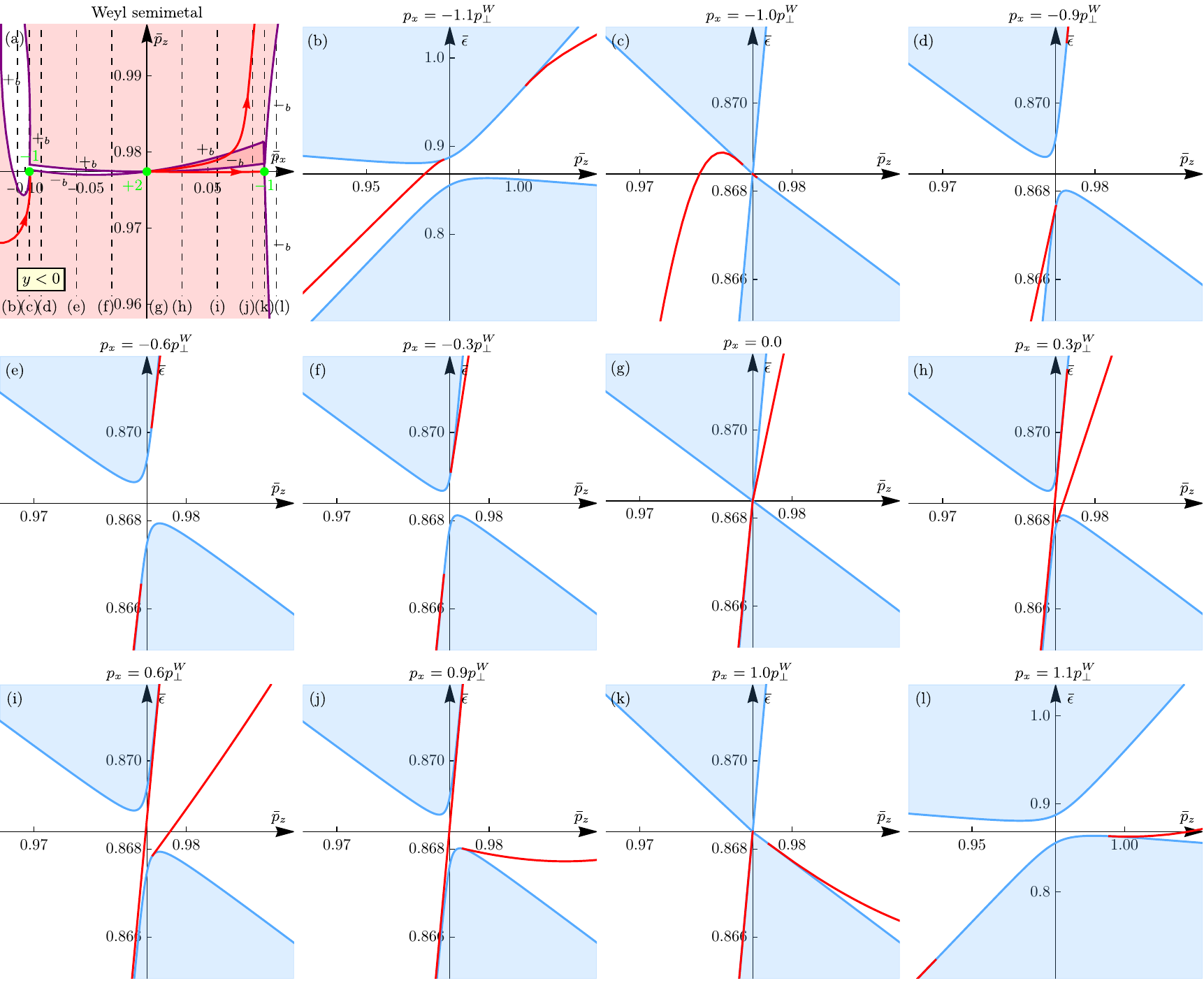}
\caption{
The surface-state spectrum $\Ec^W_{y<0}(p_x,p_z)$ of the Weyl semimetal
along the $p_x=\x{const}$ straight-line paths in the vicinity of the former $+_u$ projected line node
$p_z\approx p_z^W\approx p_z^{LN}(0)$ (additional characterization of \figr{ssWy}).
The main difference from the analogous \figr{ssLNypx} for the line-node semimetal
is that the line node gaps out everywhere expect for the Weyl points at $p_x=0,\pm p_\p^W$.
and the surface-state spectrum adjusts accordingly.
The edge-state spectra $\Ec^W_{y<0}(p_x,p_z)$ of the effective 2D QAH systems defined on $p_x=\x{const}$ planes
obey the bulk-boundary correspondence \eqn{NPx} in all four ``gapped'' regions of $p_x$.
Note that the aspect ratio in (a) differs from 1 to resolve the features along the $p_z$ direction.
}
\lbl{fig:ssWypx}
\end{figure*}

\begin{figure*}
\centering
\includegraphics[width=\linewidth]{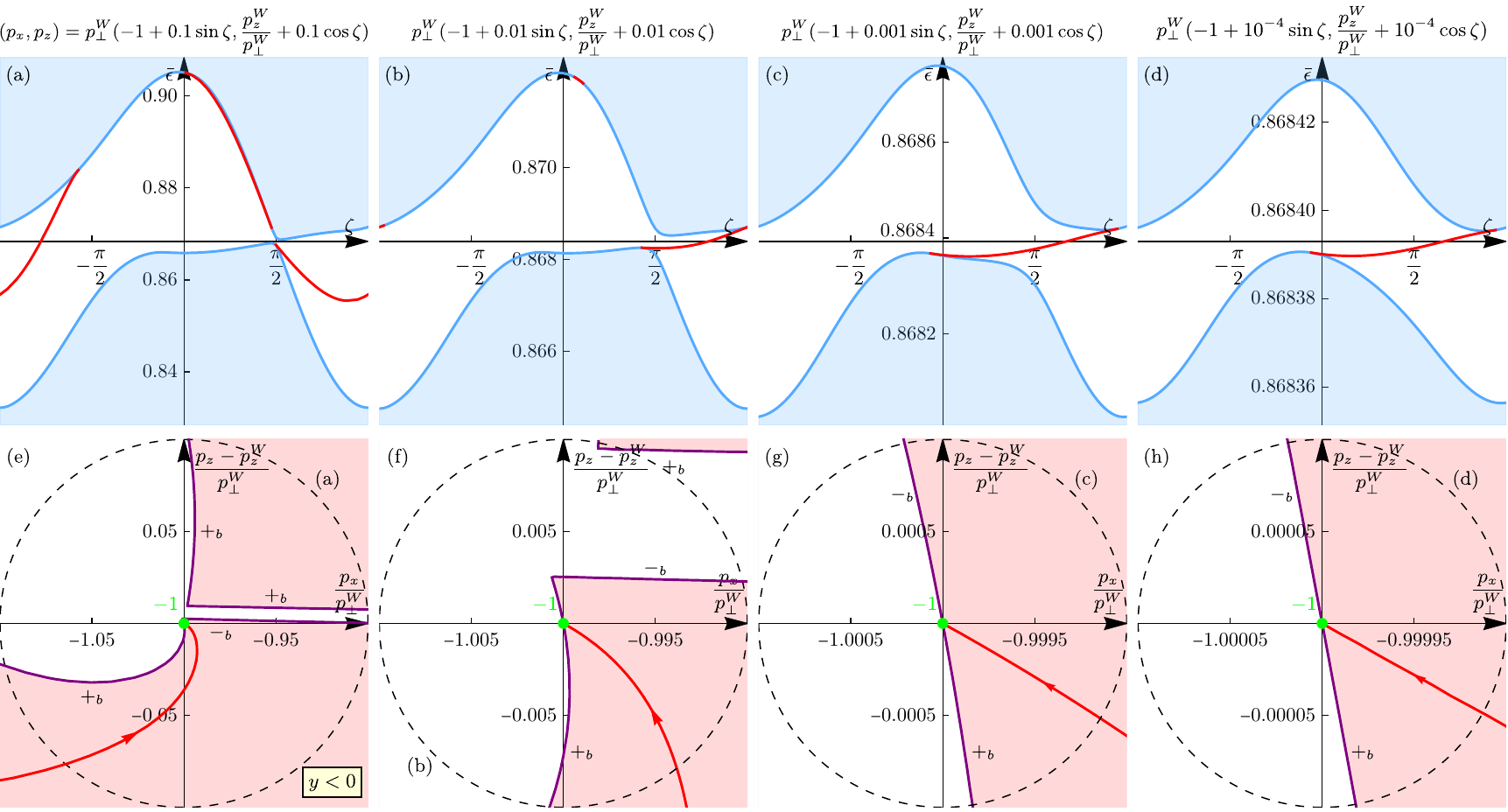}
\caption{
The surface-state spectrum $\Ec^W_{y<0}(p_x,p_z)$ of the Weyl semimetal along the circular paths
around the projected Weyl point $(p_x,p_z)=(-p_\p^W,p_z^W)$ with the Chern number $C^W=-1$ (additional characterization of \figr{ssWy}).
The bulk-boundary correspondence~\cite{KharitonovFGCM} \eqn{Nga} characterizing the vicinity of the projected Weyl point
is satisfied and, for small enough radius, the emergent linear regime \eqn{Eclinexp} is observed,
as identified by the asymptotically straight Fermi and merging contours as the projected Weyl point is approached;
the same concerns \figsr{ssWypr0}{ssWypr+}.
}
\lbl{fig:ssWypr-}
\end{figure*}

\begin{figure*}
\centering
\includegraphics[width=\linewidth]{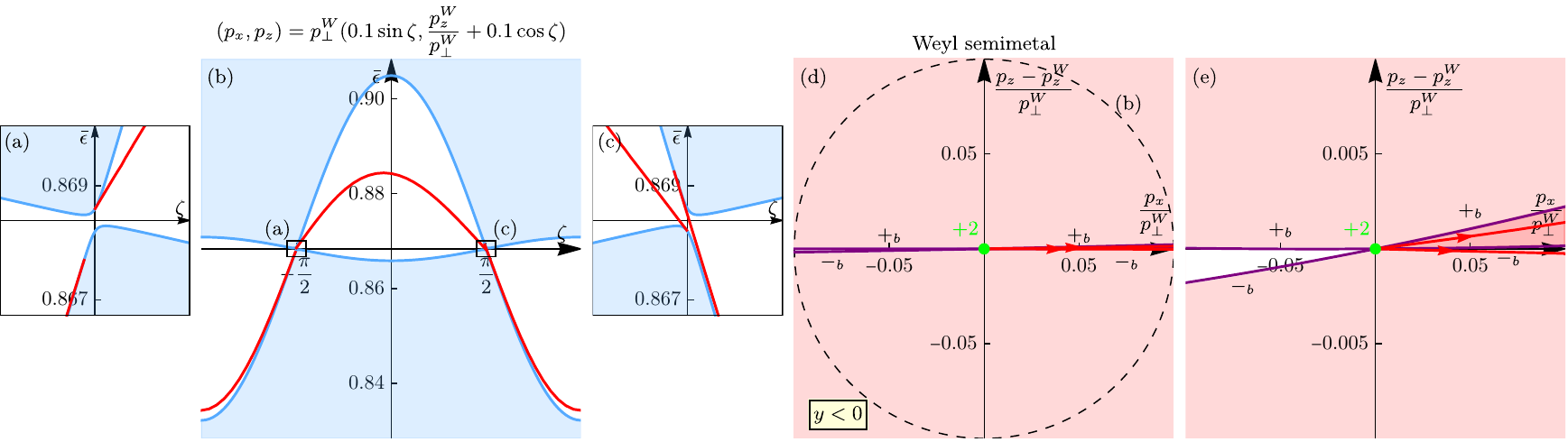}
\caption{
The surface-state spectrum $\Ec^W_{y<0}(p_x,p_z)$ of the Weyl semimetal along the circular paths
around the projected double Weyl point $(p_x,p_z)=(0,p_z^W)$ with the net Chern number $C^W=+2$ (additional characterization of \figr{ssWy}).
Note that the aspect ratio in (e) differs from 1 in order to resolve the features along the $p_z$ direction.
}
\lbl{fig:ssWypr0}
\end{figure*}

\begin{figure*}
\centering
\includegraphics[width=\linewidth]{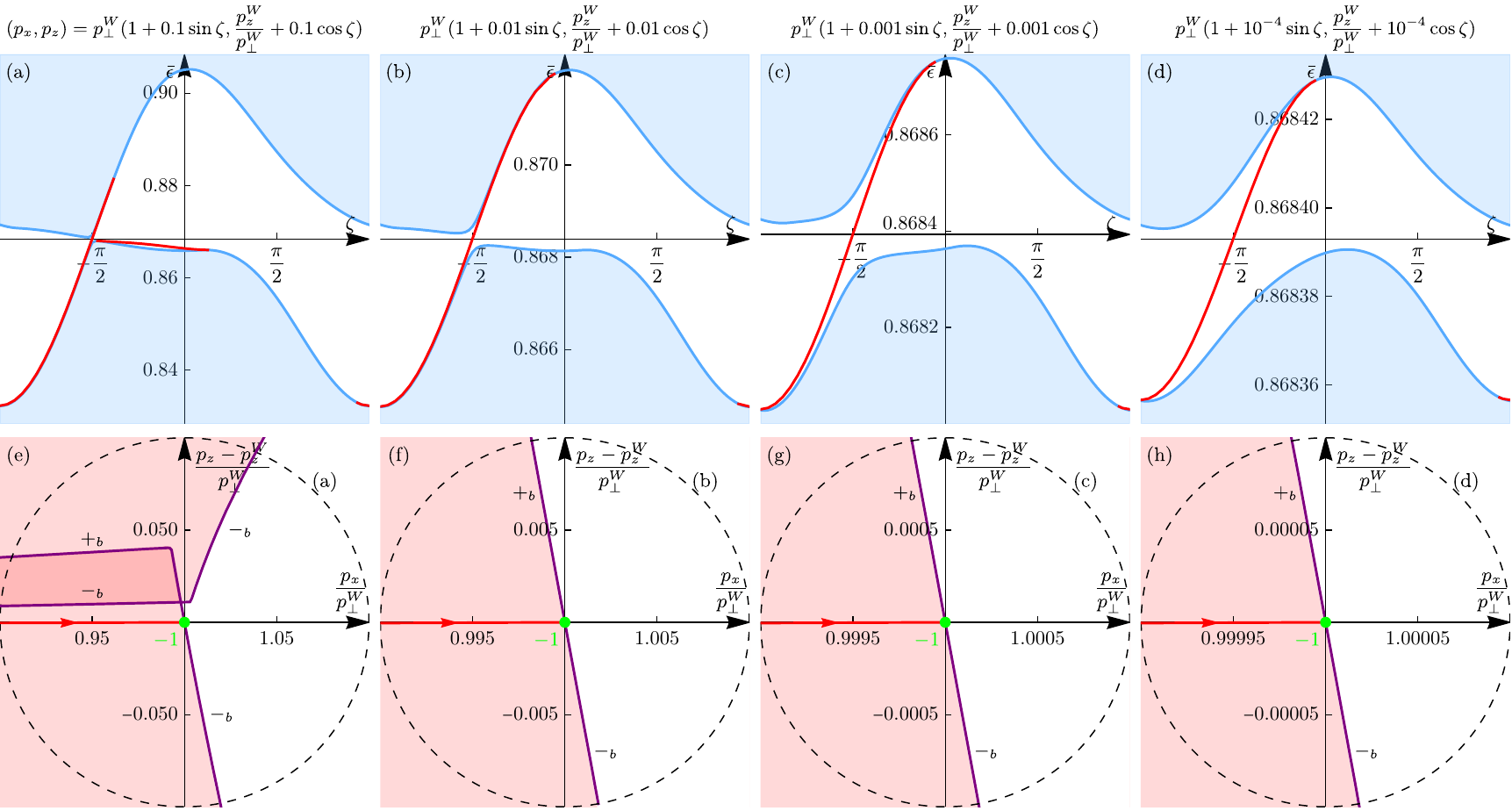}
\caption{
The surface-state spectrum $\Ec^W_{y<0}(p_x,p_z)$ of the Weyl semimetal along the circular paths
around the projected Weyl point $(p_x,p_z)=(p_\p^W,p_z^W)$ with the Chern number $C^W=-1$ (additional characterization of \figr{ssWy}).
}
\lbl{fig:ssWypr+}
\end{figure*}

We describe the surface-state spectrum $\Ec^W_{y<0}(p_x,p_z)$ of the Weyl semimetal for the $y<0$ sample
by the 2D plots presented in \figsdr{ssWy}{ssWypr+}.
\figr{ssWy} shows the overall behavior and has the same structure as \figr{ssDy17} for the Dirac semimetal and \figr{ssLNy} for the line-node semimetal.
\figsdr{ssWypx}{ssWypr+} fully characterize the vicinity of the former projected line node:
\figr{ssWypx} shows the spectrum along the series of $p_x=\x{const}$ paths, analogous to \figr{ssLNypx} for the line-node semimetal,
and \figssr{ssWypr-}{ssWypr0}{ssWypr+} -- along the circular paths around the projected Weyl points
$(p_x,p_z)=(- p_\p^W,p_z^W)$, $(0,p_z^W)$, and $(+p_\p^W,p_z^W)$ of the $+_u$ group, respectively.
The surface-state spectrum along some other characteristic paths, not presented here, can be deduced using the symmetry
$\Ec_{y<0}^W(-p_x,-p_z)=\Ec_{y<0}^W(p_x,p_z)$.

\subsubsection{Projected bulk spectrum}

As explained in \secr{bulkW}, the cubic BIA terms transform the line-node semimetal
into a Weyl semimetal by ``gapping out'' each line node in the bulk spectrum everywhere except for the four Weyl points.
For the $y=0$ surface, there is a total of six projected Weyl points in the surface-momentum plane $(p_x,p_z)$ [\figr{ssWy}(a) and \figr{ssWypx}(a)]:
two Weyl points with the same Chern numbers are projected onto each of the points $(p_x,p_z)=(0,\pm p^W_z)$, resulting in their net Chern number $C^W=+2$;
one Weyl point is projected onto each of the points $(p_x,p_z)=(\pm p^W_\p, \pm p^W_z)$, with the Chern number $C^W=-1$.
The upper and lower bulk-band boundaries $E_{y<0,\pm_b}^W(p_x,p_z)$
of the Weyl semimetal touch only at these six projection Weyl points, at the common energy $\eps^W$.
For the used parameters, the differences between the values $(p_\p^{LN}(0),p_z^{LN}(0),\eps^{LN}(0))$ of the line node
and $(p_\p^W,p_z^W,\eps^W)$ of the Weyl points are negligible.

The ``gapping out'' of the line node is best seen in the spectrum along
the straight-line path $p_z=p_z^W$ passing through the $+_u$ group of the three projected Weyl points,
shown in \figr{ssWy}(j), where the ``gap'' $E^W_{y<0,+_b}(p_x,p_z^W)-E^W_{y<0,-_b}(p_x,p_z^W)\sim \eps_{\be_3}$ [\eq{epsbe3}]
in the bulk spectrum is now due to the cubic BIA terms.
Compare this to a much smaller scale of the ``gap'' $E^{LN}_{y<0,+_b}(p_x,p_{z0})-E^{LN}_{y<0,-_b}(p_x,p_{z0})$
in the bulk spectrum of the line-node semimetal along the $p_z=p^{LN}_z(0)=p_{z0}$ path, shown in \figr{ssLNy+}(b),
which is only due to the very weak $p_z^{LN}(\phi)$ dependence.

\subsubsection{Surface-state spectrum, general properties}

In the surface-momentum plane $(p_x,p_z)$, changes occur in the regions of width $p_{\be_3}$ [\eq{pbe3}] around the former projected line nodes,
which for the considered values is much smaller than the typical radius $\sim p_{\be_1}$ [\eq{pbe1}] of the line-node.
Away, the bulk and surface-state spectrum is very similar to that of the line-node semimetal, compare \figr{ssWy} to \figr{ssLNy} and \figr{ssLNy+}.
Already for the $p_z=p_z^W\pm 0.1p_\p^W$ paths, positioned close to the $p_z=p_z^W$ line, on which the $+_u$ group of projected Weyl points resides,
the bulk-state and surface-state spectra of the Weyl semimetal are essentially identical quantitatively to those of the line-node semimetal:
compare \figr{ssWy}(i) and (k) to \figr{ssLNy+}(a) and (c).

The key changes upon introducing the cubic BIA terms are as follows. As demonstrated in \secr{ssLN},
the surface-state band $\Ec^{LN}_{y<0}(p_x,p_z)$ of the line-node semimetal merges with the bulk states at both sides of the projected line node.
Upon introducing the cubic BIA terms and opening the small gap $\sim \eps_{\be_3}$ [\eq{be3}]
in the bulk spectrum at the $+_u$ projected line node,
the two sides $p_z\gtrless p_z^{LN}(\phi(p_x))$ of the surface-state band $\Ec^{LN}_{y<0}(p_x,p_z)$ adjust,
so that in the Weyl semimetal the surface-state band
$\Ec^W_{y<0}(p_x,p_z)$ remains connected to either the upper or lower bulk-band boundary $E^W_{y<0,\pm_b}(p_x,p_z)$.
This happens differently in the $p_x\in(-p_\p^W,0)$ and $p_x\in(0,+p_\p^W)$ regions.
Namely, in the region $p_x\in(-p_\p^W,0)$, the part of the surface-state band approaching the former projected line node
from smaller $p_z$ and smaller energies connects to the lower $-_b$ bulk-band boundary
and the part thereof approaching from larger $p_z$ and higher energies connects to the upper $+_b$ bulk-band boundary.
As a result, the surface-state band does not cross the opened bulk gap in the region $p_x\in(-p_\p^W,0)$ around $p_z=p_z^W$.
In the region $p_x\in(0,+p_\p^W,0)$, the behavior is reversed.
As a result, the surface-state band crosses the opened bulk gap twice in the region $p_x\in(0,+p_\p^W)$ around $p_z=p_z^W$.

These differences in the bulk and surface-state spectra of the Weyl and line-node semimetals
are best seen in the spectrum along the series of $p_x=\x{const}$ paths,
presented in \figr{ssWypx} for the vicinity of the $+_u$ group of projected Weyl points, as compared to \figr{ssLNypx} for the line-node semimetal.
The spectrum in the vicinity of the $-_u$ group is related to $+_u$ via the symmetry $\Ec_{z>0}^W(-p_x,-p_z)=\Ec_{z>0}^W(p_x,p_z)$.

These differences are also seen in the spectrum along the circular paths of radius $p_r=0.1 p_\p^W$
around the projected Weyl points $(p_x,p_z)=(- p_\p^W,p_z^W)$, $(0,p_z^W)$, and $(+p_\p^W,p_z^W)$,
presented in \figr{ssWypr-}(a), \figr{ssWypr0}(b), \figr{ssWypr+}(a) for the Weyl semimetal, respectively,
as compared to \figr{ssLNy+}(d),(e),(f) for the line-node semimetal,
where the spectrum along the circular paths of radius $p_r=0.1 p_{\p0}$
centered at the points $(p_x,p_z)=(- p_{\p0},p_{z0})$, $(0,p_{z0})$, and $(+p_{\p0},p_{z0})$ of the projected line node
are presented.

These differences are also manifested in the Fermi [$\Ec^W_{y<0}(p_x,p_z)=\eps^W$, red]
and merging [$\Ec^W_{y<0}(p_x,p_z)=E^W_{y<0,\pm_b}(p_x,p_z)$, purple] contours of the Weyl semimetal,
presented in \figr{ssWypx}(a) and \figssr{ssWypr-}{ssWypr0}{ssWypr+}.
We see that the regions of absent surface states around the $(p_x,p_z)=(\pm p_\p^W,p_z^W)$ projected Weyl points are still present,
inherited from the line-node semimetal, controlled mainly by the linear BIA term.
However, additional Fermi and merging contours appear in the region of the former projected line node,
according to the restructuring of the bulk and surface-state bands explained above.

Away from the regions of the former projected line nodes,
the Fermi contours of the Weyl semimetal have the same shape as those of the line-node semimetal [\figr{ssWy}(a)].
As the Fermi contour connected to the $(p_x,p_z)=(-p_\p^W,p_z^W)$ projected Weyl point deviates from the $+_u$ region,
it becomes the same as that of the line-node semimetal until it eventually reaches the $-_u$ region,
where it connects to the $(0,-p_z^W)$ projected double Weyl point.
The Fermi contour connected to the $(p_\p^W,p_z^W)$ projected Weyl point
connects to the $(0,p_z^W)$ projected double Weyl point along an almost straight line.
The other Fermi contour connected to $(0,p_z^W)$ also moves towards the $(p_\p^W,p_z^W)$ point;
however, it does not connect to it, but only passes closely;
after which it deviates from the $+_u$ region and follows closely the Fermi contour of the line-node semimetal
until it reaches to $-_u$ group, which it connect to the $(p_\p^W,-p_z^W)$ point.
Altogether, there are two disconnected contours,
in accord with absent surface states crossing the $\eps^W$ level in the $p_x\in(-p_\p^W,0)$ region around $p_z=+p_z^W$
and in the $p_x\in(0,+p_\p^W)$ region around $p_z=-p_z^W$.
Note that, similar to the discussion for the Dirac semimetal in \secr{Ddiscussion},
this only means that there is no connectivity 
{\em at this energy}: {\em all six} projected Weyl points {\em are} connected by the surface-state band.

\subsubsection{Surface-state spectrum, topological properties}

We now explore the topological aspect of the Weyl semimetal, as well as of the Dirac and line-node semimetals.
As per the general topological framework presented in \secr{bbcW}, the surface-state spectrum $\Ec^W_{y<0}(p_x^\ga(t),p_z^\ga(t))$
along various paths $(p_x^\ga(t),p_z^\ga(t))$ presented in \figsdr{ssWy}{ssWypr+} (straight lines and circles),
can be regarded as edge-state spectra of effective generalized 2D QAH systems, defined on the respective cylinders $\Sig^\ga$.
All these spectra obey the bulk-boundary correspondence \eqn{bbcga}, as we now demonstrate.

According to the changes in the spectrum in the vicinity of the former projected line nodes,
the contribution from the region around $p_z=p_z^W$ to the chirality of the effective edge-state bands oriented with growing $p_z$
is $0$ in the $p_x\in(-p_\p^W,0)$ region and $+2$ in the $p_x\in(0,+p_\p^W)$ region.
By the symmetry $\Ec_{y<0}^W(-p_x,-p_z)=\Ec_{y<0}^W(p_x,p_z)$, the contribution to the chirality from the region around $p_z=-p_z^W$,
is $-2$ in the $p_x\in(-p_\p^W,0)$ region and $0$ in the $p_x\in(0,+p_\p^W)$ region.
Adding these together, the chiralities $N(\Sig_{p_x})$ of the effective edge-state spectra of the $p_x=\x{const}$ QAH systems (\figr{ssWypx})
are indeed in accord with the Chern numbers $C(\Sig_{p_x})=0,-2,2,0$,
of the four topological phases at $p_x<-p_\p^W$, $-p_\p^W<p_x<0$, $0<p_x<p_\p^W$, and $p_\p^W<p_x$, respectively,
manifesting bulk-boundary correspondence (\secr{bbcW}).
The planes with $p_x=-p_\p^W,0,p_\p^W$ that cross the projected Weyl points represent the phase transitions between these phases.

The chirality of the effective edge-state spectra of the $p_z=p_z^W\pm 0.1 p_\p^W$ QAH systems [\figr{ssWy}(i),(k),(f)] is zero,
in accord with the zero Chern number $C(\Sig_{p_z})=0$ of the topological phases $p_z<-p_z^W$, $-p_z^W<p_z<p_z^W$, $p_z^W<p_z$.
We remind that for $p_z=\x{const}$ QAH systems,
although there are phase transitions at the planes $p_z=\pm p^W_z$ containing the $\pm_u$ groups of Weyl points,
there is no change of the Chern number,
since the sum of the Chern numbers of the Weyl points at $p_z=\pm p^W_z$ is zero: $1-1-1+1=0$.

Also, we have assigned orientation according to the source-and-drain picture
to all Fermi contours in the surface-momentum planes in all \figsdr{ssWy}{ssWypr+} for the Weyl semimetal.
As explained in \secr{bbcWFermicontours},
this allows one to deduce the chirality of effective edge-state spectrum along any path.
We observe that the so-determined chiralities are indeed in full accord with the bulk-boundary correspondence \eqn{bbcga}.

We can now also comment on the topological aspect of the surface-state spectrum of the Dirac and line-node semimetals.
These systems can be regarded as special cases of Weyl semimetals
and their surface states can also be analyzed in terms of quantum Hall topology and its bulk-boundary correspondence.
They are trivial Weyl semimetals with composite, net-neutral nodes,
since for any surface not crossing the nodes the effective 2D QAH system is trivial with zero Chern number.

Of particular interest are the circular paths enclosing one projected Dirac point [(d) and (h) of \figsdr{ssDy0}{ssDy17} and \figr{ssDylin}].
The Chern number of the Dirac point is zero, and in all six presented cases of $\alr_0$,
the surface-state spectrum along such circular paths is in accord with bulk-boundary correspondence.
Although the surface-state spectrum is qualitatively different in the two regimes $|\alr_0|<\tf12$ and $\tf12<|\alr_0|<1$,
it is topologically the same and has the same zero chirality:
for $|\alr_0|<\tf12$, there are two edge-state bands, one merging with the lower bulk-band boundary $E^D_{y<0,-_b}(p_x,p_z)$,
and the other with the upper bulk-band boundary $E^D_{y<0,+_b}(p_x,p_z)$;
for $\tf12<|\alr_0|<1$, there is one edge-state band not connected to the bulk-band boundaries.
Of course, the net chirality of the effective edge-state spectrum is also zero for any other path (like straight lines)
not crossing the projected Dirac points.

This topological property of the surface states is preserved by continuity
when the Dirac semimetal is transformed into the line-node semimetal and when the latter is in turn transformed into the Weyl semimetal.
For the circular paths enclosing the $+_u$ Dirac point of the Dirac semimetal [\figr{ssDy17}(d) and (h)],
for the circular paths enclosing the $+_u$ line node of the line-node semimetal [\figr{ssLNy}(d) and (h)],
and for the circular paths enclosing the whole $+_u$ group of projected Weyl points of the Weyl semimetal [\figr{ssWy}(d) and (h)],
the effective edge-state spectra are topologically equivalent (moreover, those of the line-node and Weyl semimetals are very close quantitatively),
their chirality is zero in accord with the zero total net Chern number of the enclosed singularities (nodal points or lines),
manifesting bulk-boundary correspondence.
This is because all three contours and systems can be continuously deformed to each other
without crossing the nodes (and therefore while the effective QAH systems remain gapped).

\subsubsection{Topological properties and emergent linear scaling near projected Weyl points}

Next, we consider circular paths enclosing {\em individual} projected Weyl points.
Circular paths of sufficiently small radii centered around the projected Weyl points
are particularly useful, as they provide a direct and complete characterization of the whole surface-state spectrum in their vicinity,
serving two purposes: capturing the local topological properties and the emergent linear scaling.
For each projected Weyl point $(p_x,p_z)=(- p_\p^W,p_z^W)$, $(0,p_z^W)$, and $(+p_\p^W,p_z^W)$ of the $+_u$ group,
we present the surface-state spectra along several such circles with decreasing radii $p_r$ in \figssr{ssWypr-}{ssWypr0}{ssWypr+}, respectively.

First, we explicitly see the bulk-boundary correspondence \eqn{Nga} manifest:
the signed number (chirality) of the effective edge-state bands connecting the lower and upper bulk-band boundaries
is indeed in full accord with the net Chern number of the projected Weyl points enclosed within the circle.
Although the spectra around the same point differ quantitatively for different radii, their topological properties remain the same.

Second, as the radius $p_r$ is decreased, the emergence of the momentum scale $p_{\be_3}$ [\eq{pbe3}] of the cubic BIA terms becomes evident.
Below this scale around the projected Weyl points, a universal linear scaling behavior of the surface-state spectrum emerges, e.g.,
\beq
	\Ec_{y<0}^W(p_\p^W+p_r\sin\zeta,p_z^W+p_r\cos\zeta)=\eps^W+\Vc_{y<0}^W(\zeta) p_r +o(p_r),
\lbl{eq:Eclinexp}
\eeq
for the projected Weyl point $(p_x,p_z)=(p_\p^W,p_z^W)$ and similar for the other points.
In this regime, the spectrum is fully characterized by its velocity $\Vc^W_{y<0}(\zeta)$ as a function of the angle $\zeta$.
Therefore, for small enough radius $p_r$, the spectrum along the circle essentially maps out the velocity $\Vc^W_{y<0}(\zeta)$.

In Ref.~\ocite{KharitonovFGCM}, the general surface-state structure in the vicinity of one projected Weyl point was calculated
for the most general form of the linear-in-momentum Hamiltonian and BCs.
The main general properties are: there is one surface-state band, which exist in a half-plane of the surface momentum;
the merging half-lines with the upper and lower $\pm_b$ bulk-band boundaries form one straight line.
Hence, for any model, the surface-state properties must asymptotically reduce to this behavior in the vicinity of a single projected Weyl point.
The linear regime can be best identified by the Fermi and merging contours,
which become asymptotically straight as the projected Weyl point is approached
(and also by the linear spectrum along the straight-line paths passing through the Weyl points).
We do observe this asymptotic behavior for the considered generalized Luttinger model of the Weyl semimetal
in the vicinity of the four single projected Weyl points $(\pm p_\p^W,\pm p_z^W)$ [\figsr{ssWypr-}{ssWypr+}].
For the $(-p_\p^W,p_z^W)$ point, the linear regime is reached at slightly less than $p_r=0.001 p_\p^W$ [\figr{ssWypr-}(g)].
For the $(p_\p^W,p_z^W)$ point, the linear regime is reached at around $p_r=0.01 p_\p^W$ [\figr{ssWypr+}(f)].
For the $(0,p_z^W)$ double projected Weyl point, the linear regime is reached earlier, already at $p_r=0.1 p_\p^W$. [\figr{ssWypr0}(d) and (e)].
For the double projected Weyl point, the general linear-in-momentum model has not yet been derived and studied,
but, of course, linear scaling is also anticipated in this case.

The emergence of the linear scaling proves that the vicinities of projected Weyl points could be described by a linear-in-momentum model,
which would be the last low-energy model in the hierarchy of \secr{scales}.
Although we do not derive these models from the generalized Luttinger model here,
the clearly observed linear scaling proves that they exist and could be derived.

We also point out the following. Although the surface-state spectrum along the $p_x=0$ straight-line path
is qualitatively the same for the Dirac [\figr{ssDy17}(b)] and Weyl [\figr{ssWy}(b) and \figr{ssWypx}(g)] semimetals,
the full spectra in the vicinity of the projected Dirac point $(p_x,p_z)=(0,p_{u0})$
and the projected double Weyl point $(p_x,p_z)=(0,p_z^W)$ are topologically and qualitatively completely different,
as the spectra along the circular paths show:
for the Dirac point with the $C^W=0$ Chern number, the two branches at $p_x=0$ are cuts of the same topologically trivial band at $\zeta=0,\pi$
[\figr{ssDy17}(d) and \figr{ssDylin}];
for the projected Weyl point with the $C^W=+2$ Chern number,
the two branches are cuts of two different chiral subbands [\figr{ssWypr0}(a),(b),(c)].
Therefore, it is important to realize that single straight-line paths alone do not provide a complete picture of the surface-state spectrum
and such similarities can be misleading.
The plots along circular paths of sufficiently small radii, on the other hand, fully characterize the surface-state spectrum in the vicinity of the nodes.

\subsection{$z>0$ sample \lbl{sec:ssWz}}

\begin{figure*}
\centering
\includegraphics[width=\linewidth]{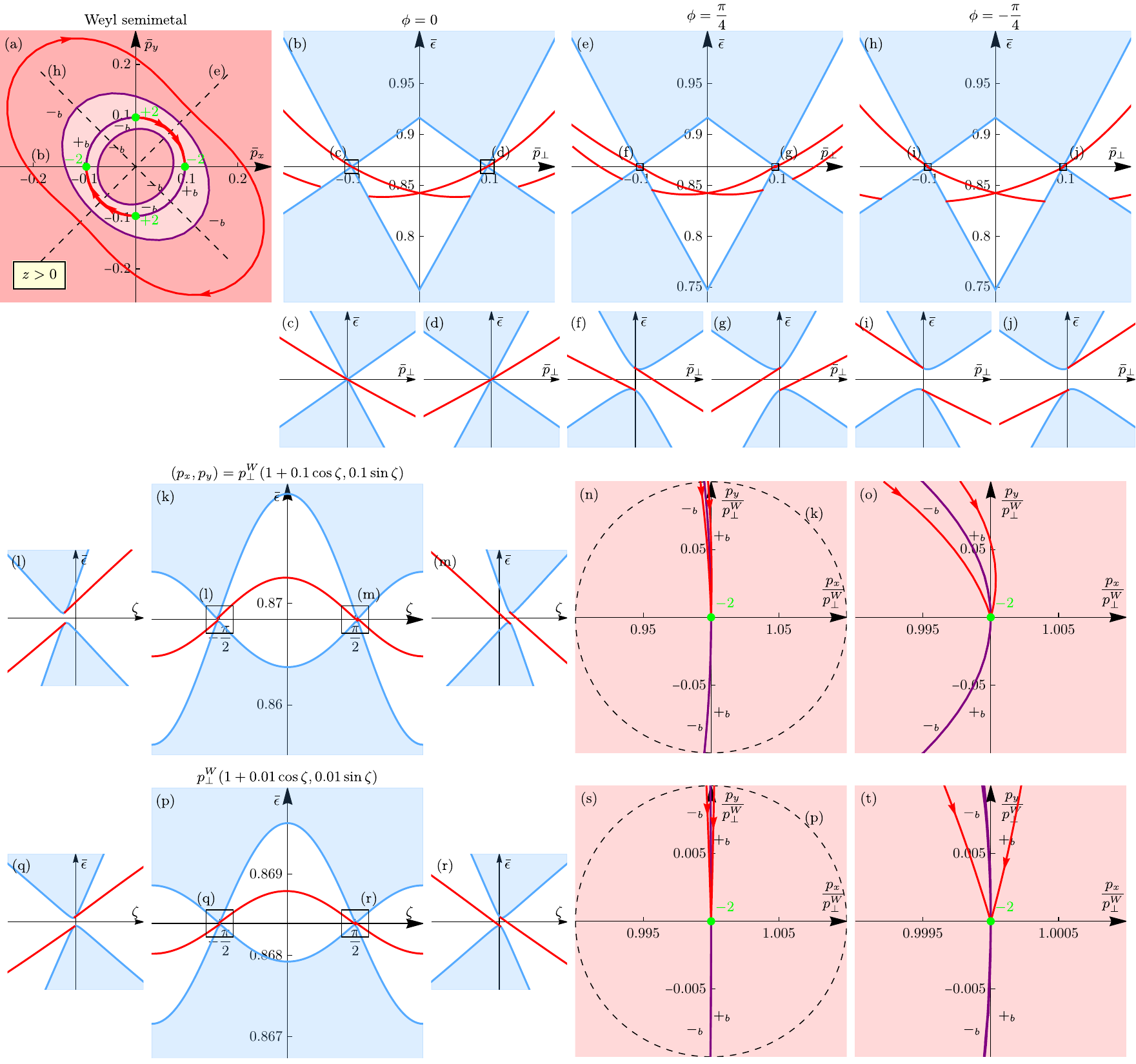}
\caption{
The surface-state spectrum $\Ec^W_{z>0}(p_x,p_y)$ of the Weyl semimetal for the $z>0$ sample.
Other conditions same as in \figr{ssLNy}.
(a) Surface-momentum plane.
(b),(e),(h) The spectrum along the $\phi=0,+\f\pi4,-\f\pi4$ paths, respectively,
with the insets around the former projected line node.
(k),(l),(m),(p),(q),(r)
The spectrum along the circular paths around the projected Weyl point $(p_x,p_y)=(p_\p^W,0)$
and (n),(o),(s),(t) the surface momentum plane at the same scale.
In (o) and (t), the aspect ratio differs from 1 to resolve the features along the $p_z$ direction.
}
\lbl{fig:ssWz}
\end{figure*}

The surface-state spectrum $\Ec^W_{z>0}(p_x,p_y)$ of the Weyl semimetal for the $z>0$ sample is presented in \figr{ssWz}.
As a consequence of the bulk band structure explained in \secr{topoW},
the Weyl points with equal Chern numbers are projected onto each other in pairs in the surface-momentum plane $(p_x,p_y)$.
This leads to four projected Weyl points
$(p_x,p_y)=(\pm p_\p^W,0)$ and $(0,\pm p_\p^W)$ [\figr{ssWz}(a)] with the net Chern numbers $C^W=\mp 2$, respectively.
Away from the region of width $p_{\be_3}$ around the former projected line node,
the spectrum is very similar to that of the line-node semimetal, compare to \figr{ssLNz}.

The same main changes upon introducing the cubic BIA terms and transforming the line-node semimetal to the Weyl semimetal
occur in the surface-state spectrum for the $z>0$ sample as for the $y<0$.
As demonstrated in \secr{ssLN}, the upper band of the surface-state spectrum $\Ec^{LN}_{z>0}(p_x,p_y)$ of the line-node semimetal
merges with the projected line node from both the inner $p_\p<p_\p^{LN}(\phi)$ and outer $p_\p^{LN}(\phi)<p_\p$ regions.
Upon introducing the cubic BIA terms and opening the gap in the bulk spectrum in the four angular sectors of $\phi$,
the two sides sides of the upper surface-state band adjusts, so that in the Weyl semimetal
the upper band of the surface-state spectrum $\Ec^W_{z>0}(p_x,p_y)$
remains connected to either the upper or the lower bulk-band boundary $E^W_{z>0,\pm_b}(p_x,p_y)$.
This happens differently in different angular sectors between the projected Weyl points.
In the sectors $0<\phi<\f\pi2$ and $\pi<\phi<\f{3\pi}2$ [\figr{ssWz}(e),(f), and (g)], the inner band growing with $p_\p$
merges with the $+_b$ upper bulk-band boundary and the growing outer band merges with the $-_b$ lower bulk-band boundary.
As a result, in these sectors, these parts of the surface-state band cross the gaps,
each contributing $+1$ chirality for any path oriented with growing $p_\p$.
In the remaining sectors $\f{\pi}2<\phi<\pi$ and $\f{3\pi}2<\phi<2\pi$ [\figr{ssWz}(h),(i), and (j)], the situation is reversed:
the inner band growing with $p_\p$ merges with the $-_b$ lower bulk-band boundary
and the growing outer band merges with the $+_b$ lower bulk-band boundary.
As a result, in these sectors, these parts of the surface-state band do not cross the gap and contribution nothing to chirality.
These properties are also manifested in the Fermi and merging contours [\figr{ssWz}(a),(n),(o),(s),(t)].
From each projected Weyl point $(p_x,p_y)=(0,\pm p_\p^W)$ with $C^W=+2$,
two inner Fermi contours originate and terminate at the projected Weyl point $(p_x,p_y)=(\pm p_\p^W, 0)$ with $C^W=-2$, respectively.

The net chirality of the effective edge-state bands can be deduced from this for any path
and will be equal to the Chern number of the effective QAH system, manifesting the bulk-boundary correspondence.
In particular, for straight-line paths passing through $\pb_\p=\nv$, with $\phi=\x{const} \neq0,\pm\f\pi2,\pi$
the chirality is always zero, in accord with the zero Chern number, but there are two different realizations.
For $0<\phi<\f{\pi}2$, there is a cancellation of $+2$ and $-2$ contributions.
For $\f{3\pi}2<\phi<2\pi$, there are no edge-state bands crossing the gap.
The zero chirality and Chern number for any $\phi=\x{const}\neq 0,\pm\f\pi2,\pi$
path can also be explained in terms of $\Tc_-$ time-reversal symmetry, see \secr{Tc-}.

In \figr{ssWz}(k),(l),(m),(p),(q),(r),
we also present the spectrum along the circular paths enclosing individual projected Weyl points for the radii $p_r=0.1p_\p^W, 0.01p_\p^W$.
The bulk-boundary correspondence is manifested.
From the plots \figr{ssWz} (n),(o),(s),(t) of the surface-momentum plane,
linear scaling, analogous to \eq{Eclinexp}, is seen emerging at about $p_r=0.001 p_\p^W$
as the Fermi and merging contours become asymptotically straight as the projected Weyl point is approached.

\section{Others aspects \lbl{sec:other}}

\subsection{Effective 2D $\Tc_-$ insulators \lbl{sec:Tc-}}

Just like any surface (in particular, any plane) in the 3D momentum space of a Weyl semimetal can be regarded as an effective 2D QAH system (class A),
for present $\Tc_-$ symmetry (which is the case for the considered model), any $\Tc_-$-invariant surface
can be regarded as an effective 2D $\Tc_-$-symmetric topological insulator (if no Weyl points are crossed) (class AII).
Since the BC \eqn{bc} is $\Tc_-$-symmetric,
the corresponding effective edge-state spectrum satisfies the class-AII bulk-boundary correspondence.

Any plane passing through $\pb=\nv$ is $\Tc_-$-symmetric.
For the $y<0$ sample, relevant $\Tc_-$-symmetric planes are those that project onto the lines $(p_x,p_z)=p_r(\sin\zeta,\cos\zeta)$ with $\zeta=\x{const}$.
There are three phases as a function of $\zeta$.
The effective edge-state spectrum,
which for $|\zeta|\ll 1$ can be deduced from the $p_x=\x{const}$ paths around $p_z=p_z^W$ in \figr{ssWypx} and symmetry,
is trivial in terms of class-AII topology in all three phases, although differs qualitatively.
For the $z>0$ sample, relevant $\Tc_-$-symmetric planes are those that project onto the lines $(p_x,p_y)=p_\p(\cos\phi,\sin\phi)$ with $\phi=\x{const}$.
There are two phases: $\phi\in(-\f\pi2,0)$ and $\phi\in(0,\f\pi2)$.
The effective edge-state spectrum is nontrivial in terms of class-AII topology in both phases [\figr{ssWz}(e) and (h)],
although differs qualitatively.

\subsection{Relation between different surfaces \lbl{sec:diffsurfaces}}

\begin{figure}
\centering
\includegraphics[width=.40\linewidth]{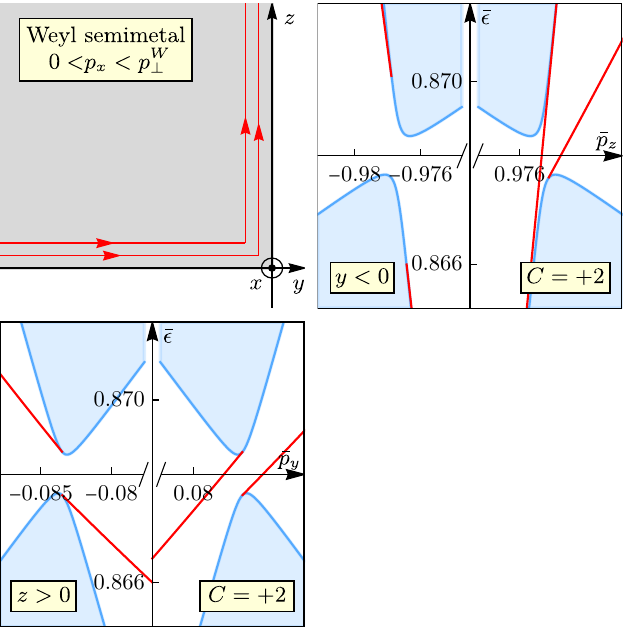}
\caption{
Relation between the surface-state spectra of the Weyl semimetal for different surface orientations.
For an effective 2D QAH system, specified by a plane in 3D momentum space of the Weyl semimetal, exemplified here with $p_x=p_\p^W/2$ plane of the $0<p_x<p_\p^W$ topological phase, the effective edge-state spectra
$\Ec^W_{y<0}(p_x,p_z)$ for the $y<0$ sample and $\Ec^W_{z>0}(p_x,p_y)$ for the $z>0$ sample
have the same chirality $N=2$, equal to the Chern number $C(\Sig_{p_x)})=2$ of the plane,
in accord with the bulk-boundary correspondence \eqn{NPx}.
This can be understood from the visualization of the chiral quantum Hall edge states following the edge of one sample of any shape,
such as a sample occupying the quadrant region specified by both $y<0$ and $z>0$ conditions.
Other conditions are the same as in \figr{ssWy}.
}
\label{fig:ssWpx}
\end{figure}

\begin{figure*}
\centering
\includegraphics[width=\linewidth]{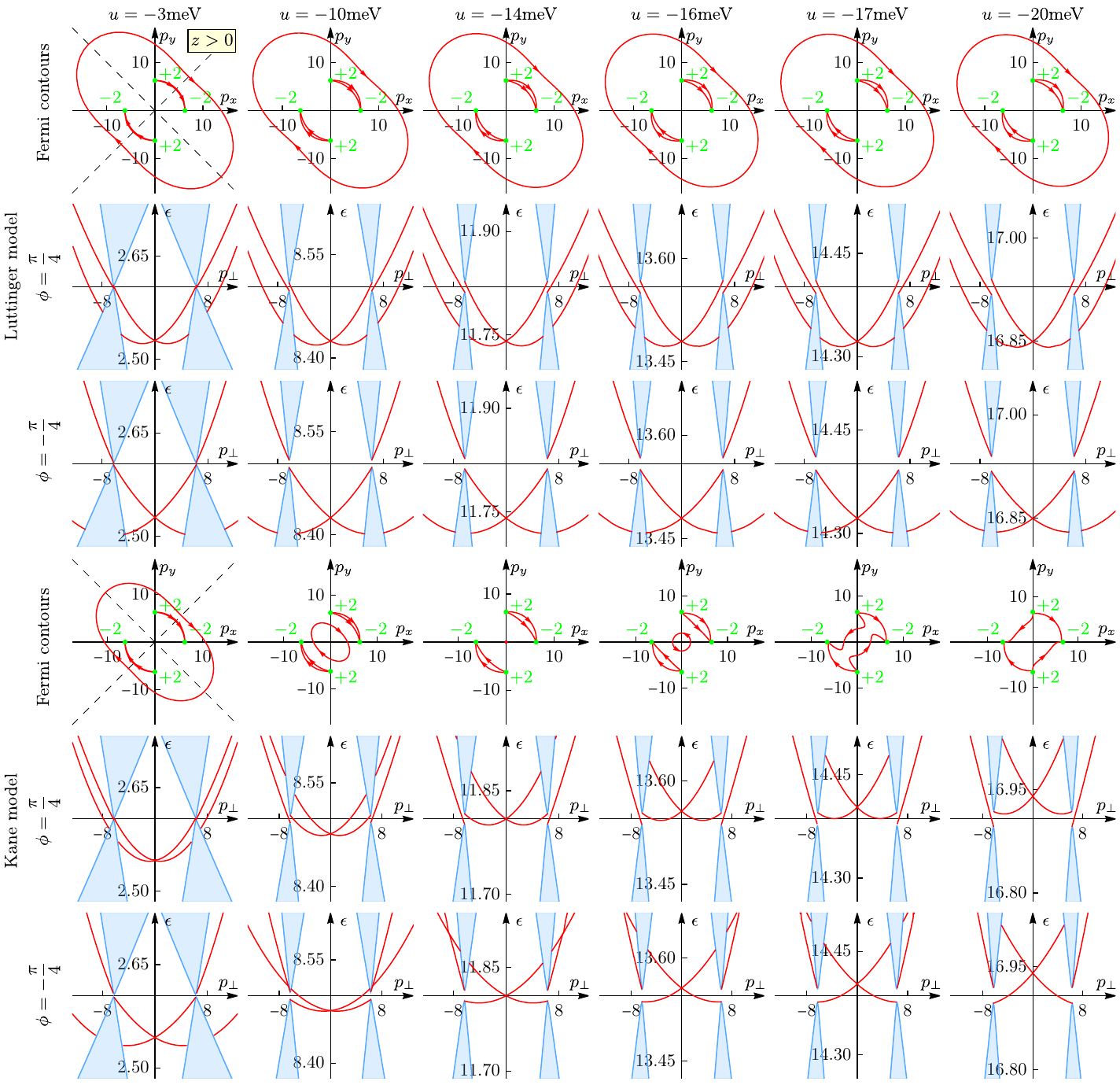}
\caption{
Comparison of the surface-state spectra of the Weyl semimetal calculated within the generalized Luttinger (first three rows)
and Kane (last three rows) models,
for the parameters of HgTe (\tabsr{LM}{KM}) and for different values of compressive strain $u$.
For each model, the first row presents the Fermi contours of the surface-state band at the Weyl-point energy $\eps^W$,
the second and third rows present the spectrum along the $\phi=\pm\f{\pi}4$ paths, respectively.
For small enough strain $u$ (like $u=-3\x{meV}$),
there is qualitative and good quantitative agreement at momenta on order $p_\p\sim p_{\be_1}$.
For larger strain, outside of the validity range of the Luttinger model,
the spectrum of the Kane model undergoes qualitative transformations,
while the spectrum of the Luttinger model remains qualitatively the same.
The units of momentum and energy in these plots are $\mu \x{m}^{-1}$ and $\x{meV}$.
}
\lbl{fig:WLMKM}
\end{figure*}

Although considering effective 2D QAH systems defined on cylinders in the 3D momentum space of a Weyl semimetal
is particularly useful for characterizing the vicinity of the Weyl points, this construction is tailored to a specific surface orientation.
Considering QAH systems defined on planes, on the other hand,
has the advantage that different surfaces become different effective edges of the same QAH system,
the topological properties of the edge-state spectra of which are related.

For instance, for $p_x=\x{const}$ planes, the surface-state spectra $\Ec_{y<0}^W(p_x,p_z)$ and $\Ec_{z>0}^W(p_x,p_y)$
can be regarded as effective edge-state spectra at fixed $p_x=\x{const}$ as functions of $p_z$ and $p_y$, respectively.
The bulk-boundary correspondence [\eq{NPx}] has to hold for any edge of a given QAH system;
hence, the chiralities of these edge-state spectra have to be the same and equal to the bulk Chern number $C(\Sig_{p_x})$.
This can be understood from the chiral properties of the edge states of a QAH system (\figr{ssWpx})
with an edge consisting of straight-line sections with different orientations:
due to their chiral nature, the edge states cannot terminate and persist with preserved chirality into the next section.

We illustrate this in \figr{ssWpx} for the effective QAH system defined on the $p_x=p_\p^W/2$ plane,
which exemplifies the $0<p_x<p_\p^W$ phase with the Chern number $C(\Sig_{p_x})=+2$.
The surface states for $y<0$ and $z>0$ samples in this case have already been presented in \figsr{ssWy}{ssWz}.
According to the considerations in \secsr{ssWy}{ssWz}, the net chirality of the effective edge-state spectra
is $N=+2$ for both surfaces, as seen in \figr{ssWpx}.
Note that it is important here that both $(-y,x,z)$ and $(z,x,y)$ form right-handed systems,
which is why the chiralities of the spectra for $y<0$ and $z>0$ samples are the same (and not opposite).

\subsection{Comparison between the Luttinger and Kane models \lbl{sec:WLMKM}}

In this section, we perform the comparison of the surface-state spectra calculated within the generalized
Luttinger and Kane models in the Weyl semimetal phase, i.e., with all perturbations present: strain and linear and cubic BIA terms.
As explained in \secsr{KM}{scales}, the Luttinger model is the rigorous low-energy limit of the Kane model,
when all the relevant energy scales are much smaller than the level spacing $\De$ between the $j=\f32$ and $j=\f12$ states.
For the hierarchy of scales we consider, the largest scale in the Luttinger model is the strain $|u|$;
therefore, agreement is expected when the condition $|u|\ll |\De|$ is satisfied.
We now explore what this means quantitatively by comparing the surface-state spectra for various values of $|u|$
while $\De$ and other parameters remains fixed and equal to those of HgTe (\tabsr{LM}{KM}).

We calculate the spectra for the $z>0$ sample along the characteristic $\phi=\pm\f\pi4$ paths and their Fermi contours, presented in \figr{WLMKM}.

We notice that for the bulk spectrum at the momentum scale $p_{\be_1}$ [\eq{pbe1}] of the linear BIA term,
there is no appreciable quantitative difference between the Luttinger and Kane models
for the presented values of strain and there are no qualitative differences as the strain is changed:
the values $p_\p^W$ of the Weyl points remain almost the same (they are determined mainly by the $p_{\be_1}$ scale, which does not depend on strain),
and only the energy position and splitting of the bulk bands have some dependence on $|u|$.

Further, there are also no qualitative changes in the surface-state spectrum of the Luttinger model upon changing strain.
This is understandable, since there is no larger scale in the Luttinger model that could compete with strain.

At the smallest presented value $u=-3\x{meV}$ (which is the value used throughout for the line-node and Weyl semimetals),
the surface-state spectra of the Kane and Luttinger models agree qualitatively and reasonably well quantitatively,
confirming that at such value the Luttinger model (with the values of other parameters those of HgTe) is applicable.
(Upon decreasing the strain value, quantitative agreement would only improve.)
The agreement is best seen in the Fermi contours:
there is a closed Fermi contour in the outer region
and two ``double'' Fermi contours connected to the pairs of the projected double Weyl points.

However, we observe that, upon increasing the strain magnitude $|u|$ and thus making it more comparable to the level spacing $|\De|$,
the surface-state spectrum of the Kane model undergoes not just quantitative, but qualitative transformations.
The main ``driver'' of these changes is the surface-state energy $\Ec^{W,K}_{z>0,\pm}(\nv)$
at the double-degeneracy point $\pb_\p=\nv$ moving upwards
relative to the bulk-band boundaries upon increasing $|u|$; this effect does not happen in the Luttinger model.
Upon this evolution, first,
the connectivity to the Weyl points switches from the lower to the upper surface-state band between $u=-3\x{meV}$ and $u=-10\x{meV}$.
The energy of the $\pb_\p=\nv$ point still remains below the Weyl-point (Fermi) level $\eps^{W,K}$.
The closed Fermi contour shrinks and passes through the (former) line-node region.
For smaller strain magnitude $|u|$,
this Fermi contour in the outer region is due to the lower surface-state band $\Ec^{W,K}_{z>0,-}(\pb_\p)$ crossing the Fermi level
and for larger $|u|$, this Fermi contour in the inner region is due to the upper surface-state band $\Ec^{W,K}_{z>0,+}(\pb_\p)$.
Further, at the strain value very close to $u=-14\x{meV}$, the surface-state energy $\Ec^{W,K}_{z>0,\pm}(\nv)$ crosses the Fermi level $\eps^{W,K}$,
where the Fermi contour shrinks into a point.
At larger strain magnitude, $u=-16,-17,-20\x{meV}$,
the closed Fermi contour in the inner region is again due to the crossing of the lower surface-state band.
The qualitative difference in the Fermi contours at $u=-16\x{meV}$ and $u=-17,-20\x{meV}$
is whether along the $\phi=\f{\pi}4$ path (and in a range of $\phi$ around it)
the lower surface-state band crosses the Fermi level twice
or not at all, respectively, whether it has a minimum in $p_\p$ below or above the Fermi level.

\section{Conclusion \lbl{sec:conclusion}}

We have presented the theoretical analysis of the surface states of the Luttinger semimetal in the presence of several perturbations:
compressive strain and linear- and cubic-in-momentum bulk-inversion-asymmetry (BIA) terms.
There are several physical, technical, and methodological points presented in this work, which we summarize and emphasize in this conclusion.

It has been demonstrated that the Luttinger semimetal without perturbations can indeed be regarded as the parent, highest-symmetry system
for multiple nontrivial phases.
For the considered perturbations, the system has a peculiar structure,
where each successive perturbation creates a new semimetal phase via the modification of the nodal structure of the previous phase,
resulting in a sequence of four semimetal phases: the parent Luttinger semimetal and Dirac, line-node, and Weyl semimetals.

Calculation of the surface states of these semimetal phases has been the main practical goal of this work.
Since already the unperturbed Luttinger semimetal exhibits surface states~\cite{KharitonovLSM},
the surface states of the new phases can be regarded as their evolution.

Among the theoretical technical points,
we have shown that the semi-analytical method of calculating surface states within the continuum models with BCs
allows for calculations for a true half-infinite system with any desired resolution.
This allowed us to capture very fine features of the surface-state spectrum, e.g., in the vicinity of the nodes.
We have also used an efficient and insightful method of characterizing surface-state spectra through a set of 2D plots,
which consists of the plot of the surface-momentum plane with Fermi and merging contours and shaded regions occupied by the bands
and 2D plots of the surface-state spectrum along a set of paths:
key straight-line paths and circular paths around the projected point nodes.
Circular paths provide a full characterization of the vicinity of the projected point nodes,
including the demonstration of bulk-boundary correspondence.
This set of 2D plots provides a comprehensive description of the whole 3D plot of the surface-state spectrum as a function
of the 2D surface momentum.

Among the interesting physical features of the system, for the considered Weyl semimetal,
we observed multiple situations where the effective edge-states bands of different QAH systems with the same Chern number
(either within one gapped phase or in different phases with the same Chern number),
while having the same chirality and being topologically equivalent,
nonetheless differ quite a bit qualitatively
(such as the situations of absent edge states crossing the bulk gap and present edge states with cancelled chiralities for the zero Chern number).

The main theoretical methodological point has been the application of continuum models with BCs for the study of surface states.
In accord with the hierarchy of the perturbations and semimetal phases they create,
there is a hierarchy of low-energy continuum models describing the vicinities of the nodal structures.
We derive most of these models and demonstrate quantitative asymptotic agreement between some of them,
which explicitly proves that continuum models with proper BCs are perfectly applicable for the study of surface states.

\begin{acknowledgments}
M. K. acknowledges the financial support by the Grant No. KH 461/1-1 of the Deutsche Forschungsgemeinschaft (DFG, German Research Foundation).
J.-B. M. and E. M. H. acknowledge funding by the DFG through SFB 1170, Project-ID 258499086,
through the W\"urzburg-Dresden Cluster of Excellence on Complexity and Topology in Quantum Matter ct.qmat (EXC2147, Project-ID 390858490),
as well as by the ENB Graduate School on Topological Insulators.
\end{acknowledgments}

\appendix
\section{Linear-in-momentum model at the line node \lbl{app:appendix}}
Here we present additional formulas of relevance to \secsr{bulkLN}{bulkW}.

The Hamiltonian of the line-node semimetal in the basis \eqn{Uphi} reads
\[
	\Hh'^{LN}(\pb)=\Uh^\dg_\phi\Hh^{LN}(\pb)\Uh_\phi=
	\lt(\ba{cc}
		\Hh_a(\pb) & \Hh_\p(\pb) \\
		\Hh_\p^\dg(\pb) & \Hh_b(\pb) \ea\rt)
	+\Hh'_\square(\pb),
\]
\[
	\Hh_a(\pb)=\lt(\ba{cc}
	(\al_0-\tf12\al_z)p_\p^2+(\al_0+\al_z)p_z^2+\tf32\be_1 p_\p+u & \sq3(\al_z p_\p-\be_1)(p_z+\tf12 i p_\p\sin2\phi) \\
	\sq3(\al_z p_\p-\be_1)(p_z- \tf12 i p_\p\sin2\phi) & (\al_0+\tf12\al_z)p_\p^2+(\al_0-\al_z)p_z^2+\tf32\be_1 p_\p-u \ea\rt),
\]
\[
	\Hh_b(\pb)=\lt(\ba{cc}
		(\al_0-\tf12\al_z)p_\p^2+(\al_0+\al_z)p_z^2-\tf32\be_1 p_\p+u & \sq3(\al_z p_\p+\be_1)(p_z-\tf12i p_\p\sin2\phi)\\
		\sq3(\al_z p_\p+\be_1)(p_z + \tf12i p_\p\sin2\phi) & (\al_0+\tf12\al_z)p_\p^2+(\al_0-\al_z)p_z^2-\tf32\be_1 p_\p-u
	\ea\rt),
\]
\[
	\Hh_\p(\pb)=\lt(\ba{cc}	0 & \tf{\sq3}2(\al_z p_\p+\be_1)p_\p\cos2\phi \\
		\tf{\sq3}2(-\al_z p_\p+\be_1)p_\p\cos2\phi & 0
	\ea\rt),
\]
\[
	\Hh'_\square(\pb)=\al_\square\lt(\ba{cccc}
	\tf35(\tf12p_\p^2-p_z^2) & \f{2\sq3}5p_\p(p_z+\tf12ip_\p\sin2\phi) & 0 & -\tf{3\sq3}5 \tf12p_\p^2\cos2\phi\\
	\f{2\sq3}5p_\p(p_z-\tf12ip_\p\sin2\phi) & -\tf35(\tf12p_\p^2-p_z^2) & \tf{3\sq3}5 \tf12p_\p^2\cos2\phi & 0\\
	0 & \tf{3\sq3}5 \tf12p_\p^2\cos2\phi & \tf35(\tf12p_\p^2-p_z^2) & \f{2\sq3}5p_\p(p_z-\tf12ip_\p\sin2\phi)\\
	-\tf{3\sq3}5 \tf12p_\p^2\cos2\phi & 0 & \f{2\sq3}5p_\p(p_z+\tf12ip_\p\sin2\phi) & -\tf35(\tf12p_\p^2-p_z^2)
	\ea\rt).
\]

Without cubic anisotropy ($\al_\square=0$),
we observe that at $p_\p=\be_1/\al_z$ the second state decouples from the other three at any $\phi$
and at $p_\p=-\be_1/\al_z$ the fourth state decouples from the other three.
Depending on the sign of $\be_1/\al_z$, the crossing of the band of one of these states with one of the bands
of the other three states forms the line node.

At $\phi=\f\pi4$, there is exact block decoupling (due to reflection symmetry in $\Db_{2d}$)
of the Hamiltonian
\[
	\Hh'^W(\pb)|_{\phi=\f\pi4}
	=\Uh^\dg_\phi\Hh^W(\pb)\Uh_\phi|_{\phi=\f\pi4}
	=\lt(\ba{cc} \Hh_a^W(p_\p,p_z) & \nm \\
		\nm & \Hh_b^W(p_\p,p_z) \ea\rt)
\]
for the Weyl semimetal, with the cubic anisotropy and cubic BIA terms.
Presenting the first block in terms of the Pauli matrices,
\[
	\Hh_a^W(p_\p,p_z)=\sum_{\al=0,x,y,z} \tauh_\al [h_{a\al}^{LN}(p_\p,p_z)+h_{a\al}^3(p_\p,p_z)],
\]
the contribution of the line-node semimetal with cubic anisotropy (all terms except for the cubic BIA terms) reads
\beqar
	h_{a0}^{LN}(p_\p,p_z)&=&\al_0(p_\p^2+p_z^2)+\tf32\be_1 p_\p \approx \eps^{LN}(\tf\pi4)+ 2\al_0(p_{\p\f\pi4} q_\p +p_{z\f\pi4} q_z)
	=\eps^{LN}(\tf\pi4)+ v_{0\p}q_\p +v_{0z}q_z,
\\
	h_{ax}^{LN}(p_\p,p_z)&=&\sq3((\al_z+\tf25\al_\square) p_\p-\be_1)p_z \approx \sq3 (\al_z+\tf25\al_\square) p_{z\f\pi4} q_\p =v_{x\p} q_\p,
\\
	h_{ay}^{LN}(p_\p,p_z)&=&
	-\sq3((\al_z+\tf25\al_\square) p_\p-\be_1)\tf12p_\p \approx \sq3 (\al_z+\tf25\al_\square) \tf12 p_{\p\f\pi4} q_\p = v_{y\p} q_\p,
\\
	h_{az}^{LN}(p_\p,p_z)&=&(\al_z-\tf35\al_\square)(-\tf12p_\p^2 +p_z^2)+u \approx 2(\al_z-\tf35\al_\square)(-\tf12p_{\p\f\pi4} q_\p +p_{z\f\pi4} q_z)
	=v_{z\p}q_\p+v_{zz}q_z,
\lbl{eq:haLN}
\eeqar
and the contribution of the cubic BIA terms reads
\beqar
	h_{a0}^3(p_\p,p_z)&=&\tf12p_\p[(\be_{31}+\tf74\be_{32}+3\be_{33}+3\be_{34})\tf12p_\p^2-(\be_{31}+\tf74\be_{32}-3\be_{33})p_z^2],
\\
	h_{ax}^3(p_\p,p_z)&=&-\sq3 p_z(\be_{33}p_\p^2+\be_{34}p_z^2),
\\
	h_{ay}^3(p_\p,p_z)&=&\tf{\sq3}2p_\p[-\tf12p_\p^2(\be_{31}+\tf74\be_{32}-\be_{33}-\be_{34})+(\be_{31}+\tf74\be_{32}+\be_{33})p_z^2],
\\
	h_{az}^3(p_\p,p_z)&=&\tf12p_\p(\be_{31}+\tf{13}4\be_{32})(\tf12p_\p^2-p_z^2).
\lbl{eq:ha3}
\eeqar

Assuming $\be_1/\al_z>0$, the terms $h^{LN}_{a\al}(p_{\p\f\pi4},p_{z\f\pi4})=0$, $\al=x,y,z$ vanish
at $p_\p=p_{\p\f\pi4}$ and $p_z=p_{z\f\pi4}$ [\eq{pLN}], which signifies the node of the line-node semimetal.
The linear-in-momentum model [\eqss{HcLN2init}{HcLN2}{HcW2}] is obtained by expanding in momentum about this point.
To leading order, the line-node-semimetal part is expanded in momentum deviation $(q_\p,q_\phi,q_z)$,
as shown in \eq{haLN}, which gives the linear-in-momentum terms,
(importantly, we also check that there is no linear in $q_\phi$ contribution when the deviation from the $\phi=\f\pi4$ plane is taken into account),
while the cubic BIA terms \eqn{ha3} are taken at the line node point to produce the energy parameters
\[
	\eps_\al=h_{a\al}^3(p_{\p\f\pi4},p_{z\f\pi4}),\spc \al=0,x,y,z.
\]
The energy parameter arising in \eq{HcW2} upon the basis change reads
\[
	\eps_x'=\f{\eps_x v_{x\p}+\eps_y v_{y\p}}{\sq{v_{x\p}^2+v_{y\p}^2}}.
\]

\end{document}